\documentclass[aps,prd,eqsecnum,epsf,nofootinbib]{revtex4}
\usepackage{amsfonts,latexsym,amsmath,amssymb,times,mathrsfs}
\usepackage[dvips]{graphicx}
\usepackage[usenames]{color}

\newcommand{\bm}[1]{\hbox{\boldmath{$#1$}}}
\newcommand{\sbm}[1]{\hbox{\boldmath{\scriptsize$#1$}}}
\newcommand{\Mp}{M_{\rm pl}}
\newcommand{\dd}{{\rm d}}
\newcommand{\gR}{{^g\!R}}
\newcommand{\sR}{{^s\!R}}
\newcommand{\gz}{{^g\!\zeta}}
\newcommand{\ggm}{\delta {^g\!\gamma}}

\newcommand{\Approx}{\stackrel{\rm  IR}{\approx}}
\newcommand{\Gbz}{{^g\!\bar{\zeta}}}
\newcommand{\Gbg}{\delta {^g\!\bar{\gamma}}}
\newcommand{\dGbz}{{^g\!\dot{\bar{\zeta}}}}
\newcommand{\dGbg}{\delta {^g\!\dot{\bar{\gamma}}}}
\newcommand{\tgij}{\delta \tilde{\gamma}_{ij\,I}}

\newcommand{\cL}{{\cal L}}
\newcommand{\cLs}{{\cal L}_s}
\newcommand{\cLsI}{{\cal L}_s^{-1}}

\newcommand{\cP}{{\cal P}}
\newcommand{\cPd}{{\cal P}^{(d)}}

\newcommand{\cLRI}{{\cal L}_R^{-1}}

\newcommand{\cLRsI}{{\cal L}_{R, s}^{-1}}

\newcommand{\cLRtI}{{\cal L}_{R, t}^{-1}}

\newcommand{\cLRaI}{{\cal L}_{R, \alpha}^{-1}}
\newcommand{\cLRkI}{{\cal L}_{R, \sbm{k}}^{-1}}

\newcommand{\s}{S}

\newcommand{\xgl}{x_{\rm{}^{_{gl}}}}
\newcommand{\gl}{{\rm{}^{_{gl}}}}
\newcommand{\xglp}{x}
\newcommand{\bxgl}{\bm{x}_{\rm{}^{_{gl}}}}
\newcommand{\bygl}{\bm{y}_{\rm{}^{_{gl}}}}
\newcommand{\bxglp}{\bm{x}}
\newcommand{\gauge}{{\em gauge} }
\newcommand{\PSH}{\zeta^c,\delta\gamma^c}

\newcommand{\PSHIt}[1]{t_{#1}; \PSH}
\newcommand{\eSH}{\left| \PSH\, \right\rangle}
\newcommand{\eSHI}{\left|{t; \PSH}\, \right\rangle\!_I}

\newcommand{\eSHIt}[1]{\left|\PSHIt{#1}\, \right\rangle\!_I}
\newcommand{\ddP}{{\cal D} \zeta^c\, {\cal D} \delta\gamma^c\,}

\newcommand{\cDsx}{{\cal D}^s_{{\raisebox{-2pt}{${}^{\! x}$}}}} 
\newcommand{\cDsxj}[1]{{\cal D}^s_{{\raisebox{-2pt}{${}^{\! x_{#1}}$}}}} 
\newcommand{\cDtx}[1]{{\cal D}^t_{{\raisebox{-2pt}{${}^{\! x}$}}#1}} 
\newcommand{\cDtxj}[2]{{\cal D}^t_{{\raisebox{-2pt}{${}^{\!
x_{#1}}$}}#2}} 
\newcommand{\ev}{{({\rm ev})}}

\begin{document}

\thispagestyle{empty}


\title{Strong restriction on inflationary vacua from the local \gauge
invariance III\\
: Infrared regularity of graviton loops}
\date{\today}
\author{Takahiro Tanaka$^{1}$}
\email{tanaka_at_yukawa.kyoto-u.ac.jp}
\author{Yuko Urakawa$^{2}$}
\email{yurakawa_at_ffn.ub.es}
\affiliation{\,\\ \,\\
$^{1}$ Yukawa Institute for Theoretical Physics, Kyoto university,
  Kyoto, 606-8502, Japan\\
$^{2}$ Departament de F{\'\i}sica Fonamental i Institut de Ci{\`e}ncies del Cosmos, 
Universitat de Barcelona,
Mart{\'\i}\ i Franqu{\`e}s 1, 08028 Barcelona, Spain}

\preprint{200*-**-**, WU-AP/***/**, hep-th/*******}


\begin{abstract}
 It has been claimed that the super Hubble modes of the graviton generated during inflation can make loop corrections
 diverge. Even if we introduce an infrared (IR) cutoff at a comoving
 scale as an {\it ad hoc} but a practical way for the regularization, we encounter the
 secular growth, which may lead to the breakdown of perturbative
 expansion for a sufficiently long lasting inflation. In this paper, we
 show that the IR pathology concerning the graviton can be attributed to
 the presence of residual \gauge degrees of freedom in the local observable universe
 as in the case of the adiabatic curvature perturbation. We will show
 that choosing the Euclidean vacuum as the initial state ensures the
 invariance under the above-mentioned residual \gauge
 transformations. We will also show that as long as we consider a \gauge
 invariant quantity in the local universe, we encounter neither the IR
 divergence nor the secular growth. The argument in this paper applies
 to general single field models of inflation up to a sufficiently high
 order in perturbation. 
\end{abstract}


\maketitle
\section{Introduction} 
The inflationary spacetime leads to the generation of gravitational
waves. Even though the amplitude of gravitational waves is
smaller than the amplitude of the adiabatic curvature perturbation, 
the detection of the primordial
gravitational waves generated during inflation is expected to be within
our reach. The measurement of the primordial gravitational waves is 
crucially important to
uncover the model of inflation, providing information
which cannot be obtained through the measurement of the
scalar perturbations. In particular, the amplitude of the
gravitational waves can directly measure the energy scale of inflation
unlike the amplitude of the adiabatic curvature perturbation, which is
sensitive also to
the detailed dynamics of the inflationary universe. We can find an
example which highlights the importance of measuring
the primordial gravitational waves also for non-linear perturbations, say in Refs.~\cite{Maldacena:2011nz,
Soda:2011am, Shiraishi:2011st}, which elucidated the impact of parity
violation in gravity sector on the bi-spectrum of the primordial
gravitational waves. Thus, it will be profitable to study the method to
predict the primordial gravitational waves also in the presence of the
non-linear interaction. In this paper, our focus is on the loop
correction due to the primordial gravitational waves.

It is known that loop corrections of a massless scalar field generated
during inflation can suffer infrared (IR) divergences~\cite{
Ford:1977in,Allen:1987tz,Kirsten:1993ug, Tagirov:1972vv,
Sasaki:1992ux,Tsamis:1993ub, TW96, Onemli:2002hr, Onemli:2004mb, Kahya:2006hc,
Boyanovsky:2004gq, Brunier:2004sb, Boyanovsky:2004ph, Boyanovsky:2005sh,
Boyanovsky:2005px, Marozzi:2012tp, Prokopec:2002jn, Miao:2005am, Prokopec:2006ue,
Prokopec:2007ak, TW962, ABM, Sloth:2006az, Sloth:2006nu, Seery:2007we,
Seery:2007wf, Urakawa:2008rb, Cogollo:2008bi, Rodriguez:2008hy,
Bartolo:2010bu, Kitamoto:2012vj, Kitamoto:2013rea, Leonard:2012fs, Leonard:2013xsa,
	Kahya:2010xh, Marozzi:2013uva, Adshead:2008gk, Gao:2009fx, RB}. Since a
massless scalar field yields the scale-invariant spectrum in the IR
limit as ${\cal P}(k) \propto 1/k^3$, a naive loop integral yields a factor
$\int \dd^3\bm{k}/k^3 \sim \int \dd k/k$, which logarithmically diverges. 
As is expected from the fact that the mode equation of the tensor 
perturbation, which we refer to also as the graviton, 
is nothing but the mode equation for
a massless scalar field, graviton loop corrections also appear to
yield IR divergences. To quantify the graviton loop corrections, we need to
provide a way to regularize them. One may propose to introduce an IR
cutoff say at a comoving scale $k_{\rm IR}$ as a practical way for the
regularization. However, this will not provide a satisfactory solution,
because the loop integral of the super Hubble modes gives 
$\int^{aH}_{k_{\rm IR}}\dd k/k \sim \ln (aH/k_{\rm IR})$, which
logarithmically increases in time. Here $a$ and $H$ are the scale factor
and the Hubble parameter of the background spacetime, respectively. 
Compared with the
tree level contribution, the loop corrections are typically suppressed
by the Planck scale as $(H/\Mp)^2$ with $\Mp^2 \equiv (8\pi G)^{-1}$.
However, this suppression may be compensated by the secular growth for a sufficiently long lasting
inflation. (A thorough overview about the possible origins of the IR
divergences can be found in the review paper~\cite{Seery:2010kh} by Seery.)

The IR behaviour of the graviton has attracted attention also as a
possible origin of the running of coupling constants~\cite{TW96, Onemli:2002hr, Onemli:2004mb, TW962,
Woodard:2005cw, Janssen:2008px, Polyakov:2007mm, Polyakov:2009nq,
Krotov:2010ma, Alvarez:2010te, KK10, KK1012, KK11,
Romania:2012av}. Tsamis and Woodard
claimed that the logarithmically growing secular effect due to the graviton 
loops can lead to the screening of the cosmological constant, 
suggesting that the screening may provide  
a dynamical solution to the cosmological constant problem~\cite{TW962}. 
More recently, Kitamoto and Kitazawa studied the IR effect on the gauge
coupling and claimed that the secular growth from the graviton loops 
can screen the gauge
coupling~\cite{Kitamoto:2012vj, Kitamoto:2013rea}. 
A related issue is discussed for the U(1) gauge field in
Refs.~\cite{Leonard:2012fs, Leonard:2013xsa}. 
If the secular growth due to the
graviton loops is actually physical, it will provide a phenomenological
impact. However, these secular growths originate from the increasing IR 
contributions, and hence the predicted secular effects significantly depend on the regularization method of IR
contributions. Therefore, to trust the secular growth as a physical
effect, its presence should be verified based on a rigorous method for the regularization. The
graviton loop corrections have been mostly discussed in the exact de
Sitter background and their regularity is still under debate
even at the linear level. In Ref.~\cite{Higuchi:2011vw}, Higuchi {\it et al}. claimed the
existence of a regular two-point function (see also
Refs.~\cite{Allen19862, Higuchi2000}), 
while Miao {\it et al}. objected against it in
Ref.~\cite{Miao:2011ng}. This issue was discussed also in Refs.~\cite{Fewster:2012bj,
Higuchi:2012vy, Morrison:2013rqa}.

The IR pathology has been studied more intensively for the adiabatic
curvature perturbation~\cite{IRsingle, IRgauge_L, IRgauge, IRNG, SRV1,
SRV2, IRreview, BGHNT10, GHT11, GS10, GS11, GS1109, SZ1203, PSZ, SZ1210,
ABG}. In the presence of gravity, we need to remove the
influence of gauge degrees of freedom in calculating the observable
fluctuations. In the cosmological perturbation theory, gauge
artifacts are usually removed by employing gauge conditions that
completely fix the coordinate choice. However,
when we consider actual observations, the gauge conditions may not be
fixed in a suitable way such that preserves the gauge invariance for an
actual observer. In fact, we can observe the fluctuations only within the region causally connected to us, which is 
limited to a finite portion of each time constant slice. Therefore, precisely
speaking, the gauge conditions used in the
conventional cosmological perturbation theory, 
which require the information all over the time constant slice of the universe, 
does not match the actual
process of observations. To regularize the IR contributions for
the curvature perturbation, it is necessary to take into account this subtle
issue~\cite{IRgauge_L, IRgauge, SRV1, SRV2, IRreview}. Since in 
actual observations we can impose the gauge conditions only in the observable
region, there inevitably appears the ambiguity associated with the degrees
of freedom in choosing coordinates in the outside of the observable
region. Such residual coordinate degrees of
freedom can be attributed to the degrees of freedom
in the boundary conditions of our observable local universe. It was shown
that requesting the invariance under the change of such residual
coordinate degrees of freedom in the local universe can ensure the IR regularity and the
absence of the secular growth~\cite{IRgauge_L, IRgauge, SRV1, SRV2,
IRreview}. This is, so to speak, because 
we can absorb the singular IR contributions by the residual coordinate degrees of freedom (see also
Refs.~\cite{GB02, GB03}). This issue is reviewed in
Ref.~\cite{IRreview}. It has been pointed
out that the residual coordinate degrees of freedom also can affect the IR
behaviour of the graviton~\cite{IRgauge, IRreview}. In the present paper, we
will show that, when we require the invariance under the residual coordinate
transformations, the IR regularity and the absence of the secular growth
are guaranteed also for the graviton loops. 
(The IR issues about the entropy
perturbation were studied in Refs.~\cite{IRmulti,IRgauge_multi} and those
about a test scalar field in de Sitter background were studied in
Refs.~\cite{SH, MM10, MMall, MM11, Higuchi:2010xt, SH11}.)

For the graviton, the relation between the IR divergence and the gauge artifact 
has been discussed during the past few decades. Allen showed that
the IR divergence in the free graviton propagator which appears for some
particular values of the gauge parameter can be understood from the fact that
the gauge fixing term does not select a unique gauge for these
particular values of the gauge parameter~\cite{Allen1986}. (See also
Refs.~\cite{Allen:1986tt, Higuchi1986, Higuchi:2000ge, Faizal:2011iv}.) Even if we
properly choose the gauge parameter, it is known that the transverse traceless mode can still suffer
from the IR divergence through the loop corrections, which is the target of this
paper. The connection between the IR divergence and the
gauge artifact was pointed out also in Ref.~\cite{GT07}, where the secular
growth predicted by Tsamis and Woodard in
Ref.~\cite{TW962} was reexamined. It was shown that the spatially
averaged Hubble expansion computed by Tsamis and Woodard is not 
invariant under the change of the time slice and hence the
screening effect which shows up in their averaged Hubble parameter 
suffers from the gauge artifact~\cite{Unruh, GT07}. 
Meanwhile, in Ref.~\cite{GT07}, it was shown that the locally defined
Hubble expansion which may mimic the observable Hubble rate becomes
time-independent. This example suggests that 
computing observable quantities, unaffected by the influence from 
the outside of the observable region, will play an important role 
to quantify the IR effects of the graviton~\cite{GT07} (see also Ref.~\cite{ReplyTW}).

This paper is organized as follows. In Sec.~\ref{Sec:overview}, we will 
briefly show how the IR divergence and the secular growth can appear
when we naively compute the loop corrections for the curvature
and graviton perturbations.  
Then, in Sec.~\ref{SSec:RDF}, we will
point out the presence of the residual coordinate degrees of freedom in the
local observable universe, which describes the influence from
the outside of the observable region. One way to preserve the invariance
under the residual coordinate transformations will be discussed in
Sec.~\ref{SSec:GIq}. Using the prescription which will be introduced in
Sec.~\ref{Sec:preparation}, we will show, in Sec.~\ref{Sec:Euclidean}
and Sec.~\ref{Sec:IRregularity}, that when we choose the Euclidean vacuum as the initial state, the IR
regularity and the absence of the secular growth are ensured. In this
paper, we will discuss the inflationary universe which contains a single
scalar field and we will not directly
discuss the pure gravity setup, although our argument may 
also provide some insight into the latter case.

\section{Overview of IR issues} \label{Sec:overview}
In this section, we will give an overview of the IR issues of the curvature
perturbation and the graviton perturbation. In this paper, we consider a
standard single field inflation model whose action is given by    
\begin{eqnarray}
 S = \frac{\Mp^2}{2} \int \sqrt{-g}~ [R - g^{\mu\nu}\phi_{,\mu} \phi_{,\nu} 
   - 2 V(\phi) ] \dd^4x~, \label{Exp:action} 
\end{eqnarray}
where $\Mp$ is the Planck mass and $\phi$ is the dimensionless scalar
field which is an ordinary scalar field divided by $\Mp$. 
We choose the time slicing by adopting the uniform field gauge
\begin{align}
 & \delta \phi=0\,. \label{GC}
\end{align} 
Under the ADM metric decomposition, which is given by
\begin{align}
 & \dd s^2 = - N^2 \dd t^2  + g_{ij} (\dd x^i + N^i \dd t) (\dd x^j + N^j
  \dd t)~, \label{Exp:ADMmetric}
\end{align}
we further decompose the spatial metric $g_{ij}$ as
\begin{align}
 & g_{ij} = e^{2(\rho+\zeta)} \gamma_{ij} \equiv e^{2(\rho+\zeta)}
 \left[ e^{\delta \gamma} \right]_{ij}\,,
\end{align}
where $a\equiv e^\rho$ is the scale factor, $\zeta$ is the
so-called curvature perturbation and $\delta \gamma_{ij}$ is the
graviton perturbation which satisfies 
the transverse and traceless condition
\begin{align}
 & \delta \gamma_{ii}=0\,, \qquad \partial_i \delta \gamma^i\!_j=0\, . 
\label{TTgauge}
\end{align}
Since $\delta \gamma_{ij}$ is traceless, we find $\det \gamma=1$.

\subsection{Various divergences from the curvature and graviton 
  perturbations} \label{SSec:divs}
In this subsection, after we briefly review the linear perturbation
theory for the scalar and tensor perturbations, we will summarize
several different origins of the divergences that possibly appear in the
loop corrections of these two perturbations. 

\subsubsection{Scalar perturbation}
The quadratic action for the scalar perturbation $\zeta$,
which describes the evolution of the interaction picture field $\zeta_I$, is given by 
\begin{eqnarray}
  S_0^s = \Mp^2 \int \dd t\,\int \dd^3 \bm{x}\, e^{3\rho}  \varepsilon_1
   \left[ \dot{\zeta}_I^2 - e^{-2\rho} (\partial_i \zeta_I)^2 \right]\,,
\end{eqnarray}
and its equation of motion is given by
\begin{eqnarray}
  \left[ \partial_t^2 + (3+\varepsilon_2) \dot{\rho}
 \,\partial_t - e^{-2\rho} \partial^2 \right]\!\zeta_I = 0\,.
\end{eqnarray}
Here, dot denotes the differentiation with respect to
cosmological time $t$. Here, for notational convenience, we introduce
the horizon flow functions as  
\begin{align}
 & \varepsilon_1\equiv \dot{\rho} \frac{\dd}{\dd \rho} \frac{1}{\dot{\rho}}\,,\qquad \quad
 \varepsilon_{n}\equiv \frac{1}{\varepsilon_{n-1}} \frac{\dd}{\dd \rho}
 \varepsilon_{n-1}~,
\end{align}
with $n\geq 2$. We do not assume that these functions are
small, leaving the background inflation model unrestricted. 

For quantization, we expand the curvature perturbation $\zeta_I(x)$ as
\begin{align}
 \zeta_I(x) &= \int \frac{\dd^3 \bm{k}}{(2\pi)^{3/2}}
    \,e^{i\sbm{k}\cdot\sbm{x}}  v^s_k(t) 
a_{\sbm{k}}+ ({\rm h.c.})\,, \label{Exp:zetaI} 
\end{align}
where $a_{\sbm{k}}$ is the annihilation operator, which satisfies
\begin{align}
 & \left[\, a_{\sbm{k}},\, a_{\sbm{k}'}^\dagger \,\right] = \delta^{(3)}
 (\bm{k} - \bm{k}')\,, \qquad \left[\, a_{\sbm{k}},\, a_{\sbm{k}}\,
 \right]  =0\,.
\end{align}
The mode function $v^s_k(t)$ should satisfy
\begin{align}
 & \left[ \partial_t^2 + (3 + \varepsilon_2)
 \dot{\rho} \partial_t + (k e^{-\rho})^2  \right] v^s_k(t) = 0~\,, \label{Modeeq}
\end{align}
and is normalized as  
\begin{align}
 & \left( v^s_k e^{i \sbm{k} \cdot \sbm{x}},\, v^s_p e^{i \sbm{p} \cdot
 \sbm{x}} \right) = (2\pi)^3 \delta^{(3)} (\bm{k} - \bm{p})\,, \label{Cond:N}
\end{align}
where the Klein-Gordon inner product is defined by 
\begin{align}
  (f_1, f_2) \equiv 
 - 2 i \Mp^2 \int \dd^3 \bm{x}\,  e^{3\rho} \varepsilon_1 
  \{ f_1 \partial_t f_2^* - (\partial_t f_1) f_2^*\}\,.
\end{align}
Notice that Eq.~(\ref{Modeeq}) states that $v^s_k(t)$ becomes time-independent in the IR limit 
$k/(e^\rho \dot{\rho}) \ll 1$. When we choose the mode function for the
adiabatic vacuum, which approaches the WKB solution in the UV limit
$k/(e^\rho \dot{\rho}) \gg 1$, the power spectrum becomes almost
scale invariant in the IR limit as  
\begin{eqnarray}
  P^s(k) \equiv |v^s_k(t)|^2 = \frac{1}{4 k^3} \frac{1}{ \varepsilon_1(t_k)}
   \left( \dot{\rho}(t_k) \over \Mp \right)^2  \left[ 1 + {\cal O} \left(
  (k / e^\rho \dot{\rho})^2 \right) \right]\,,  \label{Exp:Pk}
\end{eqnarray}
where we evaluated $v^s_k(t)$ at the Hubble crossing time $t=t_k$ with
$k=e^{\rho(t_k)} \dot{\rho}(t_k)$, since 
the curvature perturbation gets frozen rapidly after $t_k$.

When we assume that the corresponding free theory has an almost scale-invariant spectrum in
the IR limit, a naive consideration can easily lead to the IR divergence
due to loop corrections. For instance, one may expect that an interaction vertex which contains
the curvature perturbation without the 
derivative operator yields the factor
\begin{eqnarray}
 \langle  \zeta^2_I(x) \rangle = \int \frac{\dd^3 \bm{k}}{(2\pi)^3} P^s(k) \,, 
\label{Eq:zeta2I}
\end{eqnarray}
whose momentum integral logarithmically diverges 
in the IR as $\int \dd^3 \bm{k}/k^3$ for the scale-invariant spectrum. 
Even if the spectrum is not exactly scale invariant as in 
Eq.~(\ref{Exp:Pk}), the deep IR modes give a significant contribution 
to Eq.~\eqref{Eq:zeta2I}. 
Following Ref.~\cite{IRreview}, we refer to such unsuppressed
contribution due to the integral over small $k$ as {\em the IR divergence} (IRdiv).

When we introduce an IR cutoff for the regularization, say,  
at the Hubble scale at the initial time $t_i$, the super Hubble (superH)
modes in the variance of $\zeta_I$ give rise to the secular growth which
is logarithmic in the scale factor $a=e^\rho$ as  
\begin{eqnarray}
  \langle \zeta^2_I(x) \rangle_{\rm superH} \propto \int^{e^{\rho(t)}
   \dot{\rho}(t)}_{e^{\rho(t_i)} \dot{\rho}(t_i)} 
   \frac{\dd k}{k} = \ln \left( e^{\rho(t)}
   \dot{\rho}(t) \over e^{\rho(t_i)} \dot{\rho}(t_i) \right)\,.
 \label{Eq:secular}
\end{eqnarray}
Then, the loop corrections, which are suppressed by an extra
power of the amplitude of the power spectrum  
$(\dot{\rho}/\Mp)^2$, may dominate in case the inflationary epoch lasts 
sufficiently long, leading to the breakdown of the perturbative
expansion. We refer to 
the modes with
$e^{\rho(t_i)} \dot{\rho}(t_i) \alt k \alt e^{\rho(t)} \dot{\rho}(t)$
as {\it the transient IR} (tIR) modes and refer to the enhancement of the loop
contributions due to the tIR modes as {\it the IR
secular growth} (IRsec). From the definition, it is clear that the tIR modes were in the sub Hubble (subH) range at the initial
time $t_i$, but have been transmitted into the superH ones by the time $t$. 
The influences of the IRsec have been 
discussed by introducing an IR cutoff 
in Refs.~\cite{Lyth2007, Lyth2006, Bartolo:2007ti,
Dimastrogiovanni:2008af, Tasinato:2012js,
Nurmi:2013xv, Byrnes:2013qjy, Enqvist:2008kt}. We refer to the case when
both the IRdiv and the IRsec are absent as IR regular.

So far, we discussed the secular growth which originates from the momentum
integration, keeping the time coordinates of the interaction vertices fixed. 
However, the time integration also can yield the secular growth. If the
contribution from the interaction vertex in the far past
remains unsuppressed, it will diverge when we send the initial time to
the infinite past. 
We refer to the secular growth due to the temporal integral as
the SG, discriminating it from the previously discussed
IRsec. (Regarding the SG, see also Refs.~\cite{Weinberg05,
Weinberg06}.)

\subsubsection{Graviton perturbation}
The quadratic action for the graviton perturbation $\delta \gamma_{ij}$, 
which describes the evolution of the interaction picture field 
$\delta \gamma_{ij\,I}$, is given by 
\begin{eqnarray}
  S^t_0 = \frac{\Mp^2}{8} \int \dd t\,\int \dd^3 \bm{x}\, e^{3\rho}
   \left[ \delta \dot{\gamma}_{\,j\,I}^i \delta \dot{\gamma}_{\,\,i\,I}^j -
    e^{-2\rho} \partial^l \delta \gamma^i_{\,j\,I} \partial_l \delta
    \gamma^j_{\,\,i\,I} \right]\,,
\end{eqnarray}
and its equation of motion is given by
\begin{eqnarray}
  \left[  \partial_t^2 + 3 \dot{\rho} \partial_t -
 e^{-2\rho} \partial^2 \right] \delta \gamma_{ij\,I} = 0\,.  \label{Def:cLt}
\end{eqnarray}
We quantize $\delta \gamma_{ij\,I}$ as  
\begin{align}
 &  \delta {\gamma}^i_{\,j\,I}(x) = \sum_{\lambda=\pm} \int \frac{\dd^3
 \bm{k}}{(2\pi)^{3/2}} v^{t(\lambda)}_k(t)
 e^i\!_j(\bm{k},\, \lambda) e^{i \sbm{k} \cdot \sbm{x}}
 a_{\sbm{k}}^{(\lambda)} + ({\rm h.c.})\,, \label{Exp:gI}
\end{align}
where $\lambda$ labels the helicity,
$e_{ij}(\bm{k},\, \lambda)$ is the transverse and traceless 
polarization tensor which satisfies
\begin{align}
 &  e^i\!_i(\bm{k},\, \lambda) = k^i e_{ij}(\bm{k},\, \lambda) = 0
 \,, \quad  e_{ij}(\bm{k},\, \lambda) e^{ij}(\bm{k},\, \lambda') =
 \delta_{\lambda, \lambda'} \,,
\end{align}
and $a_{\sbm{k}}^{(\lambda)}$ is the annihilation operator which satisfies
\begin{eqnarray}
  \left[ a_{\sbm{k}}^{(\lambda)},\, a_{\sbm{p}}^{(\lambda')\dagger}
  \right] = 
         \delta_{\lambda \lambda'} \delta^{(3)} (\bm{k} - \bm{p})\,.
\end{eqnarray}
When the graviton perturbation is isotropic, its variance (in the coincidence limit) is given by
\begin{align}
 & \langle \delta \gamma_{ij\,I} (x) \delta \gamma_{kl\,I} (x)  \rangle 
 = \frac{1}{4} \int \frac{\dd^3 \bm{k}}{(2\pi)^3}  \left( \cP_{ik} P_{jl} + \cP_{il} \cP_{jk} -
    \cP_{ij} \cP_{kl} \right) P^t(k)\,,  \label{Exp:spectrumGW0}
\end{align}
where $\cP_{ij}$ is the transverse traceless projection tensor: 
\begin{align}
 & \cP_{ij} \equiv \delta_{ij} - \frac{k_i k_j}{k^2}\,,
\end{align}
and $P^t(k)$ is the power spectrum of the graviton perturbation:
\begin{eqnarray}
  P^t(k) = 2\, |v^t_k(t)|^2\,.
\end{eqnarray}
Here, the factor 2 counts the number of helicity. Since the equation for $v^{t(\lambda)}_k$ is identical to the one 
for a massless scalar field, 
the graviton spectrum in the adiabatic vacuum is 
almost scale invariant in the IR limit as
\begin{eqnarray}
  P^t(k) = \frac{4}{k^3} 
\left( \dot{\rho}(t_k) \over \Mp \right)^2 \left[1
 +  {\cal O} \left( (k/e^\rho \dot{\rho})^2 \right) \right]\,.
\end{eqnarray}
Integrating over the angular coordinates in Eq.~(\ref{Exp:spectrumGW0}),
we obtain
\begin{align}
 & \langle \delta \gamma_{ij\,I} (x) \delta \gamma_{kl\,I} (x)  \rangle 
 = \frac{1}{20 \pi^2} \left( \delta_{ik} \delta_{jl} + \delta_{il} \delta_{jk} -
   \frac{2}{3}\delta_{ij} \delta_{kl} \right) \int \frac{\dd k}{k} k^3 P^t(k)\,.  \label{Exp:spectrumGW}
\end{align}
Similarly to the curvature perturbation, we find that the IR and tIR modes in
Eq.~(\ref{Exp:spectrumGW}) yield the IRdiv and IRsec,
respectively. The influence of the IRsec due to graviton loops is
discussed in Ref.~\cite{Dimastrogiovanni:2008af}. Meanwhile,
the interaction vertices with the graviton 
also can yield the SG in the same way as the curvature perturbation.

\subsection{Residual \gauge degrees of freedom and IR issues} \label{SSec:RDF}
\subsubsection{Solving the constraint equations}  \label{SSSec:SCE}
Eliminating the Lagrange multipliers $N$ and $N_i$, we derive the
action written solely in terms of 
the dynamical fields $\zeta$ and $\delta
\gamma_{ij}$~\cite{Maldacena2002}. 
In the gauge defined by Eqs.~(\ref{GC}) and (\ref{TTgauge}), 
the constraint equations are given by
\begin{eqnarray}
 && \sR - 2 V -  (\kappa^{ij} \kappa_{ij} - \kappa^2 ) 
 - N^{-2}  \dot{\phi}^2 = 0 \,,  \\
 && D_j ( \kappa^j_{~i} - \delta^j_{~i} \kappa )  = 0 ~,
\end{eqnarray}
where $D_i$ is the covariant differentiation associated with the spatial
metric $g_{ij}$, and 
\begin{align}
 & \kappa_{ij} \equiv \frac{1}{2N} (\dot{g}_{ij} - D_i N_j - D_j N_i)\,
\qquad \mbox{and}
 \qquad \kappa \equiv g^{ij} \kappa_{ij}
\end{align}
are the extrinsic curvature and its trace, respectively.
We expand the metric perturbations as 
\begin{align}
  \zeta &= \zeta_I + \zeta_2 + \cdots\,, \label{Exp:zeta} \\
  \delta \gamma_{ij} &= \delta \gamma_{ij\,I} + \delta \gamma_{ij\,2} +
   \cdots\,,  \label{Exp:gij} \\
  N &= 1 + N_1 + N_2 + \cdots\,,  \\
  N_i &= N_{i\,1} + N_{i\,2} + \cdots\,, 
\end{align}
where we use the subscript $I$ to express the first order curvature and 
graviton perturbations, since they correspond to the interaction picture
fields. Then, the $n$-th order Hamiltonian
constraint and the momentum constraints are expressed in the form
\begin{eqnarray}
 & V N_n - 3 \dot{\rho} \dot{\zeta}_n + e^{-2\rho}
 \partial^2 \zeta_n + \dot{\rho} e^{-2\rho} \partial^i N_{i\,n} =H_n\,, \label{HCn}\\
& 4 \partial_i \left( \dot{\rho} N_n - \dot{\zeta}_n \right) -
 e^{-2\rho} \partial^2 N_{i\,n} + e^{-2\rho} \partial_i
 \partial^j N_{j\,n} = M_{i\,n}\,,  \label{MCn}
\end{eqnarray}
where $\partial^2$ denotes the Laplacian. For $n=1$, $H_1$ and
$M_{i\,1}$ are $0$ and for $n\geq 2$, $H_n$ and $M_{i\,n}$ consist of $n$ 
interaction picture fields (either $\zeta_I$ or $\delta
\gamma_{ij\,I}$).

Since the constraint equations (\ref{HCn}) and (\ref{MCn}) are
elliptic-type equations, we need to employ a (spatial) boundary
condition to determine a solution of $N_n$ and $N_{i\,n}$. A general solution
of Eqs.~(\ref{HCn}) and (\ref{MCn}) in the absence of the graviton 
perturbation can be found in Appendix of Ref.~\cite{SRV2}. An extension to include the
graviton perturbation 
proceeds in a straightforward manner and the general
solution is given as  
\begin{align}
 N_n & = \frac{1}{\dot{\rho}} \dot{\zeta}_n + \frac{V}{4 \dot{\rho}}
   e^{-2\rho} \left( e^{2\rho} \partial^{-2} \partial^i M_{i\,n} -   G_n
 \right) \label{Exp:Nn} \,, \\
  N_{i\,n} &= \partial_i \partial^{-2} \biggl[
 \frac{\dot{\phi}^2}{2\dot{\rho}^2} e^{2\rho}  \dot{\zeta}_n -
 \frac{1}{\dot{\rho}} \partial^2 \zeta_n 
+ \frac{e^{2\rho}}{\dot{\rho}} H_n - \frac{V}{4\dot{\rho}^2}
  \left\{ e^{2\rho}
   \partial^{-2} \partial^j   M_{j\,n}  - G_n \right\} 
  \biggr] \cr
 & \qquad  - \left( \delta_i\!^j - \partial_i \partial^{-2}
 \partial^j  \right) \left\{ e^{2\rho} \partial^{-2}  \left( M_{j\,n} -
		      \frac{4\dot{\rho}}{V} \partial_j H_n \right)  - G_{j\,n}
 \right\}\,, \label{Exp:Nin}
\end{align}
where 
$G_n(x)$ and $G_{i\,n}(x)$ are 
arbitrary solutions of the Laplace equations 
\begin{eqnarray}
 \partial^2 G_n(x) = 0\,, \qquad \quad \partial^2 G_{i\,n}(x) =0\,,  
\end{eqnarray}
and $n$-th order in perturbation. 
Since the function $G_{i\,n}(x)$ contributes
only through its transverse part, the number of introduced
independent functions is three. 
By employing appropriate boundary conditions at the spatial infinity,
the solutions of elliptic type equations are uniquely determined. Substituting thus obtained expression for $N$ and $N_i$, 
the action $S=\int \dd^4 x {\cL}[\zeta,\, \delta \gamma_{ij},\, N,\, N_i]$ 
can be, in principle, expressed only in terms of the dynamical 
fields $\zeta$ and $\delta \gamma_{ij}$. Then, the
evolution of $\zeta$ and $\delta \gamma_{ij}$ is 
governed by the {\it non-local} action with the inverse Laplacian.  

As was pointed out in Refs.~\cite{IRgauge_L, IRgauge}, the inverse
Laplacian $\partial^{-2}$ may enhance the singular behavior of
perturbation in the IR limit
by introducing a term with the factor $1/k^2$ 
where $k$ is a comoving momentum of the constituent fields. (A
detailed explanation is given in the review article \cite{IRreview}.)
We will return to this issue in Sec.~\ref{RLR}.

\subsubsection{Observable local patch}  \label{SSSec:observable}
To discuss the observable quantities, we introduce the observable region
as the region causally connected to us. We express the observable region 
on the time slice at the end of inflation 
$t_f$ and its comoving radius as ${\cal O}_{t_f}$ 
and $L_{t_f}$, respectively. 
The causality requires that $L_{t_f}$ should satisfy
$$L_{t_f} \alt \int^{t_0}_{t_f} \frac{\dd t}{e^{\rho(t)}},$$
where $t_0$ is the present time. 
The cosmological observations can measure the $n$-point functions 
of the fluctuation with the arguments $(t_f, \, \bm{x})$ 
contained within the observable region
${\cal O}_{t_f}$. For later use, we refer to 
the causal past of ${\cal O}_{t_f}$ as the
observable region ${\cal O}$ and refer to the intersection between 
${\cal O}$ and a $t$-constant slice $\Sigma_t$ as ${\cal O}_t$. 
We approximate the comoving radius of the region ${\cal O}_t$ as
\begin{eqnarray}
   L_t \equiv L_{t_f} + \int^{t_f}_t \frac{\dd t'}{e^{\rho(t')}} 
 \simeq L_{t_f} + \frac{1}{e^{\rho(t)} \dot{\rho}(t)}\,.
\end{eqnarray} 
As $L_t$ is approximated by 
$L_t \simeq 1/(e^{\rho} \dot{\rho})$ in the distant past, 
the effects of the superH modes with $k \alt e^{\rho(t)} \dot{\rho}(t)$
can be understood as the influence from the outside of the
observable region ${\cal O}$. 
These modes potentially affect the fluctuations in ${\cal O}_{t_f}$ 
by two means. 
One is due to the non-local interaction through 
the inverse Laplacian $\partial^{-2}$, while 
the other is through the Wightman functions
\begin{align}
 & G^{+\,s}(x_1,\, x_2) \equiv \langle \zeta_I(x_1) \zeta_I(x_2) \rangle 
\end{align}
and
\begin{align}
 & G^{+\,t}_{ijkl}(x_1,\, x_2) \equiv \langle \delta \gamma_{ij\,I}(x_1)
 \delta \gamma_{kl\,I}(x_2) \rangle \,.
\end{align}
Even if the spatial distance 
$|\bm{x}_1- \bm{x}_2|$ is bounded from above by confining both $\bm{x}_1$ and
$\bm{x}_2$ within the observable region, the contribution 
to the Wightman functions from the superH modes with 
$k \leq |\bm{x}_1- \bm{x}_2|^{-1}$ is not suppressed. 
The superH modes make these Wightman functions divergent for scale-invariant or red-tilted 
spectrum. (See Eqs.~(\ref{Eq:secular}) and (\ref{Exp:spectrumGW}).) 
The regularity of the contribution from the superH modes can
be verified, only if their contribution is suppressed by an additional 
factor of $k$.

\subsubsection{Residual gauge degrees of freedom} \label{SSSec:RDF}
In Sec.~\ref{SSSec:observable}, we introduced the observable region 
${\cal O}$, which is a limited portion of the whole universe. We claim
that the observable quantities must be composed of the fluctuations within
${\cal O}$. Since only the fluctuations within ${\cal O}$ are 
relevant, there is no reason to request the
regularity of the fluctuations at the spatial infinity in solving
the elliptic-type constraint equations (\ref{HCn}) and (\ref{MCn})
(at least, at the level of Heisenberg 
equations of motion). Then, there arise degrees of freedom in 
choosing the boundary conditions, which are described by arbitrary 
homogeneous solutions of the Laplace equation, 
$G_n(x)$ and $G_{i\,n}(x)$ in Eqs.~(\ref{Exp:Nn}) and (\ref{Exp:Nin}). 
These arbitrary functions in $N$ and $N_i$ can be understood as 
the degrees of freedom in choosing the coordinates. Since the time slicing 
is fixed by the gauge condition (\ref{GC}), the residual gauge degrees
of freedom can reside only in the spatial coordinates $x^i$.

Let us consider these residual coordinate
transformations associated with $G_n(x)$ and $G_{i\,n}(x)$. 
We add the subscript $gl$ to the original global coordinate for the flat Friedmann-Lema\^itre-Robertson-Walker universe with the metric perturbations in order to reserve the simple
notation $\bm{x}$ for the coordinates after transformation. As we have shown in
Refs.~\cite{IRgauge_L, IRgauge}, 
the coordinate transformations $\bxgl\to \bm{x}$ are 
specified as 
\begin{align}
 & \xgl^i = x^i - \sum_{m=1}^\infty {s^i}_{j_1 \cdots j_m}(t) 
 x^{j_1}
 \cdots x^{j_m} + \cdots\,, \label{Exp:RC} 
\end{align}
where ${s^i}_{j_1 \cdots j_m}(t)$ is symmetric over $j_s$ with
$s=1,\, \cdots,\, m$ and satisfies 
$\delta^{j j'} {s^i}_{j_1 \cdots j \cdots j'  \cdots j_m}(t)=0$. 
Here, we abbreviated the non-linear terms in Eq.~(\ref{Exp:RC}). 
These transformations diverge at the spatial
infinity, no matter how small the coefficients are. 
Nevertheless, within the local
region ${\cal O}$, the magnitude of the coordinate transformations~(\ref{Exp:RC})
can be kept perturbatively small. Since the transformations~(\ref{Exp:RC}) 
are nothing but coordinate transformations, the Heisenberg equations for
a diffeomorphism invariant theory remain unchanged under 
these transformations. 

We should note that, once we substitute the expressions for $N$ and $N_i$
to obtain the equation of motion solely written in terms of 
the curvature perturbation $\zeta$ and the graviton perturbation 
$\delta \gamma_{ij}$, 
the symmetry under the residual coordinate transformations is
lost, because $N$ and $N_i$ depend on the specified boundary conditions. 
In this sense the coordinate transformations~(\ref{Exp:RC}) 
should be distinguished from the standard gauge transformations 
that leave the overall action invariant. Therefore, to avoid confusion,
we distinguish the coordinate transformations~(\ref{Exp:RC}), writing it
in the italic font as the \gauge transformation.

Among the residual \gauge transformations, we focus on
\begin{eqnarray}
   & \xgl^i = e^{-s(t)} \left[e^{-S(t)/2}
 \right]^i\!_j\, x^j+{\cal O}\left(\s^2\right)
\,,
   \label{Exp:xres} 
\end{eqnarray}
which is concerned with the IR contributions of the
curvature and graviton perturbations. Here, $s(t)$ is a time-dependent function and $S_{ij}(t)$ is a
time-dependent traceless tensor. 
When we perform the time-dependent dilatation transformation,  
the homogeneous part of the curvature perturbation transforms as
\begin{align}
 & \zeta\to 
   \zeta - s(t)\,.
\end{align}
(Precise meaning of this transformation will be explained later.)
In Refs.~\cite{SRV1, SRV2}, we showed that preserving the invariance under 
the dilatation transformation parametrized by $s(t)$ is crucial to show
the regularity of the loops of the curvature perturbation. Intriguingly,  
the transformation (\ref{Exp:xres}) shifts the
graviton perturbation as
\begin{eqnarray}
 \delta\gamma_{ij}\to \delta\gamma_{ij}
 - S_{ij}(t)+{\cal O}\!\left(\delta\gamma\, \s,\s^2\right)
\,, \label{Eq:shiftg}
\end{eqnarray} 
at the linear level, which is analogous to the shift for $\zeta$. 
Although the non-linear extension of 
the above transformation is rather non-trivial,  
this observation suggests that analogous proof 
of the IR regularity may also work for graviton loops. 
The relation between the IRdiv due to graviton loops
and the homogeneous shift (\ref{Eq:shiftg}) was pointed out several
times. Gerstenlauer {\it et al}.~\cite{GHT11} and
Giddings and Sloth~\cite{GS10} showed that the leading IRdiv of
the graviton loops can be attributed to the change of the spatial
coordinates (\ref{Exp:xres}) with $s(t)=0$ due to the
accumulated effect of the IR graviton.

\subsection{Genuine \gauge invariance and quantization}  \label{SSec:GIq}
The observable fluctuations should not be affected by the residual
\gauge degrees of freedom, 
which were discussed in the preceding subsection.
In this subsection we discuss how to introduce a quantity which is
invariant under the residual \gauge
transformations. We call such a quantity a genuinely \gauge invariant
quantity. One may think that the genuine \gauge invariance will be
preserved by fixing the residual \gauge degrees of freedom 
completely. If we could perform a complete \gauge fixing by employing
appropriate boundary conditions for $N$ and $N_i$ at the boundary of the
observable region ${\cal O}$, the IRdiv and IRsec will not appear, because the
maximum wavelength of fluctuations in such a \gauge would be bounded
approximately by the size of ${\cal O}$. We pursued this possibility in
Refs.~\cite{IRsingle, IRreview}. When we perform the quantization
after complete \gauge fixing or equivalently perform the 
quantization within the local observable region ${\cal O}$, the global
translation invariance is not easily preserved any longer, 
leading to technical complexities. To avoid the complexities, in Ref.~\cite{IRsingle}, 
first we performed the quantization in the whole universe, and 
then we fixed the coordinates by carrying out the residual 
\gauge transformation. In this manner we showed 
that the absence of the IRdiv and
IRsec is guaranteed if the initial fluctuation does not suffer from
these IR pathologies. However, it turned out that 
the IRdiv can arise even in the initial fluctuation
after we perform the residual \gauge transformation to 
employ complete \gauge fixing~\cite{IRreview}.

Here, following Refs.~\cite{IRgauge_L, IRgauge}, 
we perform the quantization, taking into 
account the whole universe without fixing the residual \gauge degrees of
freedom, which allows us to keep the global translation invariance
manifestly. Then, we construct a genuine \gauge invariant operator and
choose an initial state which will be understood as the 
genuine \gauge invariant state. Since the time slice is uniquely specified by the gauge condition
(\ref{GC}), the genuine \gauge invariance will be ensured when
a quantity preserves the invariance under a change of spatial coordinates.

To construct a genuinely \gauge invariant operator, we consider
correlation functions for the scalar curvature of the induced metric on 
a $\phi\!=$constant hypersurface, $\sR$. 
Since $\sR$ itself transforms as a scalar quantity under spatial
coordinate transformations, the correlation functions of $\sR$ 
are not invariant. However, the $n$-point function of $\sR$ will become
gauge invariant, if we specify its $n$ arguments in a
coordinate-independent manner. The distances measured by the spatial geodesics that connect
all the pairs of $n$ points characterize the configuration 
in a coordinate independent manner. 
Here, we adopt a slightly easier way, specifying the $n$
spatial points by the geodesic distances and the 
directional cosines that are measured from 
an arbitrarily chosen reference point. 
Although the choice of the reference point and the frame 
is a part of the residual \gauge, 
this ambiguity will not matter as we will choose a quantum state 
such that does not break the spatial homogeneity and isotropy of the universe.

Our geodesic normal coordinates 
are introduced by solving the spatial geodesic equation 
on each time slice: 
\begin{eqnarray}
 \frac{\dd^2 \xgl^i}{\dd \lambda^2} +  {^s \Gamma^i}_{jk} \frac{\dd
  \xgl^j}{\dd \lambda} \frac{\dd \xgl^k}{\dd \lambda} =0~,
\label{GE}
\end{eqnarray}
where ${^s \Gamma^i}_{jk}$ is the Christoffel symbol 
for the three dimensional spatial metric $e^{2\zeta}\gamma_{ij}$, and
$\lambda$ is the affine parameter. 
The initial ``velocity'' is given by
\begin{eqnarray}
 \frac{\dd \xgl^i(\bm{x},\lambda)}{\dd \lambda}
  \bigg\vert_{\lambda=0}= e^{-\zeta(\lambda=0)} 
 \left[ \gamma(\lambda=0 ) \right]^i\!_j\, x^j\,. \label{IC}
\end{eqnarray}
A point $\bm{x}$ in the geodesic normal coordinates is identified
with the end point of the geodesic $\xgl^i(\bm{x},\lambda=1)$ 
in the original coordinates. 
Perturbatively expanding $\xgl^i$ in terms of $x^i$, we obtain 
$$
\xgl^i= x^i + \delta x^i(\bm{x}).
$$ 
Notice that the relation between $\bxgl$ and
$\bm{x}$ depends on the metric perturbations, which become quantum
operators after quantization. 
Finally, we find that, by means of the geodesic normal coordinates, 
the genuinely \gauge invariant variable is given by 
\begin{eqnarray}
 {^g\!R}(x) &\equiv \sR (t,\, \xgl^i(\bm{x})) = 
 \sR (t,\, x^i + \delta x^i (\bm{x}))\,. \label{Def:gR}
\end{eqnarray}

In order to calculate the $n$-point functions of $\gR$, 
we also need to specify the quantum state such  
that is invariant under the residual \gauge transformations. 
However, in the present approach, we cannot 
directly discuss this invariance 
as a condition for allowed quantum states. 
This is because the residual \gauge
degrees of freedom cease to exist, when we quantize fields in the whole
universe. Let us recall the discussion 
in the case of the curvature perturbation $\zeta$~\cite{SRV1, SRV2,
IRreview}. 
By construction, the operator $\gR$ is not affected by the residual
\gauge degrees of freedom. However, 
the $n$-point functions of $\gR$ can be correlated to the fields 
in the causally disconnected region. In Sec.~\ref{SSSec:RDF}, we
discussed two ways by which the outside of the observable region ${\cal
O}$ can affect the fluctuation in ${\cal O}$. 
One is through the boundary conditions for the inverse Laplacian
$\partial^{-2}$. 
Since changing these boundary conditions is nothing but performing the residual \gauge transformation 
(see Sec.~\ref{RLR}), 
the $n$-point functions of the genuine \gauge invariant curvature 
perturbation $\gR$ are not affected even if we 
restrict the integration region of $\partial^{-2}$ 
to the region ${\cal O}$.  
Therefore, as long as we consider a genuinely \gauge invariant operator,
the inverse Laplacian $\partial^{-2}$ never gives the conjunction
between the inside and the far outside of ${\cal O}$. 
The other leak of the influence from the outside of ${\cal O}$ 
is due to the long-range correlation through the Wightman
functions, which can remain even if we consider 
genuinely \gauge invariant variables. 
Therefore this long-range correlation may 
give a possible origin of the IRdiv and IRsec. In the case with the
curvature perturbation $\zeta$, it is shown that requesting the IR regularity by suppressing the long-range correlation
constrains the quantum state of the inflationary
universe~\cite{SRV1}. Interestingly, 
the IR regularity condition on quantum states can be interpreted as the condition
that requests the quantum states to be unaffected by the
time-dependent dilatation transformation, which is one of the residual
\gauge degrees of freedom~\cite{SRV1}. In Appendix \ref{Sec:SRV}, we show that also for
the graviton perturbation, 
a similar genuine \gauge invariance condition on quantum
states is derived from the IR regularity condition.  

\section{Preliminaries}
\label{Sec:preparation}
In this section, as a preparation to analyze the $n$-point functions of the genuine \gauge
invariant curvature perturbation, 
we introduce a family of canonical variables. 
First, in Sec.~\ref{SSec:CQ}, we describe the
basic formulation for the canonical quantization 
in terms of the original set of variables $\zeta$, 
$\delta \gamma_{ij}$ and their conjugate momenta. 
In Sec.~\ref{SSec:CQt},
we introduce a family of alternative sets of canonical variables,  
in terms of which the proof of the IR
regularity becomes more transparent. 

\subsection{Canonical quantization}  \label{SSec:CQ}
For notational simplicity, we suppress the subscript ``gl'' in this
subsection. In the following discussion, we express the action for the curvature perturbation
$\zeta$ and the graviton perturbation $\delta \gamma_{ij}$ derived by solving the Hamiltonian and
momentum constraint equations as
\begin{align}
 & S= \int \dd t \int \dd^3 \bxglp\, {\cL}_{dyn}\left[ \zeta(\xglp
),\, \partial_t \zeta(\xglp),\,
   \delta \gamma_{ij}(\xglp),\, \partial_t \delta \gamma_{ij}(\xglp) \right] \,,  \label{Exp:L}
\end{align}
which includes the non-local integration operator $\partial^{-2}$. Here,
${\cL}_{dyn}$ denotes the functional form of the Lagrangian density
obtained after we eliminate the Lagrange multipliers $N$ and $N_i$. 
We also introduce the Hamiltonian $H$ and the Hamiltonian 
density ${\cal H}$ as 
\begin{align}
 H(t) &\equiv \int \dd^3 \bxglp\, \pi(\xglp)
 \partial_t  \zeta(\xglp) +  \int \dd^3 \bxglp \pi^{ij}(\xglp)
 \partial_t \delta \gamma_{ij}(\xglp)  - \int \dd^3 \bxglp {\cL}_{dyn} \left[\zeta(\xglp),\, \partial_t \zeta(\xglp),\,
  \delta \gamma_{ij}(x),\, \partial_t \delta \gamma_{ij}(\xglp) \right] \cr
  & \equiv  \int
 \dd^3 \bxglp\, {\cal H}[\zeta(\xglp), \pi(\xglp), \delta \gamma_{ij}(\xglp),
 \pi_{ij}(\xglp)]\,, \label{Def:Ht}
\end{align}
where we introduced the conjugate momenta as
\begin{align}
 & \pi(\xglp)\equiv \frac{\partial {\cal L}_{dyn}(\xglp)}{\partial \left(\partial_t \zeta(\xglp)\right)}
\,, \qquad \quad
 \pi^{ij}(\xglp) \equiv \frac{\partial {\cal L}_{dyn}(\xglp)}{\partial \left(
   \partial_t \delta \gamma_{ij}(\xglp)\right)}\,. 
\end{align}
This set of canonical variables
$\Phi\equiv\{ \zeta, \pi, \delta \gamma_{ij}, \pi_{ij}\}$
should satisfy the standard commutation relations
\begin{align}
 & \left[ \zeta(t,\, \bm{x}),\, \pi(t,\, \bm{y}) \right]= i
 \delta^{(3)} (\bm{x} - \bm{y}) \,, \qquad \left[ \zeta(t,\, \bm{x}),\,
 \zeta(t,\, \bm{y}) \right] = \left[ \pi(t,\, \bm{x}),\, \pi(t,\,
 \bm{y}) \right]=0\,, \label{Exp:Comz}
\end{align}
and
\begin{align}
 & \left[ \delta \gamma_{ij}(t,\, \bm{x}),\, \pi^{kl}(t,\, \bm{y}) \right]= i
 \delta^{(3)}_{~ij}\,\!^{kl} (\bm{x} - \bm{y}) \,, \qquad \left[ \delta \gamma_{ij}(t,\, \bm{x}),\,
 \delta \gamma_{kl}(t,\, \bm{y}) \right] = \left[ \pi^{ij}(t,\,
 \bm{x}),\, \pi^{kl}(t,\, \bm{y}) \right]=0\,, \label{Exp:Comg}
\end{align}
where 
\begin{align}
 & \delta^{(3)}_{~ij}\,\!^{kl}(\bm{x} - \bm{y})\equiv \frac{1}{2} \sum_{\lambda = \pm} \int
 \frac{\dd^3 \bm{k}}{(2\pi)^3} e^{i \sbm{k} \cdot (\sbm{x} - \sbm{y})}
 e_{ij}(\bm{k}, \lambda) e^{kl}(\bm{k},\, \lambda) \label{Def:deltaT}
\end{align}
is the tensorial delta function with the 
transverse traceless projection. 

\subsection{Canonical transformation associated with residual \gauge
  transformations}  \label{SSec:CQt}
In this subsection, we introduce a family of alternative sets of 
canonical variables 
\begin{align}
 & \tilde{\Phi} \equiv \{ \tilde{\zeta},\, \tilde{\pi},\,\delta
\tilde{\gamma}_{ij},\,\tilde\pi_{ij}\}
\end{align}
whose Hamiltonian $\tilde{H}(t)$ is written only in terms of 
\begin{align}
 & \tilde{\zeta}(x) - s(t)\,,\quad \tilde{\pi}(x)\,,\quad  \delta
\tilde{\gamma}_{ij}(x) - \s_{ij}(t)\,, \quad
\tilde{\pi}^{ij}(x) \,, \label{Exp:tPhisS0}
\end{align}
where $s(t)$ and $\s_{ij}(t)$ are
arbitrary time-dependent function and symmetric-traceless matrix, 
respectively. We treat both $s(t)$ and $\s_{ij}(t)$ perturbatively,
assuming that they are as small as metric perturbations. 
We also show that 
$s(t)$ and $\s_{ij}(t)$ without time differentiation are 
contained in the Hamiltonian $\tilde{H}(t)$
only in the combination described in Eq.~(\ref{Exp:tPhisS0}). 

\subsubsection{Introducing new canonical variables}  \label{SSSec:CTti}
For illustrative purpose, we first consider a coordinate transformation
with $s$ and $\s_{ij}$ time independent, which 
induces constant shifts $\tilde{\zeta}(x) -s$ and 
$\delta\tilde{\gamma}_{ij}(t,\, \bm{x}) -\s_{ij}$. 
To be concrete, 
we consider the coordinate transformation $\bxgl\to
\bm{x}$ 
with  
\begin{align}
 & \xgl^i \equiv  e^{-s}  \Lambda^{~~i}_{T\,j}\, x^j \equiv \Lambda^i\!_j\,
 x^j\,,  \label{Exp:CT}
\end{align}
where $\Lambda^{~~i}_{T\,j}$ is a functional of $\s_{ij}$ which satisfies 
\begin{align}
 & \det \Lambda_T=1  \label{Exp:detLTc}
\end{align}
and
\begin{align}
 & \Lambda^{~~i}_{T\,j} = \delta^i\!_j - \frac{1}{2} \s^i\!_j +{\cal O}
(\s^2)\,.  \label{Exp:LTc}
\end{align}
Notice that the coordinate transformation $\bxgl \to \bm{x}$ does not
change the boundary condition of $N$ and $N_i$, and hence it is one of the
gauge transformations, that leave the action invariant. For the time being, we will not specify the terms
of ${\cal O}(\s^2)$ in $\Lambda^{~~i}_{T\,j}$.

Next, we consider the change of the spatial metric
$g_{ij}$ under the gauge transformation~(\ref{Exp:CT}). As is addressed in Ref.~\cite{SRV1},
when we neglect the graviton perturbation, setting
$\Lambda^{~~i}_{T\,j}=0$, 
the dilatation transformation 
changes the spatial metric as 
$$
e^{2(\rho+ \zeta(\xgl))} \delta_{ij} \dd \xgl^i
\dd \xgl^j
 = e^{2(\rho+ \tilde\zeta({x})-s)} \delta_{ij} \dd
 x^i \dd x^j\, , 
$$
where we have defined
$ \tilde\zeta(x) \equiv \zeta(\xgl)$. 
We find that the curvature perturbation $\zeta(\xgl)$ transforms 
to $ \tilde\zeta({x}) - s$, 
which suggests that
this scaling transformation can be used 
to find the canonical variables 
$\tilde{\Phi}$ that are subjected to the necessary constant shift.

Compared with the curvature perturbation, finding a
transformation which shifts the graviton perturbation
by $-S_{ij}$ is much more non-trivial, 
particularly at non-linear order. 
Therefore, introducing the transverse traceless tensor $\delta \tilde\gamma_{ij}$,
we express the spatial metric obtained after the coordinates transformation
(\ref{Exp:CT}) as  
\begin{align}
 & \tilde{g}_{ij}(x) \equiv  e^{2 \{\rho + \tilde\zeta(x ) - s \}} \tilde{\gamma}_{ij}(x) \equiv e^{2 \{\rho + \tilde\zeta(x) - s \}} \left[ e^{\delta \tilde{\gamma}(x) - \s} \right]_{ij}\,,  \label{Def:dg}
\end{align}
with the requested shift by $- \s_{ij}$. In the following, we
assume that $s(t)$ and $\s_{ij}(t)$ are of the same order as $\tilde{\zeta}$
and $\delta \tilde{\gamma}_{ij}$. From $g_{ij}\dd \xgl^i\dd \xgl^j=\tilde g_{ij} \dd x^i
\dd x^j$, 
we find that
$\delta \tilde \gamma_{ij}(x)$ should be related to
$\delta \gamma_{ij}(\xgl)$ as
\begin{align}
 & 
 \tilde \gamma_{ij} (x) = \gamma_{kl}(\xgl)  (\Lambda_T)^k\!_i (\Lambda_T)^l\!_j  \,. \label{Exp:gdg}
\end{align}
Once the functional form of $\Lambda^{~~i}_{T\,j}$ is determined, Eq.~(\ref{Exp:gdg}) 
specifies $\delta \tilde\gamma_{ij}$ order by order in perturbation. By
expanding the inverse matrix of $\Lambda^{~~i}_{T\,j}$ as 
$$
(\Lambda_T^{-1})_{ij} =  \delta_{ij} + \frac{1}{2}\s_{ij}
 + {\cal O}(\s^2), 
$$ 
Eq.~(\ref{Exp:gdg}) leads to
\begin{align}
 & \delta \gamma_{ij}(\xgl)=  \delta \tilde \gamma_{ij}(\xgl) +
 {\cal O} \left( \delta \gamma \s,\, \s^2 \right)\,. \label{Exp:gdg2}
\end{align}
On the right hand side, we explicitly wrote only the linear order in
perturbation. Since the left hand side of Eq.~(\ref{Exp:gdg2})
is independent of $\s_{ij}$, the field $\delta \tilde\gamma_{ij}$ should
be defined so that the $\s_{ij}$ dependence on the right hand side is canceled.
In particular, Eq.~(\ref{Exp:gdg2}) states that 
$\delta \gamma_{ij}(\xgl)$ should agree with $\delta
\tilde \gamma_{ij}(\xgl)$ at the linear order in perturbation. Since the
diffeomorphism invariance of the action 
implies that the Lagrangian densities for $g_{ij}$ and $\tilde{g}_{ij}$ should
take the same functional form, using the Lagrangian density ${\cL}_{dyn}$ in
Eq.~(\ref{Exp:L}), we can express the action for $\tilde{g}_{ij}$ as
\begin{align}
 S&= \int \dd t \int \dd^3 {\bm{x}}\, {\cL}_{dyn} \left[ 
  \tilde\zeta({x}) - s ,\, \partial_t \tilde\zeta({x}),\, 
 \delta \tilde \gamma_{ij}({x}) - \s_{ij},\,
 \partial_t \delta \tilde\gamma_{ij}({x}) \right]\,. 
\label{Exp:sym}
\end{align}

Next, we extend the above argument to 
time-dependent transformations with  
\begin{align}
 & \xgl^i \equiv  
 e^{-s(t)} \Lambda^{~~i}_{T\,j}(t)\, x^j \equiv \Lambda^i_{\,\,j}(t)\,
 x^j  \,,  \label{Exp:CTt}
\end{align}
where $\Lambda^{~~i}_{T\,j}(t)$ is a functional of 
$\s_{ij}(t)$ whose explicit form will be specified later. 
Similarly to the case of constant $\Lambda^i\!_j$, 
we introduce 
a new set of canonical variables $\tilde{\Phi}$
by 
\begin{align}
 & \tilde{\zeta}({x}) \equiv \zeta(\xgl)\,,
\label{Def:variablezeta}
\\
 & \tilde{\gamma}_{ij}(x) \equiv  \left[ e^{\delta \tilde{\gamma}(x) - \s(t)} \right]_{ij}   \equiv
  \gamma_{kl}(\xgl) (\Lambda_T)^k\!_i (\Lambda_T)^l\!_j  \,,
\label{Def:variablegamma}
\end{align}
with the formal definition of the conjugate momenta given by 
\begin{align}
 & \tilde\pi(x)\equiv \frac{\partial {\cal L}_{dyn}(x)}{\partial (\partial_t \tilde\zeta(x))}\,, \qquad \quad
 \tilde\pi^{ij}(x) \equiv \frac{\partial {\cal L}_{dyn}(x)}{\partial (\partial_t \delta \tilde\gamma_{ij}(x))}\,.  \label{Def:variable2}
\end{align}
In general, the residual \gauge transformation is not well defined in
the whole universe, since the transformation can diverge at
the spatial infinity. However, the residual \gauge transformations
\eqref{Exp:CTt}, exceptionally, keep the variables finite in the whole
universe as is manifest from the above relations. Therefore, we can
consistently discuss quantum theory in terms of the set of canonical
variables $\tilde{\Phi}$ as well.

\subsubsection{Commutation relations}
Next, we will show that the variables
$
\tilde{\Phi}= \{ \tilde{\zeta},\, \tilde{\pi},\,\delta
\tilde{\gamma}_{ij},\,\tilde\pi_{ij}\},
$
defined in Eqs.~\eqref{Def:variablezeta}, \eqref{Def:variablegamma}
and (\ref{Def:variable2}),
satisfy the standard commutation relation. 
Because of the time variation of $\Lambda^i_{\,j}$,
the partial time derivative with the original global spatial coordinates 
$\bxgl$ fixed differs from the
one with the new coordinates $\bm{x}$ fixed. 
We choose the transformation matrix
$\Lambda^i_{T\,j}$ such that the difference between the two partial
time derivative operations does not give $\s_{ij}(t)$ 
without the time derivative. 
Then, we find $\Lambda^{~~i}_{T\,j}$ 
should satisfy
\begin{align}
 & \frac{\dd}{\dd t} \Lambda_T[\s(t)]^i\!_j = - \frac{1}{2} \Lambda_T[\s(t)]^i\!_k
 \dot{\s}(t)^k\!_j \,, \label{Eq:LT}
\end{align}
or equivalently
\begin{align}
 & \frac{\dd}{\dd t} \Lambda^{-1}_T[\s(t)]^i\!_j =  \frac{1}{2} 
 \dot{\s}(t)^i\!_k \Lambda^{-1}_T[\s(t)]^k\!_j \,. \label{Eq:LTi}
\end{align}
In fact, with this choice of $\Lambda^{~~i}_{T\,j}$, we obtain
\begin{align}
 &  \partial_t \zeta(t,\, \bxgl) = 
 \partial_t \tilde{\zeta}({x})  +
  \biggl[\dot{s}(t) {\bf 1} + \frac{\dot{S}(t)}{2} \biggr]{}
\!\raisebox{3pt}{${}^n$}\!\raisebox{-3pt}{${}_{\,m}$} {x}^m
 \frac{\partial}{\partial {x}^n}
 \tilde{\zeta}({x}) \,, \label{Exp:dzeta} \\
 & \partial_t \gamma_{ij}(t,\, \bxgl) =
 \Biggl\{ \partial_t \tilde{\gamma}_{kl}({x}) + \dot{\s}_{(k}^{~~m}(t)  \tilde{\gamma}_{l)m}({x})
  + \biggl[\dot{s}(t) {\bf 1}  +
 \frac{\dot{S}(t)}{2} \biggr]
\!\raisebox{3pt}{${}^n$}\!\raisebox{-3pt}{${}_{m\,}$} {x}^m
 \frac{\partial}{\partial {x}^n} \tilde{\gamma}_{kl}({x})
 \Biggr\}  (\Lambda_T^{-1})^k\!_i\, (\Lambda_T^{-1})^l\!_j  \,, \label{Exp:dggm}
\end{align}
where ${\bf 1}$ denotes the unit matrix and the round brackets on 
indices represent symmetrization. The terms with spatial
derivative on the right hand sides of Eqs.~(\ref{Exp:dzeta}) and
(\ref{Exp:dggm}) stem from the difference 
between the two partial time derivative operators. Since Eq.~(\ref{Eq:LT}) implies 
\begin{equation}
 \frac{\dd}{\dd t} \det \Lambda_T[\s(t)] =0\,,
\label{Eq:consistentdet}
\end{equation} 
the condition \eqref{Exp:detLTc} can be extended to the time-dependent
case as
\begin{align}
 & \det \Lambda_T[\s(t)] = 1\,.
\label{Exp:detLTc2} 
\end{align}

Using Eq.~(\ref{Exp:dzeta}), we find that the conjugate momentum $\tilde{\pi}$
is related to $\pi$ as
\begin{align}
 & \tilde{\pi}({x}) = \frac{\partial 
 {\cL}_{dyn}({x})}{\partial [\partial_t 
  \tilde{\zeta}({x})]} = e^{-3s(t)} \frac{\partial
 {\cL}_{dyn}(\xgl)}{\partial [\partial_t \zeta(\xgl)]} = e^{-3s(t)}
 \pi(\xgl) \,. \label{Rel:piz}
\end{align}
In the second equality we used
\begin{align}
 & [\det \Lambda^{-1}(t)] {\cL}_{dyn}({x}) = e^{3s(t)}
 {\cL}_{dyn}({x})  = {\cL}_{dyn}(\xgl)\,, \label{Ref:twoL}
\end{align}
which is derived by changing the spatial coordinates in the action from
${\bm{x}}$ to $\bxgl$. Deriving the relation between $\tilde{\pi}^{ij}$ and
$\pi^{ij}$ is more non-trivial, but using 
\begin{align}
 & \frac{\partial  (\partial_t \delta \gamma_{ij}(\xgl))}{\partial 
 (\partial_t \delta \tilde{\gamma}_{kl}(x))} = \frac{\partial  (\partial_t
  \tilde{\gamma}_{mn}(x))}{\partial (\partial_t \delta
 \tilde{\gamma}_{kl}(x))} \frac{\partial (\partial_t \gamma_{pq}(\xgl))}{\partial  (\partial_t
  \tilde{\gamma}_{mn}(x))} \frac{\partial  (\partial_t
 \delta \gamma_{ij}(\xgl))}{\partial (\partial_t \gamma_{pq}(\xgl))}
= \frac{\partial  \delta \gamma_{ij}(\xgl)}{\partial \delta
  \tilde{\gamma}_{kl}(x)}\,,
\end{align}
where in the second equality we used 
\begin{align}
 & \frac{\partial  (\partial_t \gamma_{ij}(\xgl))}{\partial (\partial_t \delta
 \gamma_{kl}(\xgl))}=  \frac{\partial \gamma_{ij}(\xgl)}{\partial \delta
 \gamma_{kl}(\xgl)}\,, \qquad \quad \frac{\partial (\partial_t
 \tilde{\gamma}_{ij}(x))}{\partial (\partial_t \delta
 \tilde{\gamma}_{kl}(x))}=  \frac{\partial \tilde{\gamma}_{ij}(x)}{\partial \delta
 \tilde{\gamma}_{kl}(x)}\,,
\end{align}
and
\begin{align}
 & \frac{\partial  (\partial_t \gamma_{ij}(\xgl))}{\partial (\partial_t
 \tilde{\gamma}_{kl}(x))} =  \frac{\partial   \gamma_{ij}(\xgl)}{\partial
 \tilde{\gamma}_{kl}(x)} = (\Lambda_T^{-1})^k\!_{(i}\, (\Lambda_T^{-1})^l\!_{j)} \,,
\end{align}
which can be derived by using Eq.~(\ref{Exp:dggm}), we obtain 
\begin{align}
 & \tilde{\pi}^{ij}({x}) = \frac{\partial {\cL}_{dyn}({x})}{\partial (\partial_t \delta
 \tilde{\gamma}_{ij}({x}))} = e^{-3s(t)} \frac{\partial (\partial_t \delta \gamma_{kl}(\xgl)) }{\partial (\partial_t \delta
 \tilde{\gamma}_{ij}({x}))} \frac{\partial {\cL}_{dyn}(\xgl)}{\partial (\partial_t \delta
 \gamma_{kl}(\xgl)} = e^{-3s(t)} \frac{\partial \delta \gamma_{kl}(\xgl) }{\partial \delta
 \tilde{\gamma}_{ij}({x})}  \pi^{kl}(\xgl) 
 \,. \label{Rel:pig}
\end{align}
Here, we simply assume that the operator ordering is properly chosen.

Once we establish the relations between the two sets of the canonical
variables, $\Phi$ and $\tilde{\Phi}$, the commutation relations
for $\Phi$ yield the commutation relations for
$\tilde{\Phi}$. Using Eqs.~(\ref{Exp:Comz}), (\ref{Def:variablezeta})
and (\ref{Rel:piz}), we obtain
\begin{align}
\left[ 
 \tilde{\zeta}(t, {\bm{x}}),\, \tilde{\pi} (t,
 {\bm{y}}) \right]
= i e^{-3s(t)} \delta^{(3)} (\bxgl - \bygl)
=  i\delta^{(3)} \left({\bm{x}} - {\bm{y}} \right)
\,.
\end{align}
Similarly, using Eqs.~(\ref{Exp:Comg}), (\ref{Def:variablegamma}) 
and (\ref{Rel:pig}), we find
\begin{align}
\left[ \delta \tilde{\gamma}_{ij}(t,\,
 {\bm{x}}),\, \tilde{\pi}^{kl} (t,\,
 \bm{y}) \right]  
 & 
=  i e^{-3s(t)} \frac{\partial \delta
 \tilde{\gamma}_{ij}(t,\, \bm{x})}{\partial \delta \gamma_{mn}(t,\
 \bxgl) }  \frac{\partial \delta \gamma_{pq}(t,\, \bygl) }{\partial \delta
 \tilde{\gamma}_{kl}(t,\, \bm{y})} \delta^{(3)~~pq}_{\gl mn}
 (\bxgl - \bygl)   \cr
&=  i \delta^{(3)kl}_{~ij} \left( \bm{x} - \bm{y} \right) 
\,.
\end{align}
In the second equality we noted the tensorial delta function
$\delta^{(3)~kl}_{{\gl}\, ij} (\bxgl - \bygl)$, 
given in Eq.~(\ref{Def:deltaT}), can be expressed as
$$
\delta^{(3)~kl}_{{\gl}\, ij} (\bxgl - \bygl) = 
\frac{1}{2} \sum_{\lambda= \pm} \int \dd^3 \bm{k} \frac{\partial \delta
\gamma_{ij}(t,\, \bxgl)}{\partial \Gamma^{(\lambda)}(t,\, \bm{k})}
\frac{\partial  \Gamma^{(\lambda)}(t,\, \bm{k})}{\partial \delta 
\gamma_{kl}(t,\, \bygl)} = \frac{1}{2} \frac{\delta ( \delta
\gamma_{ij}(t,\, \bxgl))}{ \delta (\gamma_{kl}(t,\, \bygl))}\,,
$$
by expanding $\delta \gamma_{ij}(\xgl)$ as
$$
 \delta \gamma_{ij}(t,\, \bxgl) = \sum_{\lambda=\pm} \int \frac{\dd^3
 \bm{k}}{(2\pi)^{3/2}} e^{i \sbm{k} \cdot \sbm{x}_{\gl}} e_{ij}(\bm{k},\,
 \lambda) \Gamma^{(\lambda)}(t,\, \bm{k})\,,
$$
and we used the factor $e^{-3s(t)}$ to change the argument from $\bxgl$
to $\bm{x}$. The remaining commutation relations can be shown in
a similar way and hence we can verify that $\tilde{\Phi}$ is actually
qualified as a set of canonical variables.

Solving Eq.~(\ref{Eq:LT}), we can determine the transformation matrix
$\Lambda^{~~i}_{T\,j}(t)$. 
As a boundary condition to solve the first order
differential equation, we employ the condition
\begin{align}
 & \Lambda^{~~i}_{T\,j}(t_f) =  \left[ e^{- \s(t_f)/2} \right]
\!\raisebox{2pt}{${}^i$}\!\raisebox{-2pt}{${}_{\,j}$}
\,,  \label{Exp:LTtf}
\end{align}
at the end of inflation $t_f$. Since we have chosen $\Lambda^{~~i}_{T\,j}(t)$ so as to satisfy
$\det \Lambda_T(t_f)=1$, 
Eq.~\eqref{Eq:consistentdet} guarantees that the condition
\eqref{Exp:detLTc2} holds for all $t$. 
Notice that we can formally solve Eq.~(\ref{Eq:LT}) as
\begin{align}
 &  \Lambda^{~~i}_{T\,j}(t)  
= \left[  \Lambda_T(t_f)\, T e^{\frac{1}{2}\int^{t_f}_t \dd t'
 \dot{S}(t')} \right]
\!\raisebox{2pt}{${}^i$}\!\raisebox{-2pt}{${}_{\,j}$}\,,  
\end{align}
using the time ordered product denoted by the operator $T$. 
Perturbatively expanding $\Lambda^{~~i}_{T\,j}(t)$ 
with respect to $\s_{ij}(t)$ to the next to leading order,
we obtain
\begin{align}
 &  \Lambda^{~~i}_{T\,j}(t) = \delta^i\!_j - \frac{1}{2} \s^i\!_j(t) + {\cal O} \left(
 \s^2 \right)\,.  \label{Exp:LTt}
\end{align}

\subsubsection{Hamiltonians}
Next, we compute the Hamiltonian for $\tilde{\Phi}$ defined by 
\begin{align}
 \tilde{H}(t) &\equiv \int \dd^3 {{\bm x}}\,
  \tilde{\pi}({x})\,
  \partial_t  \tilde{\zeta}({x}) +  \int \dd^3 {\bm{x}}\,
 \tilde{\pi}^{ij}({x})\,
 \partial_t \delta \tilde{\gamma}_{ij}({x})
   - \int \dd^3 \bm{{x}}\, {\cL}_{dyn}({x}) \,.
\end{align}
Using Eqs.~(\ref{Exp:dzeta}), (\ref{Exp:dggm}), (\ref{Rel:piz}) and
(\ref{Rel:pig}), we can relate the Hamiltonian $\tilde{H}(t)$ to $H(t)$ as
\begin{align}
 \tilde{H}(t) &= H(t) -  \biggl[\dot{s}(t) {\bf 1} +
 \frac{\dot{S}(t)}{2} \biggr]
{}\!\raisebox{2pt}{${}^l$}\!\raisebox{-2pt}{${}_{\,k}$}
  \int \dd^3 {\bm{x}} \left[
 \tilde{\pi}({x})  {x}^k  \frac{\partial}{\partial
 {x}^l} \tilde{\zeta}({x}) +
 \tilde{\pi}^{ij}({x})  {x}^k
 \frac{\partial}{\partial {x}^l} \delta \tilde{\gamma}_{ij}({x})  \right] \cr 
 & \qquad \qquad - \int \dd^3 {\bm{x}}\,
 \tilde{\pi}^{ij}({x})\left[ \dot{S}_k\!^m(t) \frac{\partial \delta
 \tilde{\gamma}_{ij}(x)}{\partial \tilde{\gamma}_{kl}(x)}
 \tilde{\gamma}_{ml}(x) - \dot{S}_{ij}(t) \right] \,. \label{Hamiltonians}
\end{align}
Equation (\ref{Hamiltonians}) reveals that, when $s(t)$ or $S_{ij}(t)$ is time-dependent, the
Hamiltonian $\tilde{H}(t)$ differs from $H(t)$. However, this
difference does not appear in the quadratic terms of the perturbed
variables. In fact, the linear terms in the square brackets on the second line
are canceled with each other. Using Eqs.~(\ref{Def:variablezeta}),
(\ref{Def:variablegamma}), (\ref{Rel:piz}), and (\ref{Rel:pig}), we can
express the Hamiltonian $H(t)$ in terms of $\tilde{\Phi}$ as 
\begin{align}
 H(t) = \int
 \dd^3 {\bm{x}}\, {\cal H} \left[\tilde{\zeta}({x}) -
 s(t), \tilde{\pi}({x}),\, \delta \tilde{\gamma}_{ij}({x}) -S_{ij}(t) ,\, \tilde{\pi}^{ij}({x}) \right]\,,  \label{Exp:HbytPhi}
\end{align}
where ${\cal H}$ is the Hamiltonian density defined in
Eq.~(\ref{Def:Ht}). Rewriting the graviton part is slightly non-trivial,
but this can be confirmed as follows. When we express $H(t)$ in terms of
$$
\gamma_{ij}(\xgl) \qquad {\rm and} \qquad   \frac{\partial {\cL}_{dyn}(\xgl)}{\partial
 (\partial_t \gamma_{ij}(\xgl))}= \frac{\partial \delta \gamma_{kl}(\xgl) }{\partial
 \gamma_{ij}(\xgl)} \pi^{kl}(\xgl)\,,
$$
these two variables transform as standard tensors in three dimensions into
$$
\tilde{\gamma}_{ij}(x) \qquad {\rm and} \qquad   \frac{\partial {\cL}_{dyn}(x)}{\partial
 (\partial_t \tilde{\gamma}_{ij}(x))}= \frac{\partial \delta \tilde{\gamma}_{kl}(x) }{\partial
 \tilde{\gamma}_{ij}(x)} \tilde{\pi}^{kl}(x)\,,
$$
leaving aside the factor $e^{3s(t)}$, 
which will be absorbed to make the combination
$\tilde{\zeta}({x})- s(t)$ (see Ref.~\cite{SRV2}). Then, since $\gamma_{ij}$
and $\tilde{\gamma}_{ij}$ are given by 
$[\exp({\delta \gamma})]_{ij}$ and $[\exp(\delta \tilde\gamma -\s)]_{ij}$, respectively, 
we can verify Eq.~(\ref{Exp:HbytPhi}).

Next, collecting the quadratic terms in $\tilde{\Phi}$ 
from the Hamiltonian, we identify the non-interacting Hamiltonian, 
\begin{align}
 & \tilde{H}_0 (t) = \int \dd^3 {\bm{x}}\,{\cal H}_0
 \left[\tilde{\zeta}(x),\,
 \tilde{\pi}(x),\, \delta \tilde{\gamma}_{ij}(x),\,
 \tilde{\pi}^{ij}(x) \right]\,, 
\end{align}
which is exactly the same form as the one for $\Phi$.  
Since both $\tilde{\zeta}$ and $\delta \tilde\gamma_{ij}$,
which are massless fields, appear with spatial derivative operators in the 
non-interacting Hamiltonian, the shifts by $-s(t)$ and
$-\s_{ij}(t)$, respectively, are eliminated. 
We also introduce the interaction Hamiltonian as 
\begin{align}
 & \tilde{H}_I(t) \equiv \tilde{H}(t) - \tilde{H}_0(t) \equiv
  \int \dd^3 {\bm{x}}\,
 \tilde{{\cal H}}_I [ \tilde\zeta(x) - s(t),\, \tilde{\pi}(x),\, \delta
 \tilde\gamma_{ij}(x) - \s_{ij}(t),\, \tilde{\pi}^{ij}(x) ] \,, \label{Def:tHI}
\end{align}
with
\begin{align}
 & \tilde{{\cal H}}_I \left[ \tilde\zeta(x) - s(t),\,
 \tilde{\pi}(x),\, \delta
 \tilde\gamma_{ij}(x) - \s_{ij}(t),\, \tilde{\pi}^{ij}(x) \right] \cr
 & = {\cal H}_I
\left[ \tilde\zeta(x) - s(t),\, \tilde{\pi}(x),\, \delta 
 \tilde\gamma_{ij}(x) - \s_{ij}(t),\, \tilde{\pi}^{ij}(x) \right]  -  \biggl[\dot{s}(t) {\bf 1} +
 \frac{\dot{S}(t)}{2} \biggr]
{}\raisebox{2pt}{${}^l$}\!\raisebox{-2pt}{${}_{\,k}$}
\left[
 \tilde{\pi}(x) x^k {\partial\over \partial x^l} \tilde{\zeta}(x) +
 \tilde{\pi}^{ij}(x) x^k {\partial\over \partial x^l} 
\delta
 \tilde{\gamma}_{ij}(x)  \right] \cr 
 & \qquad \qquad \qquad -  \tilde{\pi}^{ij}({x})\left[ \dot{S}_k\!^m(t) \frac{\partial \delta
 \tilde{\gamma}_{ij}(x)}{\partial \tilde{\gamma}_{kl}(x)}
 \tilde{\gamma}_{ml}(x) - \dot{S}_{ij}(t) \right]\,, \label{Exp:tHI}
\end{align}
where
\begin{align}
 & {\cal H}_I[\Phi] \equiv {\cal H}[\Phi] - {\cal H}_0[\Phi]
\end{align}
is the interaction Hamiltonian density for $\Phi$.

\subsubsection{Short summary of the strategy}
It will be instructive to note the two important properties of the interaction
Hamiltonian density $\tilde{{\cal H}}_I$, given in Eq.~(\ref{Exp:tHI}),
which will become crucial in the discussion of the IR regularity: First,
the fields $\tilde{\zeta}(x)$ and $\delta \tilde{\gamma}_{ij}(x)$ always
accompany the time dependent parameters $s(t)$ and $\s_{ij}(t)$ as
$\tilde{\zeta}(x) - s(t)$ and $\delta \tilde\gamma_{ij}(x) - \s_{ij}(t)$. Second, $s(t)$ and
$\s_{ij}(t)$ which are not accompanied by $\tilde{\zeta}(x)$ and $\delta
\tilde{\gamma}_{ij}(x)$, respectively, are always differentiated with
respect to time. When we consider only the adiabatic perturbation, we can
provide the new set of canonical variables which fulfills these two
properties simply by considering the time-dependent dilatation
transformation  
$\bxgl\to \bm{x}$ with $\bxgl = e^{-s(t)} {\bm{x}}$, which is one of the residual \gauge
transformation~\cite{SRV1, SRV2}. At the linear order, the
residual \gauge transformation with Eq.~(\ref{Exp:xres}) shifts the
spatially homogeneous part of the graviton perturbation. However, at
non-linear order, due to the non-commutativity between matrices, the
residual \gauge transformation with (\ref{Exp:xres}) does not
immediately introduce the shift of all the graviton perturbation in the
action.  

To provide the new set of canonical variables with their 
homogeneous parts shifted, we introduced a more non-trivial
transformation \eqref{Exp:CTt}. By choosing $\delta \tilde{\gamma}_{ij}$
as in Eq.~(\ref{Def:variablegamma}), the first property can be ensured.  
To guarantee the second property, we determined (the time
dependence of) $\Lambda^{~~i}_{T\,j}(t)$, requesting Eq.~(\ref{Eq:LT}). Then, $s(t)$ and $\s_{ij}(t)$ without the
time derivative appear neither on the right hand sides of
Eqs.~(\ref{Exp:dzeta}) and (\ref{Exp:dggm}) nor in 
the Hamiltonian $\tilde{H}(t)$. Using these properties, we later show the
IR regularity of graviton loops in a parallel way to the case of
the curvature perturbation.

\subsection{Coarse grained \gauge invariant operator}
\label{SSec:Quantization}
In Sec.~\ref{SSec:GIq}, we introduced the genuine \gauge invariant
variable $\gR$, using the geodesic normal coordinates. 
Changing the spatial coordinates to the geodesic normal coordinates
also modifies the UV contributions. Tsamis and
Woodard~\cite{Tsamis:1989yu} pointed out that using the geodesic 
normal coordinates can introduce an additional origin of UV
divergence, yielding contributions which may not be
renormalized by local counter terms~\cite{Miao:2012xc}.
This is presumably because specifying the spatial distance precisely 
in the presence of the gravitational perturbation requires taking
account of all short wavelength modes. In any realistic
observations, what we actually observe is a smeared 
field with a finite resolution. 
However, it is not so trivial how to describe a realistic 
smearing in a genuinely gauge invariant manner. 
Here, in order to remove the UV contribution in the measurement 
of the position, we replace 
the geodesic normal coordinates with approximate ones which 
are not affected by the UV contributions.

In this paper, we compute the $n$-point functions at the end of
inflation $t=t_f$. Then, in place of the geodesic normal coordinates, we
use the ``smeared'' coordinates $x^i$ 
which are related to the global coordinates
$ \hat{x}_{\gl}^i $ as\footnote{Precisely speaking, the coordinates ``$\bm{x}$'',
which will be used in the rest of paper, are not the geodesic normal
coordinates $\bm{x}$. However, for notational simplicity, we use the
same symbol $\bm{x}$ also for the coarse grained version of the geodesic
coordinates.} 
\begin{align}
 & \hat{x}_{\gl}^i  \equiv
e^{-\Gbz(t_f)} \left[ e^{- \Gbg(t_f)/2} \right]
\!\raisebox{2pt}{${}^i$}\!\raisebox{-2pt}{${}_{\,j}$}\,
x^j\,,
\label{Eq:glcoordi}
\end{align}
where we replaced $s(t_f)$ and $\s_{ij}(t_f)$ in the transformation matrix 
$\Lambda^i\!_j(t_f)$ with the smeared metric perturbations,
\begin{align}
 & \Gbz(t_f) \equiv \frac{\int \dd^3 \bm{x}\, W_{t_f}(\bm{x})
\, \zeta(t_f,\, \hat{\bm{x}}_{\gl})
 }{\int \dd^3 \bm{x}\, W_{t_f}(\bm{x})}\,, \label{Def:Gbz}  \\
  & \Gbg_{ij}(t_f) \equiv \frac{\int \dd^3 \bm{x}\, W_{t_f}(\bm{x})\, 
\delta\gamma_{S\,ij}[t_f,\, \hat{\bm{x}}_{\gl}, \Gbg(t_f)]
}{\int \dd^3 \bm{x}\, W_{t_f}(\bm{x})}\,. \label{Def:Gbg}
\end{align} 
Here, $W_t(\bm{x})$ is the window function which takes the
non-vanishing value in the local region ${\cal O}_t$ and
\begin{align}
\delta\gamma_{S\,ij}[t,\, \bxgl; \s]
 \equiv   \left[ 
\ln  \left(\gamma(t,\, \bxgl)
\Lambda_T(t) \Lambda_T(t)
\right) \right]_{ij}+\s_{ij}\,, 
\end{align}
which implicitly depends on the values of $\s_{ij}(t')$ with $t\leq t' \leq t_f$ through $\Lambda^{~~i}_{T\,j}(t)$. 
Notice that $\Lambda^{~~i}_{T\,j}$ at $t=t_f$ is exceptionally 
determined by the value of $S_{ij}$ only at $t=t_f$
owing to the boundary condition~\eqref{Exp:LTtf}.
Although $\Gbz$
and $\Gbg_{ij}$ appear on the right-hand sides of Eqs.~(\ref{Def:Gbz})
and (\ref{Def:Gbg}), we can iteratively define $\Gbz$ and $\Gbg_{ij}$ by
these expressions. Using the quantities introduced above, 
we define $\gz(t_f,\, \bm{x})$
and $\ggm_{ij}(t_f,\, \bm{x})$ as 
\begin{align}
 & \gz (t_f, \bm{x}^i) \equiv 
\zeta (t_f, \hat{\bm{x}}_{\gl}) \,, \label{Exp:gz} \\
 & \ggm_{ij} (t_f, \bm{x}^i) \equiv \delta\gamma_{S\,ij}[t_f,\, \hat{\bm{x}}_{\gl}; \Gbg(t_f)]
\,. \label{Exp:ggm}
\end{align}
Notice that $\bxgl$ includes $\zeta$ and $\delta \gamma_{ij}$ but
does not include their canonical conjugate momenta. Hence, we can define $\Gbz(t_f)$,
$\Gbg_{ij}(t_f)$, $\gz(t_f,\, \bm{x})$, and $\ggm_{ij} (t_f,\, \bm{x})$
without ambiguity of the operator ordering.

The fields $\gz(t_f,\, \bm{x})$ and $\ggm_{ij} (t_f,\,\bm{x})$ 
introduced above are not genuinely \gauge invariant. However, we can show that 
${\cal R}_x \gz(t_f,\, \bm{x})$ 
and ${\cal R}_x \ggm_{ij}(t_f,\, \bm{x})$ preserve the invariance under the
particular residual \gauge transformation given in
Eq.~(\ref{Eq:glcoordi}), where
\begin{eqnarray}
 {\cal R}_x \ni\, \frac{\partial_t}{\dot{\rho}},\quad \frac{\partial_{\sbm{x}}}{e^{\rho(t)}
 \dot{\rho}(t)},\quad \biggl(1 - \frac{\int \dd^3 \bm{x}
 W_t(\bm{x})}{\int \dd^3 \bm{y} W_t(\bm{y})} \biggr), \quad 
\cdots
\end{eqnarray}
represents an operator such that manifestly suppresses 
the IR contributions by acting on the fields  
$\gz (t,\, \bm{x})$ and $\ggm_{ij}(t,\, \bm{x})$.  
Here, the subscript associated with ${\cal R}_x$ specifies
the argument of the fields on which the operator acts. 
In the following, we will
address the IR regularity of the $n$-point functions of 
${\cal R}_x \gz(t_f,\, \bm{x})$ and 
${\cal R}_x \ggm_{ij}(t_f,\,\bm{x})$.   

\section{Euclidean vacuum and regularization scheme}  \label{Sec:Euclidean}
In order to compute genuinely \gauge invariant quantities, we also need to specify the quantum state 
so as not to be affected by the residual \gauge
degrees of freedom. However, as we mentioned in Sec.~\ref{SSec:GIq}, 
the genuine \gauge invariance of the quantum state cannot be directly
discussed in our current approach. Hence, focusing on the invariance under the 
restricted class of transformations \eqref{Exp:CTt}, we discuss the equivalence among quantum states 
specified in terms of the set of variables $\tilde\Phi$ 
with various values of $s(t)$ and $\s_{ij}(t)$. As is discussed in Ref.~\cite{SRV2}, 
the boundary condition of the Euclidean vacuum selects the unique quantum state 
irrespectively of the choice of canonical variables 
connected by the dilatation transformation. 
Namely, as long as we choose the Euclidean vacuum, the quantum state is unaltered 
by the dilatation scaling. In this section, we extend this argument to include the
graviton perturbation, using different sets of the canonical variables
$\Phi$ and $\tilde{\Phi}$, introduced in the previous section by
performing the residual \gauge transformation from $\xgl^i$ to $x^i$.
We will show that, employing the boundary condition of the
Euclidean vacuum, we can select the unique quantum state irrespective of
the choice of the canonical variables. In
Sec.~\ref{SSec:np}, using this property of the Euclidean vacuum, 
we will reformulate the perturbative expansion.

\subsection{Euclidean vacuum}  \label{SSec:Euclidean}
In this subsection, we briefly summarize the basic properties
of the Euclidean vacuum. In the case of a massive scalar field in de
Sitter space, 
the boundary condition specified by rotating the time path on 
the complex plane can be understood as 
requesting the regularity of correlation functions on the Euclidean
sphere which can be obtained by the analytic continuation from those on
de Sitter space. The vacuum state defined in this way is called the Euclidean vacuum
state. Here, we denote by the Euclidean vacuum the state 
specified by a similar boundary condition in more general spacetime.

To be more precise, 
we define the Euclidean vacuum by requesting the regularity of the
$n$-point functions,  
\begin{eqnarray}
 &  \langle T_c\, 
  \delta \gamma_{i_1 j_1}( {x}{}_{{\gl}1}) \cdots \delta \gamma_{i_m
  j_m}({x}{}_{{\gl}m}) 
  \zeta({x}{}_{{\gl}m+1}) \cdots \zeta({x}{}_{{\gl}n}) 
\rangle_E < \infty\, \qquad {\rm for}\,\,
  \quad \eta(t_a) \to  - \infty (1 \pm i \epsilon)\,,  \label{Def:EV}
\end{eqnarray}
where $a=1, \cdots, n$ and $T_c$ represents 
the path ordering along the closed time path, 
$-\infty(1-i\epsilon)\to \eta(t_f) \to -\infty(1+i\epsilon)$, in 
terms of conformal time 
\begin{align}
 & \eta(t) \equiv \int^t \frac{\dd t}{e^{\rho(t)}}\,.
\end{align}
For simplicity, here we assume that 
$e^{\rho(t)} \dot\rho(t)$ is rapidly increasing in time so that
\begin{eqnarray}
  |\eta(t)|={\cal O} \left( 1/e^{\rho(t)} \dot{\rho}(t) \right)\,.
 \label{eta}
\end{eqnarray}
We added the subscript $E$ to the expectation values for the Euclidean
vacuum defined in terms of the canonical variables $\Phi$.  

For the canonical variables $\tilde{\Phi}$, the boundary condition of
the Euclidean vacuum is similarly given by  
\begin{eqnarray}
 &  \langle T_c\, 
  \delta \tilde\gamma_{i_1 j_1}(x_{1}) \cdots \delta
  \tilde\gamma_{i_m j_m}(x_m) 
  \tilde\zeta(x_{m+1}) \cdots \tilde\zeta(x_n) 
  \rangle_{\tilde E} < \infty\, \qquad {\rm for}\,\,
  \quad \eta(t_a) \to  - \infty (1 \pm i \epsilon)\,.  \label{Def:EVt}
\end{eqnarray}
The Euclidean vacuum is expected to be invariant under the
residual \gauge transformations, since the above boundary conditions 
of the Euclidean vacuum are formally independent of the choice of canonical 
variables. In fact, we can show the equivalence between the expectation values, 
\begin{eqnarray}
 & \langle T_c\, {\cal O}
\rangle_E 
= \langle T_c\, \tilde{\cal O} \rangle_{\tilde E}\,,
\label{Exp:EV}
\end{eqnarray}
where the operators ${\cal O}$ and $\tilde{\cal O}$ are related 
with each other 
by the relations \eqref{Def:variablezeta} and \eqref{Def:variablegamma}. A more detailed explanation
regarding the uniqueness of the Euclidean vacuum can be found in
Ref.~\cite{SRV2} and the argument there can be extended to include the
graviton modes in a straightforward manner. We will find that the distinctive 
property~(\ref{Exp:EV}) is crucial in showing the IR regularity for the
Euclidean vacuum.

\subsection{Rewriting the $n$-point functions}  \label{SSec:np}
In this subsection, we rewrite the expression for the 
$n$-point functions into 
a more suitable form to examine the regularity of the IR contributions. 
Namely, we perform the perturbative expansion of the $n$-point functions
of $\gz(t_f, \bm{x}_a)$ and 
$\ggm_{ij}(t_f, \bm{x}_a)$ with $a=1,\, \cdots,\, n$, using the
canonical variables $\tilde{\Phi}$. In this subsection, we adopt the Schr\"odinger picture. Since
all the operators will be in Schr\"odinger picture, they do not have time dependence. 
Introducing the unitary operator of the time evolution
\begin{align}
 U(t, t') &\equiv 
T_c \exp \left[-
 i\int^t_{t'} \dd t\, H(t)\right]\,,
\end{align}
the $n$-point functions are expressed as 
\begin{align}
 &\langle 0| U(- \infty (1 + i\epsilon), t_f)
\gz(\bm{x}_1) \cdots \gz(\bm{x}_n) 
U(- \infty (1 - i\epsilon), t_f) |0 \rangle\,.
\label{npointfn}
\end{align}

{}Here, we introduce the eigen state of $\zeta$ and $\delta \gamma_{ij}$, $\eSH$. 
For given values of $s$ and $\s_{ij}$, $\eSH$ 
also becomes the eigen state of $\Gbz$ and $\Gbg$. 
 {\it i.e.}, 
\begin{align}
 &  \Gbz(t) \eSH 
=\frac{\int \dd^3 \bm{x} W_t(\bm{x}) \zeta(\bxgl)}{\int \dd^3
 \bm{x} W_t(\bm{x})} \eSH = s^\ev[t, \PSH, s; \s] 
\eSH\,, \label{Def:sts0} \\
 &  \Gbg_{ij}(t) \eSH = \frac{\int \dd^3 \bm{x} W_t(\bm{x}) \delta
 \gamma_{S\,ij}[t, \bxgl; \s]}{\int \dd^3 \bm{x} W_t(\bm{x})} \eSH =
 S^\ev_{ij}[t, \PSH, s; \s] 
\eSH\,, \label{Def:stt0}
\end{align}
where $\zeta^c$ and $\delta \gamma^c$ denote the eigenvalues of $\zeta$
and $\delta \gamma_{ij}$. Here the time dependence of the operators $\Gbz$ and $\Gbg_{ij}$ appears through 
$W_t(\bm{x})$, $\bxgl$ and $\s_{ij}$, while $\zeta(\bm{x})$ and $\delta\gamma_{ij}(\bm{x})$ are 
time-independent Schr\"odinger operators. 
Since $\bxgl$ and $\delta \gamma_{S,\,ij}$ 
depend on the value of $s(t)$ and the path for
picked up values of $S_{ij}(t')$ with $t \leq t' \leq t_f$, the
eigenvalues of the operators $\Gbz$ and $\Gbg_{ij}$, $s^\ev(t)$ and $S^\ev_{ij}(t)$, also depend on 
$s(t)$ and $S_{ij}(t')$.

Using the eigen state $\eSH$, we construct a decomposition of unity:
\begin{align}
 {\bf 1} &=  
 \int \ddP
  \Bigl\vert \PSH
 \Bigl\rangle \Bigr\langle  \PSH \Bigr\vert  \,. \label{Def:unit}
\end{align}
Discretizing the time coordinate along the closed time path in
Eq.~(\ref{npointfn}) as is usually done in the path integral, we insert
the unit operator (\ref{Def:unit}) at each intermediate time step as
\begin{align}
\mbox{Eq.}~\eqref{npointfn}
& =  \Biggl\langle\,0 \Bigg|  T_c \left( \prod_{a=0}^{\infty} \int \ddP
 U(t_{a+1}, t_{a})  \Bigl\vert 
 \PSH \Bigl\rangle \Bigr\langle  \PSH  
\Bigr\vert \right) \cr
& \qquad   \qquad \quad \times 
 \zeta(\hat {\bm{x}}{}_{{\gl}1} ) \cdots
 \zeta(\hat {\bm{x}}{}_{{\gl}n} )  
 \left( \prod_{b=-\infty}^{0} \int \ddP 
    \Bigl\vert \PSH  
\Bigl\rangle \Bigr\langle  \PSH   
\Bigr\vert
 U(t_{b}, t_{b-1}) 
 \right)  \! \Bigg|  0\, \Biggr\rangle\,, \label{tempnpz}
\end{align}
where we labeled the discretized time coordinate
from the distant past to $t_f$ by negative integers and 
the one from $t_f$ to the distant past by positive integers, 
with $t_0=t_f$. 
For the time being, we focus on the $n$-point functions for a 
particular time path of $\zeta^c$ and $\delta\gamma^c$, picking up, at
each time step, one representative state among the summed up eigen
states in the unit operator (\ref{Def:unit}) in Eq.~(\ref{tempnpz}). 
Namely, we consider the expectation value 
\begin{align}
\Biggl\langle\,0 \Bigg|  T_c 
\left( \prod_{a=0}^{\infty} 
 U(t_{a+1}, t_{a}) 
  \Bigl\vert \PSH  
\Bigl\rangle \Bigr\langle  \PSH 
\Bigr\vert \right)
 \zeta(\hat {\bm{x}}{}_{{\gl}1} ) \cdots
 \zeta(\hat {\bm{x}}{}_{{\gl}n} )  
 \left( \prod_{b=-\infty}^{0}     \Bigl\vert \PSH 
\Bigl\rangle \Bigr\langle \PSH  
\Bigr\vert U(t_{b}, t_{b-1}) 
 \right)  \! \Bigg|  0\, \Biggr\rangle\,. \label{tempnpz2}
\end{align}
Once the path of $\zeta^c$ and $\delta\gamma^c$ is specified, we can
choose the self-consistent values of $s(t)$ and $\s_{ij}(t)$ such that satisfy
\begin{align}
 s(t)=s^\ev [t,\PSH, s; \s]\,, \qquad 
 \s_{ij}(t)= S^\ev_{ij}[t,\PSH, s; \s]\,,
\end{align}
for all $t$ order by order. Then, using the corresponding values of $s$ and $S_{ij}$,
we introduce the canonical variables $\tilde{\Phi}(x)$ as defined in the preceding section. 
Using $\tilde{\Phi}$, we can replace $\zeta(\hat{\bm{x}}_\gl)$ in Eq.~(\ref{tempnpz2}) with $\tilde \zeta(\bm{x})$. 
Notice that in the canonical system with $\tilde{\Phi}$, we should use the unitary operator of time evolution
defined by the Hamiltonian $\tilde{H}(t)$, which differs from
$H(t)$, as
\begin{align}
\mbox{Eq.}~\eqref{tempnpz2}
&=
\Biggl\langle\,0 \Bigg|  T_c 
\left( \prod_{a=0}^{\infty} 
 \tilde U(t_{a+1}, t_{a}) 
  \Bigl\vert \PSH 
\Bigl\rangle \Bigr\langle \PSH  
 \Bigr\vert \right)
\cr
&\qquad\qquad \quad \times \tilde\zeta(\bm{x}_1) \cdots
 \tilde\zeta(\bm{x}_n)  
 \left( \prod_{b=-\infty}^{0}     \Bigl\vert \PSH 
\Bigl\rangle \Bigr\langle \PSH  
 \Bigr\vert \tilde U(t_{b}, t_{b-1}) 
 \right)  \! \Bigg|  0\, \Biggr\rangle\,, 
\label{tempnpz3}
\end{align}
with 
\begin{align}
 \tilde{U}(t, t') &\equiv 
T_c \exp \left[-
 i\int^t_{t'} \dd t\, \tilde H(t)\right]\,.
\end{align}
Here, $s$ and $\s_{ij}$ are different between the forward 
and backward paths, and hence the new canonical variables
$\tilde{\Phi}(x)$ will differ between them. Furthermore, Eqs.~(\ref{Def:sts0}) and
(\ref{Def:stt0}) imply 
\begin{align}
 & \Gbz(t_a) \eSH  = s(t_a) \eSH \,, \label{Def:sts} \\
 &  \Gbg_{ij}(t_a) \eSH = S_{ij}(t_a) \eSH 
 \,, \label{Def:stt}
\end{align}
with
\begin{align}
 & \Gbz(t) \equiv  \frac{\int \dd^3 \bm{x}\, W_t(\bm{x})\,
 \tilde{\zeta}(\bm{x})}{\int \dd^3 \bm{x}\, W_t(\bm{x})}  
\,, \label{Def:Gbz2} \\
 & \Gbg_{ij}(t) \equiv  \frac{\int \dd^3 \bm{x}\, W_t(\bm{x}) \delta
 \tilde{\gamma}_{ij}(\bm{x})}{\int \dd^3 \bm{x}\, W_t(\bm{x})} \,.  \label{Def:Gbg2}
\end{align}

Next, we write down the expression~(\ref{tempnpz2}) 
in the interaction picture. Using the unitary
operator 
\begin{align}
 \tilde{U}_0(t) &\equiv T_c \exp \left[-
 i\int^t \dd t \int \dd^3 \bm{x} \, \tilde{\cal H}_0 \right]\,,
\end{align}
with an appropriate choice of a lower boundary of the $t$-integration,  
the Schr\"odinger picture fields $\tilde{\zeta}(\bm{x})$ and 
$\delta \tilde{\gamma}_{ij}(\bm{x})$ are related to the interaction picture
fields $\tilde{\zeta}_I(t, \bm{x})$ and $\tgij(t, \bm{x})$, respectively, as
\begin{align}
 &\tilde{\zeta}(\bm{x}) = \tilde{U}_0
(t)\, \tilde{\zeta}_I (t, \bm{x})\, 
 \tilde{U}^\dag_0(t)\,, \\
 & \delta \tilde{\gamma}_{ij}(\bm{x}) 
 = \tilde{U}_0(t)\, 
 \tgij (t, \bm{x})\, \tilde{U}^\dag_0(t)\,.
\end{align}
Similarly to Eqs.~(\ref{Def:sts}) and (\ref{Def:stt}), we define the
eigenstate for the interaction picture fields as
\begin{align}
\eSHI =  \tilde{U}^\dag_0(t) \eSH  \,.
\end{align}
In the interaction picture, we obtain 
\begin{align}
\mbox{Eq.}~\eqref{tempnpz3}
=
\Biggl\langle\,0 \Bigg|  T_c 
\left( \prod_{a=0}^{\infty} 
 \tilde U_I(t_{a+1}, t_{a}) 
  \Bigl\vert
 \PSHIt{a}\Bigl\rangle_{\!I\,\,I}\!\Bigr\langle  \PSHIt{a}\Bigr\vert
 \right)
 \tilde\zeta_I(t_f,\,\bm{x}_1) \cdots
 \tilde\zeta_I(t_f,\,\bm{x}_n) 
\cr\quad\qquad\times 
 \left( \prod_{b=-\infty}^{0}     
\Bigl\vert \PSHIt{b}\Bigl\rangle_{\!I\,\,I}\!\Bigr\langle \PSHIt{b}\Bigr\vert
 \tilde U_I(t_{b}, t_{b-1}) 
 \right)  \! \Bigg|  0\, \Biggr\rangle\,, 
\label{tempnpz4}
\end{align}
with 
\begin{align}
 \tilde{U}_I(t, t') &\equiv 
T_c \exp \left[-
 i\int^t_{t'} \dd t\, \tilde H_I(t)\right]\,.
\end{align}
With $\Gbz_I(t)$ and $\Gbg_{ij\,I}(t)$ defined as
\begin{align}
 & \Gbz_I(t) \equiv  \frac{\int \dd^3 \bm{x}\,
 W_t(\bm{x}) \tilde{\zeta}_I(t, \bm{x})}{\int \dd^3 \bm{x}\,
 W_t(\bm{x})}\,, \\
 & \Gbg_{ij\,I}(t) \equiv  \frac{\int \dd^3 \bm{x}\, W_t(\bm{x}) \delta
 \tilde{\gamma}_{ij\,I}(t, \bm{x}) }{\int \dd^3 \bm{x}
 W_t(\bm{x})} \,,
\end{align}
Eqs.~(\ref{Def:sts}) and (\ref{Def:stt}) indicate 
\begin{eqnarray}
& & \Gbz_I(t_a) \eSHIt{a} 
= s(t_a) \eSHIt{a} \,, 
\label{Def:stsI} \\
& &  \Gbg_{ij\,I}(t_a) \eSHIt{a} = 
S_{ij}(t_a)  \eSHIt{a} 
\label{Def:sttI} \,. 
\end{eqnarray}

Next, we will show that, when we choose the Euclidean vacuum as the initial state, the
$n$-point functions for $\gz(x)$ and $\ggm_{ij}(x)$ can be expanded only in terms
of the interaction picture fields $\tilde{\zeta}_I(x)$ and $\tgij(x)$
with the IR suppressing operators
${\cal R}_x$.   
While the interaction Hamiltonian density $\tilde{\cal H}_I$
is messy, the IR regularity can be shown
just by using the fact that, as is given in
Eq.~(\ref{Exp:tHI}), $\tilde{\cal H}_I$ is expressed only in terms of 
\begin{eqnarray}
 \tilde{\zeta}_I(x) -s(t),\,\qquad 
 \tgij(x)- S_{ij}(t),\,\qquad \label{Exp:tPhisS}
\end{eqnarray}
and 
\begin{align}
 & \tilde{\pi}_I(x)= 2 \Mp^2 e^{3\rho} \varepsilon_1
 \dot{\tilde{\zeta}}_I(x),\, \qquad \tilde{\pi}^{ij}_{\,\,I}(x) =
 \frac{\Mp^2}{4} e^{3\rho} \delta \dot{\tilde{\gamma}}_I^{ij}(x)\,,
\end{align}
and also with the parameters,
\begin{eqnarray}
  \dot{s}(t),\,\quad \dot{\s}_{ij}(t)\,.
\end{eqnarray}
Notice that the terms in (\ref{Exp:tPhisS}) are not suppressed by ${\cal R}_x$ 
and also that the inverse Laplacian $\partial^{-2}$, 
which arises from $N$ and $N_i$, 
may decrease the power of $k$ by $1/k^2$, depending on the choice of the 
boundary conditions.

\subsubsection{Interaction picture fields without the IR suppressing
   operator}  \label{SSec:npEuclidean}
We begin with discussing the first term in Eq.~(\ref{Exp:tHI}), {\it i.e.},
\begin{align}
 & {\cal H}_I[\tilde{\zeta}_I(x)- s(t),\,\tilde{\pi}_I(x),\,\tgij(x) - \s_{ij}(t),\, \tilde{\pi}^{ij}_I(x) ]
\,.  \label{Exp:HttPhi}
\end{align}
If we can simply replace
$s(t)$ and $\s_{ij}(t)$ with $\Gbz_I(t)$
and $\Gbg_{ij\,I}(t)$, respectively in the above expression,   
$\tilde{\zeta}_I(x) - s(t)$ and
$\tgij(x) - \s_{ij}(t)$ are reduced to 
$\tilde{\zeta}_I(x)- \Gbz_I(t)$ and 
$\tgij (x) - \Gbg_{ij\,I}(t)$, which are combinations 
suppressed by the IR suppressing operator ${\cal R}_x$. 
We will show that the distinctive
property of the Euclidean vacuum given in Eq.~(\ref{Exp:EV}) allows us
to perform this replacement just by adding
terms that are composed only of
${\cal R}_x \tilde{\zeta}_I(x)$ and 
${\cal R}_x \tgij(x)$.

To perform the replacement, we notice that the operator 
$\vert \PSHIt{a} \rangle_{I\,\,I}\langle  \PSHIt{a}\vert$ is 
located next to the interaction Hamiltonian
$\tilde H_I(t_a)$ as
\begin{align*}
 &  \cdots\, \tilde H_I(t_a)
\Bigl\vert \PSHIt{a}\Bigl\rangle_{\!I\,\,I}\!\Bigr\langle \PSHIt{a}\Bigr\vert
\, \cdots \,,
\end{align*}
where the abbreviation on the right hand side of $\tilde{\cal H}_I$
denotes operators in the past of $t_a$ along the closed time path and
the one on the left hand side denotes operators in the future of $t_a$. 
For notational simplicity, we abbreviate the subscript $a$ in the 
following discussion.
Picking up a single $\tilde{\zeta}_I(x) - s(t)$ or 
$\delta \tilde{\gamma}_{ij\,I}(x) - \s_{ij}(t)$ from the first term of
$\tilde{{\cal H}}_I$, given in Eq.~\eqref{Exp:HttPhi}, 
and using Eqs.~(\ref{Def:stsI}) and (\ref{Def:sttI}), we rewrite each term as 
\begin{align}
 & \left( \tilde\zeta_I(x) - s (t) \right)  
 {\cal A}(x) \eSHI  
 = \left(
 \tilde\zeta_I(x) - \Gbz_I(t) \right)  {\cal A}(x) \eSHI + \left[
 \Gbz_I(t),\, {\cal A}(x) \right] \eSHI\,,
 \label{formula0}
\end{align}
or
\begin{align}
 & \Bigl( \delta \tilde{\gamma}_{ij\,I}(x) - \s_{ij} (t)
 \Bigr)\,  {\cal A}(x)  \eSHI
  = \Bigl( \delta \tilde{\gamma}_{ij\,I}(x)  -
 \Gbg_{ij\,I}(t) \Bigr)\, {\cal A}(x) \eSHI 
 + \left[ \Gbg_{ij\,I}(t),\, {\cal A}(x) \right] \eSHI\,,
 \label{formula2}
\end{align}
where using ${\cal A}(x)$, we schematically expressed the operators sandwiched between 
$\tilde{\zeta}_I(x) - s(t)$ or $\delta \tilde{\gamma}_{ij\,I}(x)-\s_{ij}(t)$ and $\eSHI$. 
Since ${\cal A}(x)$ is a 
part of the Hamiltonian density in Eq.~(\ref{Exp:HttPhi}), it 
can be expressed solely in
terms of the combinations in Eq.~(\ref{Exp:tPhisS}) and the conjugate
momenta $\tilde{\pi}_I$ and $\tilde{\pi}^{ij}_I$. Since $\Gbz_I(t)$
commutes with $\tilde{\zeta}_I(x)- s(t)$, $\delta \tilde{\gamma}_{ij\,I}(x)-\s_{ij}(t)$,
and $\tilde{\pi}^{ij}_I(x)$, the non-vanishing contributions in $[\Gbz_I(t),\, {\cal A}(x)]$
arise only from the commutator 
\begin{align}
 \left[ \Gbz_I(t),\,\tilde\pi_I(t, \bm{x}) \right]
 &= \frac{1}{\int \dd^3 \bm{x}~W_t(\bm{x})} \int \dd^3
 \bm{y}~W_t (\bm{y}) \left[ \tilde{\zeta}_I(t,\, \bm{y}),\,
 \tilde{\pi}_I(t,\, \bm{x})\right] = i \frac{W_t(\bm{x})}{\int
 \dd^3 \bm{x}~W_t(\bm{x})} \,,
\end{align}
which yields a local function whose Fourier mode is regular in the IR
limit. Repeating this procedure, we can rewrite
$( \tilde\zeta_I(x) - s (t) )  {\cal A}(x)$ solely in terms of 
\begin{align}
 & \tilde{\zeta}_I(x) - \Gbz_I(t)\,, \quad \tilde{\pi}_I(x)\,, \quad \delta
 \tilde{\gamma}_{ij\,I}(x) - \Gbg_{ij\,I}(t)\,, \quad \tilde{\pi}^{ij}_I(x)\,. \label{Exp:tPhisSrep}
\end{align}
The same argument can apply to 
$ ( \delta \tilde{\gamma}_{ij\,I}(x) - \s_{ij} (t)) {\cal A}(x)$. 
In this way all the interaction picture fields in the
first term of $\tilde{\cal H}_I$ are now expressed 
by ${\cal R}_x \tilde{\zeta}_I$ and 
${\cal R}_x \delta \tilde{\gamma}_{ij\,I}$.

Next, we consider the second term of the interaction Hamiltonian
(\ref{Exp:tHI}) with $\dot{s}$ and $\dot{S}_{ij}$. When we
discretize the time coordinate, the time derivative should be regarded 
as the difference between the values at two adjacent time steps. 
We can express the second term in Eq.~\eqref{Exp:tHI} sandwiched 
by ${}_I\langle \PSHIt{a+1}\vert$ and $\vert \PSHIt{a}\rangle_I$ as
\begin{align*}
 &{\raisebox{-5pt}{${}_{I\!}$}}\Bigr\langle \PSHIt{a+1}\Bigr\vert
 \left[ \tilde{\pi}_I(x_a) x^l \partial_m \tilde{\zeta}_I(x_a) +
 \tilde{\pi}^{ij}_{\,\,\,I}(x_a) x^l \partial_m \delta
 \tilde{\gamma}_{ij\,I}(x_a)   \right] 
\left(
 \dot s(t_a) \delta^m\!_l + \dot S^m\!_l(t_a)/2 \right)
   \Bigl\vert \PSHIt{a}\Bigl\rangle_{\!I}  \cr
 & =
{\raisebox{-5pt}{${}_{I\!}$}}\Bigr\langle \PSHIt{a+1}\Bigr\vert 
 \left[ \tilde{\pi}_I(x_a) x^l \partial_m \tilde{\zeta}_I(x_a) +
 \tilde{\pi}^{ij}_{\,\,\,I}(x_a) x^l \partial_m \delta
 \tilde{\gamma}_{ij\,I}(x_a)   \right] \cr 
 & \qquad \qquad \qquad  \qquad \qquad \times 
\frac{\{s(t_{a+1}) - s(t_a)\} \delta^m\!_l + \{S^m\!_l(t_{a+1}) -
 S^m\!_l(t_a)\}/2  }{t_{a+1} - t_a}   
 \Bigl\vert\PSHIt{a}\Bigl\rangle_{\!I} \,,  
\end{align*} 
with $x_a=(t_a,\, \bm{x})$. Here, we neglect the terms irrelevant in the continuous limit.
Using Eqs.~(\ref{Def:stsI}) and (\ref{Def:sttI}), we can replace 
$s(t_a)$ and $S_{ij}(t_a)$ with $\Gbz_I(t_a)$ and $\Gbg_{ij\,I}(t_a)$ placed next to 
$ \vert\PSHIt{a}\rangle_I$, while 
$s(t_{a+1})$ and $S_{ij}(t_{a+1})$ with $\Gbz_I(t_{a+1})$ and $\Gbg_{ij\,I}(t_{a+1})$ next to 
$ {}_{I}\langle\PSHIt{a+1}\vert$. For the same reason as in the previous case, 
the terms coming from the commutator between 
$[ \tilde{\pi}_I(x_a) x^l \partial_m \tilde{\zeta}_I(x_a)
+\cdots ]$ and $\Gbz_I(t_a)$ or $\Gbg_{ij\,I}(t_a)$ only give the
IR regular contributions. While, the remaining part becomes
\begin{align*}
 &  {\raisebox{-5pt}{${}_{I\!}$}}
 \Bigl\langle \PSHIt{a+1}\Bigr\vert
 \left[\dGbz_I(t_a) \delta^m\!_l +  \dGbg^m_{\,\,\,l\,I}(t_a)/2 \right] 
 \left[ \tilde{\pi}_I(x_a) x^l \partial_m \tilde{\zeta}_I(x_a) +
 \tilde{\pi}_{\,\,\,I}^{ij}(x_a) x^l \partial_m \delta
 \tilde{\gamma}_{ij\,I}(x_a)   \right] 
 \Bigl\vert\PSHIt{a}\Bigr\rangle_{\!I} \,. 
\end{align*} 
Similarly, we can replace $\dot{S}_{ij}(t)$ in the third term of the
interaction Hamiltonian (\ref{Exp:tHI}) with $\dGbg_{ij\,I}(t)$. Notice
that ${^g\!\dot{\bar{\zeta}}}_I(t)$ is recast into
\begin{align}
  {^g\!\dot{\bar{\zeta}}}_I(t) &=  \int \dd^3 \bm{x}\, \partial_t \left\{
 W_t(\bm{x}) \over \int \dd^3 \bm{x} W_t(\bm{x})  \right\}
 \tilde{\zeta}_I(x) +  {\int \dd^3 \bm{x} W_t(\bm{x})\, \partial_t\tilde{\zeta}_I(x) \over
 \int \dd^3 \bm{x} W_t(\bm{x})} \cr 
 &=  
\int \dd^3 \bm{x}\,  \partial_t \left\{
 W_t(\bm{x}) \over \int \dd^3 \bm{x} W_t(\bm{x})  \right\} \left\{  \tilde{\zeta}_{I}(x)
 -\Gbz_I(t) \right\}+ {\int \dd^3 \bm{x} W_t(\bm{x}) \partial_t\tilde{\zeta}_I(x) \over
 \int \dd^3 \bm{x} W_t(\bm{x})}\,,
\end{align}
which is manifestly expressed in the IR suppressed form, 
${\cal R}_x \tilde{\zeta}_I(x)$. 
To make the IR regularity manifest, in the last equality we added 
$0 = \Gbz_I(t)  \partial_t \left\{ \int \dd^3 \bm{x}
 \,W_t(\bm{x}) / \int \dd^3 \bm{x}\, W_t(\bm{x})  \right\}$. 
In a similar manner we can show 
$\dGbg_{ij\,I}(t)$ is also in the IR suppressed form. 

In this way, we can show that all $\tilde{\zeta}_I$s and $\delta \tilde{\gamma}_{ij\,I}$s in the
interaction vertices are multiplied by the IR suppressing operator 
${\cal R}_x$. The argument so far proceeds irrespective of the choice of
the initial quantum states. Now, we focus on the distinctive  
property of the Euclidean vacuum given in Eq.~(\ref{Exp:EV}), 
which states that the initial states chosen by the
boundary condition of the Euclidean vacuum are specified uniquely and
independent of which canonical variables are used for quantization. 
Therefore, requesting 
the Euclidean vacuum uniquely determines the initial state irrespective
of the picked-up particular path of $\zeta^c$ and $\delta \gamma^c$. Therefore, 
after the above-mentioned replacements, the possible dependence of the
$n$-point functions on the picked-up path remains only in 
$ \vert \PSHIt{} \rangle_{I} {}_{I\!}\langle \PSHIt{} \vert$, and hence we 
can remove the decomposition of unity.

\subsubsection{Restricting the interaction vertices to the local region}
\label{RLR}
Next, we will address the inverse Laplacian $\partial^{-2}$. If we
choose the boundary conditions for $\partial^{-2}$ in $N$ and $N_i$
appropriately, $N$ and $N_i$ with their argument $(t,\, \bm{x})$ in the
region ${\cal O}_t$ can be specified by the fluctuations only within 
${\cal O}_t$. 
In the general solutions of $N$ and $N_i$ given in
Eqs.~(\ref{Exp:Nn}) and (\ref{Exp:Nin}), the residual \gauge
degrees of freedom are expressed by arbitrary homogeneous 
solutions of the Laplace equation, $G_n(x)$ and 
$(\delta_i\!^j - \partial_i \partial^{-2} \partial^j)G_{j\, n}(x)$. 
We determine the homogeneous solution $G_n(x)$ such that
the solution in the observable region ${\cal O}_t$ is given by the convolution between the Green function 
and the source restricted to the local region, {\it i.e.},
\begin{align}
 &   - \frac{1}{4 \pi} \int \frac{\dd^3 \bm{y}}{|\bm{x} - \bm{y}|}
W_t(\bm{y}) \partial^i M_{i, n}(t,\, \bm{y}) =  
\partial^{-2}\partial^i M_{i, n}(x) 
-   e^{-2\rho} G_n (x) \,.
\label{Exp:LBC}
\end{align}
Similarly, using the transverse part of $G_{i\, n}(x)$, we can determine the
boundary conditions for the remaining $\partial^{-2}$ 
so as to shut off the influence from the region 
far outside of ${\cal O}_t$. 
(For a detailed explanation, see Refs.~\cite{SRV2, IRreview}.) 
Then, since all the interaction vertices are confined to the
neighborhood of ${\cal O}$, the operation of the non-local operator
$\partial^{-2}$ no longer reduces the power law index with respect to ${k}$. Thus, when we choose
the Euclidean vacuum as the initial states, we can expand the $n$-point
functions for ${\cal R}_x \gz(t_f,\, \bm{x})$ and ${\cal R}_x \ggm_{ij}(t_f,\, \bm{x})$ only
in terms of the interaction picture fields ${\cal R}_x \tilde{\zeta}_I(x)$ and
${\cal R}_x \delta \tilde{\gamma}_{ij\,I}(x)$.

Since ${\cal R}_x \gz(x)$ and ${\cal R}_x \ggm_{ij}(x)$ are not
invariant under all the residual \gauge transformations,  
their $n$-point functions can depend 
on the boundary conditions of $N$ and $N_i$.
However, if we calculate $n$-point functions for the genuinely
\gauge invariant operator $\gR$, 
changing the boundary conditions should not affect the result.

\section{Regularity of loops}  \label{Sec:IRregularity}
In this section, we will show that when we choose the Euclidean vacuum
as the initial state, the $n$-point functions of ${\cal R}_x \gz$ and
${\cal R}_x \ggm_{ij}$ no longer suffer from the IRdiv, IRsec, and SG. 
The discussion in this section goes almost in parallel with the one in Sec.~IV
of Ref.~\cite{SRV2}, where the regularity of the scalar loops is
shown. Here, we briefly highlight the discussion, deferring the more
detailed discussion to Ref.~\cite{SRV2}. 

\subsection{Euclidean vacuum from the $i\epsilon$ prescription}
\label{sec:iepsilon}
In the preceding section, we introduced the Euclidean vacuum, which
satisfies the boundary conditions 
(\ref{Def:EV}). 
Here, following Ref.~\cite{SRV2}, we show that these conditions lead to
the $i\epsilon$ prescription in the ordinary perturbative description. 
For our current purpose, the explicit form of the interaction
Hamiltonian density 
$\tilde{\cal H}_I$ is not necessary. We simply note that all the interaction
vertices in $\tilde{\cal H}_I$ can be formally expressed as
\begin{align}
 & \Mp^2 e^{3\rho(t)} \dot{\rho}^2(t)  \lambda(t)
 \prod_{m^s=1}^{N^s} {\cal R}^{(m^s)}_x \tilde{\zeta}_I(x)
 \prod_{m^t=1}^{N^t} {\cal R}^{(m^t)}_x  \delta \tilde{\gamma}_{i_{m^t}
 j_{m^t}\,I}(x)\,, \label{Exp:tHItemp}
\end{align}
where $N^s$ and $N^t$ are non-negative integers with $N^s + N^t \geq
3$. Here, $\lambda(t)$ is a dimensionless time-dependent function which
can be expressed only in terms of the horizon flow functions. To
discriminate different IR suppressing operators ${\cal R}_x$s, 
we added a subscript $(m^s)$ or $(m^t)$ to ${\cal R}_x$. The spatial indices $i_{m^t}$ and $j_{m^t}$ will be
contracted with other indices $i_{{m^t}'}$ and $j_{{m^t}'}$ or with indices in ${\cal R}_x$, which are
abbreviated for notational simplicity. In the following, we use the
formal expression (\ref{Exp:tHItemp}) as the interaction vertices.

Since the boundary conditions for the Euclidean vacuum should also hold at the
tree level, the asymptotic form of the positive frequency mode function
$v^\alpha_k(t)$ with $\alpha=s$ or $t$, in the limit $\eta\to -\infty$, should be proportional to $e^{-ik\eta(t)}$. 
Factoring out this time-dependence, we express $v^\alpha_k(t)$ as 
\begin{align}
 & v^\alpha_k(t) = \frac{{\cal A}^\alpha(t)}{k^{3/2}} f^\alpha_k(t)
 e^{-i k\eta(t)}\,, \label{Exp:vk} 
\end{align}
where we introduced 
\begin{align}
 & {\cal A}^s(t) \equiv \frac{\dot{\rho}(t)}{\sqrt{\varepsilon_1(t)}
 \Mp}\,, \\
 & {\cal A}^t(t) \equiv \frac{\dot{\rho}(t)}{\Mp}\,,
\end{align}
as approximate amplitudes of the curvature perturbation and the graviton
perturbation. 
The function $f^\alpha_k(t)$ satisfies the regular second order differential 
equation with the boundary condition 
\begin{equation}
 f^\alpha_k(t) \propto {k\over 
e^\rho \dot\rho}\qquad 
\mbox{for}\quad -k\eta(t) \to \infty~.  
\end{equation}
Since both the differential equation and the boundary condition of $f^\alpha_k(t)$
are analytic in $k$ for any $t$, the resulting function should be
analytic as well. In fact, $f^\alpha_k(t)$ does not have any singularity
such as a pole on the complex $k$-plane as a consequence of the
boundary conditions of the Euclidean vacuum.

On the other hand, in the limit $-k\eta(t_k) \ll 1$ the
function $f^\alpha_k(t)$ is proportional to $
 {{\cal A}^\alpha(t_k)/ {\cal A}^\alpha(t)},
$
where $t_k$ is the Hubble crossing time defined by 
$-k\eta(t_k)=1$, because the curvature and graviton perturbations should be constant in this limit. 
Hence, the expansion for small $k$ is in general given by  
\begin{equation}
 {\cal A}^\alpha(t)f^\alpha_k(t)= {\cal A}^\alpha(t_k)\left[1+{\cal O} (k^2|\eta(t)|^2)\right]~.
\label{asymptoticexp}
\end{equation}
By using Eq.~(\ref{Exp:vk}), the Wightman function for the curvature perturbation is given by
\begin{align}
  G^{+\,s}(x,\, x') &= \int \frac{\dd^3 \bm{k}}{(2\pi)^3} e^{i \sbm{k} \cdot
 (\sbm{x} - \sbm{x}')} v^s_k(t) v^{s\,*}_k(t')  \cr
 &= {\cal A}^s(t) {\cal A}^s(t') \int \frac{\dd^3 \bm{k}}{(2\pi)^3} \frac{1}{k^3} e^{i \sbm{k} \cdot
 (\sbm{x} - \sbm{x}')} f^s_k(t) f^{s\,*}_k(t') e^{ik
 (\eta(t')-\eta(t))} \,,
\end{align}
and the Wightman function for the graviton is given by
\begin{align}
  G_{ijkl}^{+ t}(x,\, x') &= \sum_{\lambda= \pm} \int \frac{\dd^3 \bm{k}}{(2\pi)^3} e^{i \sbm{k} \cdot
 (\sbm{x} - \sbm{x}')} e_{ij}^{(\lambda)}(\bm{k}) e_{kl}^{(\lambda)}(\bm{k}) v^t_k(t) v^{t\,*}_k(t')  \cr
&= \frac{{\cal A}^t(t) {\cal A}^t(t')}{2}  \int \frac{\dd^3
 \bm{k}}{(2\pi)^3}  \left( \cP_{ik} \cP_{jl} + \cP_{il} \cP_{jk} -
  \cP_{ij} \cP_{kl} \right) \frac{1}{k^3} e^{i \sbm{k} \cdot
 (\sbm{x} - \sbm{x}')} f^t_k(t) f^{t\,*}_k(t') e^{ik
 (\eta(t')-\eta(t))} \,,
\end{align}
where in the second equality we assumed that the quantum state is
isotropic.

Using the in-in formalism, the $n$-point functions can be expanded by
the Wightman functions $G^{+\,s}(x,\,x')$, $G_{ijkl}^{+ t}(x,\, x')$,
and their complex conjugates. When we impose the boundary conditions of the
Euclidean vacuum, we need to start the vertex integrals at $\eta=-\infty$. Although the vertex integrals are
infinitely oscillating in the limit $\eta\to -\infty$, the time integration can be made
convergent by adding a small imaginary part to the time coordinate, which is nothing but the 
ordinary $i\epsilon$ prescription. To see the convergence of the time
integration more explicitly, using the formal expression for the
interaction vertex (\ref{Exp:tHItemp}), we first consider the integral for the
vertex which is closest to the past infinity $\eta \to -\infty (1-i\epsilon)$. 
The interaction picture field $\tilde{\zeta}_I(x)$ included in this
vertex is contracted with $\tilde{\zeta}_I(x_{m^s})$ in vertices labelled by $m^s=1, 2, \cdots, N^s$, and give the
Wightman function $G^{+\,s}(x_{m^s},\, x)$. Similarly, the interaction picture
field $\delta \tilde{\gamma}_{I\,i_{m^t} j_{m^t}}(x)$ included in the
vertex is contracted with $\delta \gamma_{k_{m^t} l_{m^t}\,I}(x_{m^t})$ 
in vertices labelled by $m^t=1, 2, \cdots, N^t$, and give the
Wightman function $G^{+\,t}_{k_{m^t} l_{m^t} i_{m^t} j_{m^t}}(x_{m^t},\, x)$. 
Then, the vertex integration with $N^s$ $\tilde{\zeta}_I$s and $N^t$
$\delta \tilde{\gamma}_{ij\,I}$s gives
\begin{align}
 V^{(1)}(t',\, \{x_{m^\alpha}\})&\equiv \Mp^2 \int^{t'}_{t_i} \dd t \int \dd^3 \bm{x}\,
 e^{3\rho(t)} \dot{\rho}(t)^2  \lambda(t)
 \prod_{m^s=1}^{N^s} {\cal R}_{x_{m^s}} {\cal R}_{x}^{(m^s)}
 G^{+\,s}(x_{m^s}, x) \cr
 & \qquad \qquad \times  \prod_{m^t=1}^{N^t} {\cal R}_{x_{m^t}} 
{\cal R}_{x}^{(m^t)} G^{+\,t}_{k_{m^t} l_{m^t} i_{m^t} j_{m^t}}(x_{m^t},\, x) ~,
\label{V1vertex}
\end{align}
where $x_{m^\alpha}$ denotes either $x_{m^s}$ or $x_{m^t}$. The
Euclidean vacuum condition requires the convergence of this integral in
the limit $\eta(t_i)\to -\infty$. Since the Wightman functions
contain the exponential factor 
$$e^{i \eta(t) ({\sum}_{m^s} k_{m^s} + {\sum}_{m^t} k_{m^t})},$$ 
the integral can be made convergent by adding $+i\epsilon$ to $\eta(t)$, 
which is again exactly what is known as the $i\epsilon$ prescription. Here,
$k_{m^s}$ denotes the momentum of $G^{+\,s}(x_{m^s}, x)$
and $k_{m^t}$ denotes the momentum of $G^{+\,t}_{k_{m^t} l_{m^t} i_{m^t} j_{m^t}}(x_{m^t},\, x)$.

The vertex integration next to the closest to
the past infinity 
\begin{align}
  V^{(2)}(t'', \{x_{m^\alpha}\}, \{x_{{m^\alpha}'}\}) 
 &\equiv \Mp^2 \int^{t''}_{t_i} \dd t' \int \dd^3 \bm{x}' e^{3\rho(t')}
  \dot{\rho}(t')^2 \lambda(t')
 \prod_{{m^s}'=1}^{{N^s}'} {\cal R}_{x_{{m^s}'}} {\cal R}_{x'}^{({m^s}')} G^{+\,s}(x_{{m^s}'},
 x')  \cr
 & \qquad   \times  \prod_{{m^t}'=1}^{{N^t}'} {\cal R}_{x_{{m^t}'}}
 {\cal R}_{x'}^{({m^t}')}
G^{+\,t}_{k_{{m^t}'} l_{{m^t}'} i_{{m^t}'} j_{{m^t}'}}(x_{{m^t}'}, x')  V^{(1)}(t',\, \{x_{{m^\alpha}'}\})
\,,  \label{Exp:V2}
\end{align}
can be done in a similar manner, where 
${N^s}'$ and ${N^t}'$ are numbers of scalar and graviton propagators that connect between this second 
vertex and the vertices other than the first one. 
If we perform the integration over the time coordinate of the first vertex $t$ up to $t'$,  
the exponential factor in $G^{+\,s}(x_{m^s}, x)$ or $G^{+\,t}_{k_{m^t} l_{m^t} i_{m^t} j_{m^t}}(x_{m^t},\, x)$
can be replaced as 
\begin{align}
 & e^{i k_{m^\alpha} (\eta(t)-\eta(t_m))}\,\,
 \to 
 e^{i k_{m^\alpha} (\eta(t')-\eta(t_m))}\,. 
\end{align}
Therefore all the Wightman functions connecting the vertices 
at $t'$ or in the past of $t'$ with the vertices in the future of $t'$ give an 
exponential factor which is suppressed by adding $+i\epsilon$ to 
$\eta(t')$. (Here, we mean the future and past in the chronological
sense, and not those in the sense of the CTP.) This is again consistent with the boundary condition of the
Euclidean vacuum. The same argument can be made for the other vertices
as well.

In this subsection, considering the time integration at vertices with
fixed momenta of the Wightman propagators, we showed that the
boundary condition of the Euclidean vacuum can be imposed in
perturbative expansion by employing the $i\epsilon$ prescription. 
However, as we will describe in the next subsection, in our proof of the
IR regularity, we will perform the momentum integration of the
propagator ahead of the vertex integration.

\subsection{IR/UV suppressed Wightman function}       
\label{SSec:Wightman}
Since all $\tilde{\zeta}_I(x)$s and $\tgij(x)$s in the interaction
Hamiltonian are multiplied by the IR suppressing operators 
${\cal R}_x$, the $n$-point function of ${\cal R}_x \gz(x)$ and ${\cal R}_x \ggm_{ij}(x)$ can be
expanded by the Wightman function 
${\cal R}_x {\cal R}_{x'} G^{+\,s}(x,\,x')$ and 
${\cal R}_x {\cal R}_{x'} G^{+\,t}_{ijkl}(x,\,x')$ and their complex conjugates.
In this subsection, we calculate the Wightman functions multiplied by
the IR suppressing operator, ${\cal R}_x {\cal R}_{x'} G^+(x,\, x')$ and
${\cal R}_x {\cal R}_{x'} G^{+\,t}_{ijkl}(x,\,x')$ for $t>t'$. 
After integration over the
angular part of the momentum, the Wightman function 
${\cal R}_x {\cal R}_{x'} G^{+\,s}(x,\,x')$ is given as
\begin{align}
  {\cal R}_x {\cal R}_{x'} G^{+\,s}(x,\, x') 
 &= {1 \over 2\pi^2} \int^{\infty}_0 \frac{\dd
  k}{k}\,  {\cal R}_x{\cal R}_{x'} {\cal A}^s(t) f^s_k(t)
{\cal A}^s(t') f_k^{s*}(t')
\left[{e^{i k \sigma_+(x, x')} -e^{i k \sigma_-(x, x')} \over
  i k (\sigma_+(x, x') -  \sigma_-(x, x'))}
\right]\,, \label{Exp:DG+s}
\end{align}
where we introduced
$$
\sigma_{\pm}  (x,\, x') \equiv \eta(t')-\eta(t) \pm  |\bm{x} - \bm{x}'|\,. 
$$
The Wightman function ${\cal R}_x {\cal R}_{x'} G^{+\,t}_{ijkl}(x,\,x')$
is given by a similar expression as
\begin{align}
 {\cal R}_x {\cal R}_{x'} G_{ijkl}^{+ t}(x,\, x') &=  {1 \over 4\pi^2}
 \int^{\infty}_0  \frac{\dd k}{k}\, {\cal R}_{x'} \left( \cPd_{ik} \cPd_{jl} + \cPd_{il} \cPd_{jk} - 
  \cPd_{ij} \cPd_{kl} \right) \cr
 & \qquad \quad \times  {\cal R}_x {\cal A}^t(t) f^t_k(t)
{\cal A}^t(t') f_k^{t*}(t') \left[{e^{i k \sigma_+(x, x')} -e^{i k \sigma_-(x, x')} \over
  i k (\sigma_+(x, x') -  \sigma_-(x, x'))}
\right]\,. \label{Exp:DG+t}
\end{align}
Here, before we integrate over the angular coordinates, we replaced the
projection tensor $\cP_{ij}$ with the derivative form: 
\begin{align}
 & \cPd_{ij} = \delta_{ij} - \partial_{{x'}^i} \partial_{{x'}^j}  \partial_{\sbm{x}'}^{-2}\,,
\end{align} 
which commutes with ${\cal R}_x$. 

We first show the regularity of the $k$ integration in
Eqs.~(\ref{Exp:DG+s}) and (\ref{Exp:DG+t}). The regularity of the
Wightman function $G^{+\,s}(x, x')$ is shown in Ref.~\cite{SRV2}. We
will see that the same argument also leads to the regularity of
$G^{+\,t}_{ijkl}(x, x')$. Since the functions
$f^\alpha_k(t)$ with $\alpha=s,\,t$ are not singular, the
regularity can be verified if the integration converges both in the IR
and UV limits. The regularity in the IR limit is guaranteed by the
presence of the IR suppressing operator. The IR suppressing 
operators ${\cal R}_x$ add at least one extra factor of 
$k|\eta(t)|$ or eliminate the leading
$t$-independent term in the IR limit, and yield
\begin{align}
 {\cal R}_x  {\cal A}^s(t) f^s_k(t) \left[{
e^{i k \sigma_+(x, x')} -e^{i k \sigma_-(x, x')} \over
  i k (\sigma_+(x, x') -  \sigma_-(x, x'))} \right] 
  &={\cal A}^s(t_k) e^{i k \eta(t')} \, {\cal O} \left( k |\eta(t)| \right)
\cr
&= {\cal A}^s(t) e^{i k \eta(t')} \, {\cal O}
\left( \left\{  k |\eta(t)|
 \right\}^{(n_s+1)/2} \right)\,, \label{Eq:orders}
\end{align}
and
\begin{align}
 {\cal R}_x  {\cal A}^t(t) f^t_k(t) \left[{
e^{i k \sigma_+(x, x')} -e^{i k \sigma_-(x, x')} \over
  i k (\sigma_+(x, x') -  \sigma_-(x, x'))} \right] 
&= {\cal A}^t(t) e^{i k \eta(t')} \, {\cal O}\left( \left\{  k |\eta(t)|
 \right\}^{(n_t+2)/2} \right)\,, \label{Eq:ordert}
\end{align}
where we have introduced the spectral indices $n_s$ and $n_t$ as
\begin{align}
 & n_s-1 \equiv \dd \ln(|{\cal A}^s(t_k)|^2)/\dd \ln k \,, \\
 & n_t \equiv \dd \ln(|{\cal A}^t(t_k)|^2)/\dd \ln k \,.
\end{align}
Thus, the operation of ${\cal R}_x$ makes the $k$ integration in Eqs.~(\ref{Exp:DG+s}) and (\ref{Exp:DG+t}) regular in
the IR limit, ensuring the IR regularity. Next, we consider the convergence in the UV limit. 
When we choose the Euclidean vacuum, the $i\epsilon$
prescription facilitates the regularization of the UV modes in
Eqs.~(\ref{Exp:DG+s}) and (\ref{Exp:DG+t}), because adding a small imaginary part to all the time coordinates as
$\eta\to \eta\times (1-i\epsilon)$ leads to the replacement 
$$\eta(t')-\eta(t)\to \eta(t')-\eta(t)+ i\epsilon |\eta(t') - \eta(t)|$$ 
with $\eta(t')-\eta(t) < 0$. Then, the manifest exponential suppression
factor is introduced for large $k$. 
This UV regulator makes the integral finite for the large $k$
contribution except for the case 
$\sigma_{\pm}(x, x') = 0$, where 
$x$ and $x'$ are mutually light-like. 
Since the expression of the
Wightman functions obtained after the $k$ integration is independent of
the value of $\epsilon$, this regulator makes the UV contributions
convergent even after $\epsilon$ is sent to zero. 
For $\sigma_{\pm}(x, x') = 0$, 
the integral becomes divergent in the limit $\epsilon\to 0$, 
but the divergence related to the behavior of the Wightman functions 
in this limit is to be interpreted as the ordinary 
UV divergences, whose contribution to the vertex integrals 
must be renormalized by 
introducing local counter terms. Thus, the Wightman functions 
${\cal R}_x {\cal R}_{x'} G^{\pm\,s}(x,\, x')$ and ${\cal R}_x {\cal
R}_{x'} G^{\pm\,t}_{ijkl}(x,\, x')$ are now shown to be regular
functions.

Since the amplitudes of the Wightman functions with the IR suppressing
operator are bounded from above, we can show the regularity of the $n$-point
functions, if the non-vanishing support of 
the integrands of the vertex integrals is effectively 
restricted to a finite spacetime region. Since the causality has been
established with the aid of the residual \gauge degrees of freedom (see Sec.~\ref{RLR}), the
question to address is whether contributions from the distant past are shut off
or not. In short, this question can be rephrased as whether the SG due to the
time integral exists or not. To address such a long-term
correlation, we discuss the asymptotic behavior of 
the Wightman functions ${\cal R}_x {\cal R}_{x'} G^{\pm\,s}(x,\, x')$
and ${\cal R}_x {\cal R}_{x'} G^{\pm\,t}_{ijkl}(x,\, x')$,
sending $t'$ to a distant past. Recall that when $\sigma_{\pm}(x,x')\ne 0$, 
we can rotate the integration contour with respect to $k$ even toward the direction parallel 
to the imaginary axis, making $\epsilon$ finite. Rotating the 
direction of the path appropriately depending on the sign of
$\sigma_{\pm}(x,x')$, the integrand shows an exponential decay for 
$k \agt 1/|\sigma_\pm(x,x')| \simeq 1/|\eta(t')|$. 
Since we send $t'$ to the past infinity, where $|\eta(t')| \gg |\eta(t)|$, $\sigma_\pm(x,x')$ becomes
${\cal O}(|\eta(t')|)$, except for the region where the two points are
mutually light-like~(see Ref.~\cite{SRV2} regarding the estimation of the contribution
from this region). The rotation of the $k$ integration contour can be
done without hitting any singularity in the complex $k$-plane, 
because the functions $f^\alpha_k(t)$ are guaranteed to be analytic by 
construction. If we choose other vacua, this operation yields 
extra contributions from singularities. 
After the rotation, the integrations of $k$ on the right-hand sides of 
Eqs.~(\ref{Exp:DG+s}) and (\ref{Exp:DG+t}) are totally dominated
by the wavenumbers with $k \alt 1/|\eta(t')| \ll 1/|\eta(t)|$.
Using Eq.~(\ref{Eq:orders}) which gives the asymptotic expansion in the
limit $k |\eta(t)| \ll 1$, we obtain
\begin{align}
  {\cal R}_x {\cal R}_{x'} G^{+\,s}(x,\, x') 
 &= {\cal A}^s(t) \times {\cal O} \left[ 
  \int^{\infty}_0 \frac{\dd k}{k} \left\{  k |\eta(t)| \right\}^{(n_s+1)/2} {\cal R}_{x'} {\cal A}^s(t') f_k^{s\,*}(t')
 e^{i k \eta(t')}  \right] \cr
 & = {\cal A}^s(t) {\cal A}^s(t') \,
 {\cal O}\!\left(\left( |\eta(t)| \over |\eta(t')| \right)^{{n_s+1 \over 2}}\right)
\label{OoMs},
\end{align}
where in the second equality we performed the $k$ integration, rotating
the integration contour. Similarly, using Eq.~(\ref{Eq:ordert}), we
obtain
\begin{align}
  {\cal R}_x {\cal R}_{x'} G^{+\,t}_{ijkl}(x,\, x') 
 &= {\cal A}^t(t) {\cal A}^t(t') \,
 {\cal O}\!\left(\left( |\eta(t)| \over |\eta(t')| \right)^{{n_t+2 \over
 2}}\right)
\label{OoMt}.
\end{align}
We should emphasize that we do not employ the
long wavelength approximation regarding the Hubble scale at $t'$ to
properly evaluate the modes $k$ of ${\cal O}(1/|\eta(t')|)$ as well.

\subsection{Secular growth (SG) due to the time integral}
\label{SSec:regularity}
In this subsection, focusing on the long-term correlation, we discuss
the convergence of the vertex integrals of the $n$-point functions for
the Euclidean vacuum. We start with the integration of the $n$-point interaction 
vertex which is the closest to $\eta=- \infty(1-i\epsilon)$.
By inserting the expression of the Wightman functions 
${\cal R}_x {\cal R}_{x'} G^+(x,\, x')$ and 
${\cal R}_x {\cal R}_{x'} G^{+\,t}_{ijkl}(x,\, x')$ with $t \gg t'$, given in Eqs.~(\ref{OoMs})
and (\ref{OoMt}) into Eq.~(\ref{V1vertex}), the vertex integral $V^{(1)}$ can be estimated as 
\begin{align}
  V^{(1)}(t',\,\{x_{m^\alpha}\}) &= {\cal O}\!\Biggl[ \Mp^2
 \int^{t'}_{t_i} \dd t \int \dd^3 \bm{x}\,
 e^{3\rho(t)} \dot{\rho}(t)^2  \lambda(t) 
\{{\cal A}^s(t)\}^{N^s} \prod_{m^s=1}^{N^s} {\cal A}^s(t_{m^s}) \left(
 \eta(t_{m^s})  \over \eta(t) \right)^{{n_s+1 \over 2}} \cr
 & \qquad \qquad \times \{{\cal A}^t(t)\}^{N^t} \prod_{m^t=1}^{N^t} {\cal A}^t(t_{m^t}) \left(
 \eta(t_{m^t})  \over \eta(t) \right)^{{n_t+2 \over 2}} \Biggr]\,. 
\end{align}
As we have explained in Sec.~\ref{RLR}, 
the interaction vertices are confined within the
observable region, {\it i.e.}, the non-vanishing support of 
the integrand is bounded by $|\bm{x}| \alt L_t$, where
$L_t$ can be approximated by $|\eta(t)|$ in the distant past. Thus, we obtain
\begin{align}
V^{(1)}(t',\,\{x_{m^\alpha}\}) &= {\cal O}\Biggl[
 \int^{\eta(t')}_{-\infty}  \frac{\dd \eta}{\eta} \lambda(\eta)  
\{{\cal A}^s(\eta)\}^{N^s}  \{{\cal A}^t(\eta)\}^{N^t-2} \cr
 & \qquad \qquad \times \prod_{m^s=1}^{N^s} {\cal A}^s(t_{m^s}) \left(
 \eta(t_{m^s})  \over \eta \right)^{{n_s+1 \over 2}} \prod_{m^t=1}^{N^t} {\cal A}^t(t_{m^t}) \left(
 \eta(t_{m^t})  \over \eta \right)^{{n_t+2 \over 2}} \Biggr] \,.  \label{Exp:V1f}
\end{align}
Since we have performed momentum integral first, 
the exponential suppression for large $|\eta|$, required for the
Euclidean vacuum, no longer remains. However, picking up $\eta$-dependence of the integrand 
of Eq.~(\ref{Exp:V1f}), we still find that the contribution 
from the distant past is suppressed if 
\begin{align}
 &  \left| \lambda(\eta) \left\{ 
{\cal A}^s(\eta)\right\}^{N^s} \left\{ 
{\cal A}^t(\eta)\right\}^{N^t-2}  \eta^{-{N^s (n_s +1) +N^t (n_t +2) \over 2} }  \right| \to 0
 \qquad {\rm as} \quad  \eta \to - \infty\,.  \label{Exp:Cond}
\end{align}
When this condition is satisfied, the time integral converges, and 
the amplitude of $V_n^{(1)}(\eta',\,\{x_m\})$ is 
estimated by the value of the integrand at 
the upper end of the integration as
\begin{align}
 &  V^{(1)}(t',\,\{x_{m^\alpha}\}) \cr
 & = {\cal O}\left[ \lambda(t') \{{\cal A}^s(t')\}^{N^s}  \{{\cal
 A}^t(t')\}^{N^t-2} \prod_{m^s=1}^{N^s} {\cal A}^s(t_{m^s}) \left(
 \eta(t_{m^s})  \over \eta(t') \right)^{{n_s+1 \over 2}} \prod_{m^t=1}^{N^t} {\cal A}^t(t_{m^t}) \left(
 \eta(t_{m^t})  \over \eta(t') \right)^{{n_t+2 \over 2}}
\right]
\,.
\end{align}
Then, when one of the Wightman propagators is connected to a vertex located in the future of $x'$, 
 {\it i.e.}, $t_m>t'$, the $t$-integration yields the suppression factor 
$\{\eta(t_{m^s})/ \eta(t') \}^{{n_s+1\over2}}$ or $\{\eta(t_{m^t})/ \eta(t') \}^{{n_t+2\over2}}$.   
We denote the numbers of such scalar and graviton propagators by $\tilde{N}^s$ and
 $\tilde{N}^t$, respectively.

Similarly, we can evaluate the amplitude of $V^{(2)}$ as 
\begin{align}
  &V^{(2)}(t'',\, \{x_{m^\alpha}\}, \{x_{{m^\alpha}'}\}) \cr
 & ={\cal O}\Biggl[
   \int^{\eta(t'')}_{-\infty} \frac{\dd \eta'}{\eta'} \lambda'(\eta')
 \{{\cal A}^s(\eta')\}^{{N^s}'}  \{{\cal A}^t(\eta')\}^{{N^t}' - 2} \cr
 & \qquad \qquad \times \prod_{{m^s}'=1}^{{N^s}'}  {\cal A}(t_{{m^s}'})
 \left( \eta(t_{{m^s}'}) \over \eta' \right)^{{n_s+1 \over 2}} \prod_{{m^t}'=1}^{{N^t}'}  {\cal A}(t_{{m^t}'})
 \left( \eta(t_{{m^t}'}) \over \eta' \right)^{{n_t+2 \over 2}}  \,
 V^{(1)}(t(\eta'),\,\{x_m\}) \Biggr] \,. 
\end{align}
Extracting the $\eta'$-dependent part in the above expression, 
we obtain  
\begin{align}
   \int^{\eta(t'')}_{-\infty} \frac{\dd \eta'}{\eta'} \lambda(\eta') \lambda'(\eta')  
  \{{\cal A}^s(\eta')\}^{N^s+{N^s}'} \{{\cal A}^t(\eta')\}^{N^t+{N^t}' -
 4} |\eta'|^{-{n_s+1\over 2}( {N^s}' + \tilde{N}^s) -{n_s+2\over 2}( {N^t}' + \tilde{N}^t)} \,. 
\end{align}
Now the generalization proceeds in a straightforward way. For the $N_v$-th vertex, the
temporal integration becomes 
\begin{equation}
    \int \frac{\dd \eta_{N_v}}{\eta_{N_v}} \hat\lambda(\eta_{N_v})
     \{{\cal A}^s(\eta_{N_v})\}^{N_f^s}  \{{\cal A}^t(\eta_{N_v})\}^{N_f^t - 2
     N_v}  |\eta_{N_v}|^{-{(n_s+1)M^s + (n_t +2)M^t\over 2}}\,, \label{Exp:Ccond0}
\end{equation}
where $N^s_f$ and $N^t_f$, respectively, 
denote the numbers of $\tilde{\zeta}_I$s and
$\tgij$s 
contained in the vertices which have been integrated
before the $N_v$-th vertex, $M^s$ and $M^t$ denote the numbers of the Wightman 
propagators connected to a vertex with $\eta>\eta_{N_v}$, and
$\hat{\lambda}$ is the product of all the interaction
coefficients contained in the integrated vertices. 
Thus, the convergence condition is derived as
\begin{align}
 &  \left| \hat\lambda(\eta)  \{{\cal A}^s(\eta)\}^{N_f^s} 
 \{{\cal A}^t(\eta)\}^{N_f^t - 2 N_v} \eta^{-{(n_s+1)M^s + (n_t +2)M^t\over 2}}
 \right| \to 0  \qquad {\rm as} \quad  \eta \to - \infty\,. \label{Exp:Ccond}
\end{align}

As a simple example, we consider the case where $\varepsilon_1$ is
constant. In this case, $\hat{\lambda}$ is expressed only in terms of
$\varepsilon_1$ and takes a constant value. By assuming $M=1$ and using
$n_s-1 = - 2\varepsilon_1$, the convergence condition yields 
\begin{align}
   (1- \varepsilon_1)^2 M -  \varepsilon_1 N > 0~,
 \label{Exp:Ccondg2}
\end{align}
where $N\equiv N_f^s + N_f^t - 2N_v$ and $M \equiv M^s + M^t$. In the
slow roll limit $\varepsilon_1 \ll 1$, the above condition is recast
into 
\begin{equation}
  N  < {\cal O} \left( \frac{M}{\varepsilon_1} \right)\,. \label{Exp:Ccondg}
\end{equation}
Since all interaction vertices contain at least one propagator which is
connected to a vertex in their future, $M$ should be $M \geq 1$. 
Therefore, unless an extremely higher order in perturbation with $N >
{\cal O}(1/\varepsilon_1)$ is concerned, the contributions from the
distant past are suppressed and hence the time integrals at the
interaction vertices do not yield the SG.

The presence of the above suppression can be intuitively understood in the
same way as in the discussion for the loops of the curvature
perturbation~\cite{SRV2}. When we choose the Euclidean vacuum as the
initial state, both
the IR and UV modes in the Wightman functions are suppressed and then
only the contributions around the Hubble scale at each time are left
unsuppressed. Being affected only by the modes around the Hubble scale,
{\it i.e.}, 
$k|\eta| \simeq k/e^{\rho} \dot{\rho}  = {\cal O}(1)$, 
the Wightman functions ${\cal R}_x {\cal R}_{x'} G^{\pm\,s}(x,\, x')$ 
and ${\cal R}_x {\cal R}_{x'} G^{\pm\,t}_{ijkl}(x,\, x')$ 
are necessarily suppressed when $\eta(t)/\eta(t') \ll 1$. 
This is because if the spacetime points $x$ and $x'$ are largely
separated in time, any Fourier mode 
in the Wightman function cannot be of order of the Hubble scale 
simultaneously at $t$ and $t'$. 
When we consider the contribution of vertices located far in the past,  
at least one Wightman function should satisfy $\eta(t)/\eta(t') \ll 1$,
and therefore it is suppressed. Equation (\ref{Exp:Ccond}) shows 
such suppression by $M^s$ scalar propagators and $M^t$ graviton
propagators. As is shown in Eq.~(\ref{Exp:Ccond}), as we increase the
number of operators included in or connected to the interaction vertex,
denoted by
$N^s_f$ and $N^t_f$, the contributions from the distant past come to less
suppressed. On the other hand, as we increase the number of propagators
connected to the vertices around the observation time, labelled by $M^s$
and $M^t$, the contributions from the distant past are more suppressed. 
When $N$ is sufficiently large, {\it i.e.,} $N > 
{\cal O}(M/\varepsilon_1)$, the
suppression due to $M$ propagators can be overwhelmed by 
the large amplitude of the fluctuation, which increases when
the energy scale of inflation increases as in the far past. 
However, we should also stress that the SG never appears in the slow
roll inflation, 
unless the order of perturbative expansion $N$ takes an extremely large
value such as $1/\varepsilon_1 \simeq {\cal O}(10^2)$.

\section{Concluding remarks}
In this paper, we addressed the regularity of the graviton
loops. We showed that when we choose the
Euclidean vacuum as the initial state, similarly to the curvature
perturbation, the graviton perturbation does not cause the IRdiv
and IRsec in the $n$-point functions of genuine \gauge invariant
operators. In the absence of the graviton, simply performing the dilatation
transformation provides the new set of canonical variables 
in which all $\zeta$s in the Hamiltonian are shifted by the free 
parameter $s$. Presence of this new set of canonical variables is important to show 
the IR regularity for the Euclidean vacuum. 
Extending this previous result to the graviton perturbation, 
we provided the new set of canonical variables whose Hamiltonian
includes the curvature perturbation and the graviton perturbation with
the shifts by arbitrary time-dependent parameters $s$ and $\s_{ij}$,
respectively. Then, following a similar argument to the one in Ref.~\cite{SRV2}, we 
established the IR regularity, {\it i.e.}, the absence of the IRdiv and
IRsec for the Euclidean vacuum to any order of perturbation.  
We also showed the absence of the SG in the slow roll inflation, 
at least, unless the extremely high orders in perturbation are concerned.

As is argued also in Ref.~\cite{SRV2}, when we evaluate the SG,
considering only the superH modes is not sufficient, 
because all modes are subH modes when we send the initial time $t_i$ to 
the past infinity. In Sec.~\ref{SSec:regularity}, to evaluate the SG,
we used the Wightman functions obtained in Sec.~\ref{SSec:Wightman}. 
These Wightman
functions ${\cal R}_x {\cal R}_{x'} G^{\pm\,s}(x,\, x')$ and 
${\cal R}_x {\cal R}_{x'} G^{\pm\,t}_{ijkl}(x,\,x')$ 
are shown to take finite values, as long as the two arguments
$x$ and $x'$ are not mutually light-like. In this paper and also in Ref.~\cite{SRV2},
assuming that these UV divergences will be renormalized by
local counter terms, we did not explicitly examine the
contributions from the singular UV modes. We leave
a detailed discussion about the UV renormalization for future study. 
(See Refs.~\cite{SZ09, ABG}, where the UV regularization is
discussed. )

In this paper, we considered the inflationary universe as the background
spacetime. When we take an exact de Sitter space as the background
spacetime without introducing a scalar field, the curvature perturbation
will disappear, while the graviton perturbation can still
exist. For the pure gravity in the de Sitter limit, 
the accumulation of residual \gauge degrees of freedom is still 
an issue of debate. It has been claimed that the IR graviton
can become a trigger of the running of the coupling constant. For
instance, in Ref.~\cite{TW96}, Tsamis and Woodard claimed that the IR
graviton can screen the cosmological constant, suggesting a possibility
that the cosmological constant problem might be dynamically solved. In our
forthcoming publication, we will address the IR issues of the graviton
in the exact de Sitter background and discuss whether the screening of
the cosmological constant can still exist even if we request the genuine
\gauge invariance.

Finally, we make several comments on the 
quantum states allowed from the IR regularity conditions. We have seen that when
we choose the Euclidean vacuum as the initial state, the $n$-point functions of
the genuine \gauge invariant operator become IR regular. Then, the
question is whether 
the regularity can be maintained for other initial states or not. In
the simple setup adopted in Appendix \ref{Sec:SRV}, which immediately
ensures the standard commutation relations for the interaction picture
fields, we found that requesting the IR regularity of the graviton
loops yields the same condition on the mode function $v_k^s$ that was 
requested from the IR regularity of the loop corrections due to the curvature
perturbation. (In Ref.~\cite{IRgauge}, we claimed that the IR regularity
of the graviton loops does not yield any condition on $v_k^s$. However,
as is mentioned in Appendix \ref{Sec:SRV}, in
Ref.~\cite{IRgauge}, we chose an alternative heuristic
iteration scheme which does not immediately guarantee the standard
commutation relations for the interaction picture fields. 
Therefore there is no contradiction with the current result.) It will be
intriguing to elaborate how strictly the IR
regularity condition constrains the quantum state in the
inflationary universe. We will also leave this issue for future study. 
(See also the studies for the scalar field by Einhorn and
Larsen in Refs.~\cite{EL02, EL03} and by Marolf {\it et al}. in
Ref.~\cite{Marolf:2012kh}.)

\acknowledgements 
This work is supported by the Grant-in-Aid for the Global COE Program
"The Next Generation of Physics, Spun from Universality and Emergence"
from the Ministry of Education, Culture, Sports, Science and Technology
(MEXT) of Japan. The discussion during the workshop YITP-X-13-06 was useful to
complete this paper. T.~T. is supported by Monbukagakusho Grant-in-Aid for
Scientific Research Nos.~24103006, 24103001, 21244033,
21111006. Y.~U. is supported by the JSPS under Contract No.~21244033, MEC
FPA2010-20807-C02-02, and AGAUR 2009-SGR-168. Y.~U. would like to thank A.~Higuchi and
H.~Kodama for their valuable comments.

\appendix
\section{Constraining the initial states from the IR regularity}
\label{Sec:SRV}
When we set the initial state to the Euclidean vacuum, as we
mentioned in Sec.~\ref{SSec:Euclidean}, the equivalence between the
two sets of canonical variables which are connected by the residual \gauge
transformations is ensured. Making use of the privileged property of
the Euclidean vacuum, we can write the perturbative expansion in a way
that all the interaction picture fields are associated with the IR
suppressing operator ${\cal R}_x$, which plays the crucial role in
showing the IR regularity. This consideration suggests that the IR
regularity will not be guaranteed for an arbitrary choice of the initial
state. In this section, we will show that the requirement of IR regularity
actually yields a non-trivial restriction on the quantum state, choosing
a simple setup where the interaction is turned on at a finite initial
time $t_i$. In this appendix all field variables without the subscript $I$  
are supposed to be those in the Heisenberg picture.

\subsection{Solving the equations of motion}   \label{SSec:SRIC}
In this subsection, we compute the two-point function of 
${\cal R}_x \gz(x)$ up to the one-loop order to derive the IR regularity condition 
on the initial state. 
Assuming that the interaction is turned on at the initial time $t_i$, we set 
the relation between the Heisenberg fields and the interaction picture
fields as 
\begin{align}
 & \zeta(t_i,\, \bm{x}) = \zeta_I(t_i,\, \bm{x})\,,\qquad \quad
  \pi(t_i,\, \bm{x}) = \pi_I(t_i,\, \bm{x})\,, \label{IC}
\end{align}
and
\begin{align}
 &  \delta \gamma_{ij}(t_i,\, \bm{x}) = \delta \gamma_{ij\, I}(t_i,\, \bm{x})\,,\qquad \quad
  \pi_{ij}(t_i,\, \bm{x}) = \pi_{ij\, I}(t_i,\, \bm{x})\,, \label{IC2}
\end{align}
where $\pi_I$ and $\pi_{ij\,I}$ are the conjugate momenta of the interaction picture fields
$\zeta_I$ and $\delta \gamma_{ij\,I}$, respectively. The advantage of choosing this initial condition is that the
commutation relations for the Heisenberg field $\Phi$, given in
Eqs.~(\ref{Exp:Comz}) and (\ref{Exp:Comg}), immediately guarantee the
standard commutation relations also for the interaction picture fields,
{\it i.e.},
\begin{align}
 & \left[ \zeta_I(t,\, \bm{x}),\, \pi_I(t,\, \bm{y}) \right]= i
 \delta^{(3)} (\bm{x} - \bm{y}) \,, \qquad \left[ \zeta_I(t,\, \bm{x}),\,
 \zeta_I(t,\, \bm{y}) \right] = \left[ \pi_I(t,\, \bm{x}),\, \pi_I(t,\,
 \bm{y}) \right]=0\,, \label{Exp:ComzI}
\end{align}
and
\begin{align}
 & \left[ \delta \gamma_{ij\,I}(t,\, \bm{x}),\, \pi^{kl}_I(t,\, \bm{y})
 \right]= i \delta^{(3)}_{~~ij}\,\!^{kl}(\bm{x} - \bm{y}) \,, \quad \left[ \delta \gamma_{ij\,I}(t,\, \bm{x}),\,
 \delta \gamma_{kl\,I}(t,\, \bm{y}) \right] = \left[ \pi^{ij}_I(t,\,
 \bm{x}),\, \pi^{kl}_I(t,\, \bm{y}) \right]=0\,. \label{Exp:ComgI}
\end{align}

Here, we compute the two-point function of ${\cal R}_x \gz(x)$ by solving the
Heisenberg operator equations of motion for $\zeta$ and 
$\delta \gamma_{ij}$. Using the retarded Green functions $G_R(x,\, x')$ and 
$G_{R\,ijkl}(x, x')$ given by
\begin{align}
 & G_R(x,\, x') = - i \theta(t - t') \left[ \zeta_I(x),\, \zeta_I(x')
 \right]\,, \\
 & G_{R\,ijkl}(x,\, x') = - i \theta(t - t') \left[ \delta
 \gamma_{ij\,I}(x),\, \delta \gamma_{kl\,I}(x')
 \right]\,,
\end{align}
we can solve the equations of motion for $\zeta$ and $\delta
\gamma_{ij}$, employing the initial conditions (\ref{IC}) and (\ref{IC2}) as
\begin{align}
 &  \zeta (x) = \zeta_I(x)  +  \cLRsI {\cal S}_{\rm NL}(x)\,,
\label{SolR}  \\
 &  \delta \gamma_{ij} (x) = \delta \gamma_{ij\,I}(x)  +  \cLRtI {\cal S}_{{\rm NL}\,ij}(x)\,,
\label{SolR2}
\end{align}
with 
\begin{align}
 & \cLRsI {\cal S}_{\rm NL}(t,\, \bm{x})  \equiv - 2 \Mp^2 \int
  \dd^4 x' \varepsilon_1(t') e^{3\rho(t')}  G_R(x, x') {\cal S}_{\rm NL}(x')\,,
 \label{Exp:SolR} \\
 & \cLRtI {\cal S}_{{\rm NL}\,ij}(t,\, \bm{x})  \equiv -  \frac{\Mp^2}{4} \int
  \dd^4 x' e^{3\rho(t')}  G_{R\,kl}\!\,^{kl}(x, x') {\cal S}_{{\rm NL}\,ij}(x')\,,
 \label{Exp:SolR2} 
\end{align}
where the explicit form of the non-linear source terms
${\cal S}_{\rm NL}(x)$ and ${\cal S}_{{\rm NL}\,ij}(x')$
will be derived later. Evaluating Eqs.~(\ref{Exp:SolR}) and (\ref{Exp:SolR2}) iteratively, 
we can obtain expressions for $\zeta$ and $\delta \gamma_{ij}$,
respectively.

Inserting thus obtained solution
$\zeta$ and $\delta \gamma_{ij}$ into Eq.~(\ref{Def:Gbz}), we can perturbatively compute $\gz(x)$ as
\begin{eqnarray}
 \gz(x) = \zeta_I(x) + \gz_2(x) + \gz_3(x) + \cdots \,,  \label{Exp:gzp}
\end{eqnarray}
where $\gz_n(x)$ represents the term that consists of $n$ interaction 
picture fields. 
Expanding the interaction
picture fields $\zeta_I$ and $\delta \gamma_{ij\,I}$ as in Eqs.~(\ref{Exp:zetaI}) and (\ref{Exp:gI}), the initial
vacuum state is defined by
\begin{align}
 & a_{\sbm{k}} |0 \rangle = a^{(\lambda)}_{\sbm{k}} |0 \rangle =0 \,.  \label{Def:vacuum}
\end{align}
Notice that the $n$-point functions computed by using the Heisenberg
operator solved with the retarded Green function can be formally shown
to agree with those calculated in the in-in formalism (see, for
instance, Appendix of Ref.~\cite{SRV1}).

Using Eq.~(\ref{Exp:gzp}), the one-loop contributions to the two-point
function of ${\cal R}_x \gz(x)$ are given by
\begin{align}
 &\langle {\cal R}_{x_1}\! \gz(x_1) {\cal R}_{x_2}\! \gz(x_2)
   \rangle_{1 {\rm loop}}  \cr
  &= \langle {\cal R}_{x_1}\! \gz_2(x_1) {\cal R}_{x_2}\! \gz_2(x_2)
  \rangle  + \langle {\cal R}_{x_1}\! \zeta_I(x_1) {\cal R}_{x_2}\! \gz_3(x_2)
  \rangle + \langle {\cal R}_{x_1}\! \gz_3(x_1) {\cal R}_{x_2}\! \zeta_I(x_2)
  \rangle\,.  \label{Exp:1loop}
\end{align} 
As is discussed in Sec.~\ref{RLR}, after we choose the boundary
conditions for $\partial^{-2}$ appropriately, the inverse Laplacian does
not enhance the singular behaviour of the superH modes, and hence the
IRdiv and IRsec can appear only from the variances of $\zeta_I(x)$ and 
$\delta \gamma_{ij\,I} (x)$, whose superH contributions give
\begin{align}
 & \langle \bar{\zeta}_I^2(t) \rangle =  \int_{k \leq 1/L_t} \frac{\dd^3
  \bm{k}}{(2\pi)^3} P^s(k)  \propto
 \int_{k \leq 1/L_t} \frac{\dd k}{k} \,,
\end{align}
and
\begin{align}
 & \langle \delta \bar{\gamma}_{ij\,I} (t) \delta \bar{\gamma}_{kl\,I} (t)  \rangle 
 = \left( \delta_{ik} \delta_{jl} + \delta_{il} \delta_{jk} -
   \frac{2}{3}\delta_{ij} \delta_{kl} \right) \langle \delta
 \bar{\gamma}_I^2 (t)  \rangle \,,  \label{Exp:spectrumdg}
\end{align}
with
\begin{align}
 &  \langle \delta \bar{\gamma}_I^2 (t)  \rangle  \equiv  \frac{1}{20
 \pi^2} \int_{k \leq 1/L_t} \frac{\dd k}{k} k^3 P^t(k) \propto \int_{k \leq 1/L_t} \frac{\dd k}{k} \,.
\end{align}
Here $\bar{\zeta}_I(t)$ and $\delta \bar{\gamma}_{ij\,I}(t)$,
respectively, denote
$\zeta_I$ and $\delta \gamma_{ij\,I}$ with only the superH modes, which
mimic their spatially averaged values in ${\cal O}_t$.

When $\gz_2$ includes terms with $\zeta_I$ or $\delta \gamma_{ij\,I}$ without
differentiation, 
the first term in the second line of Eq.~(\ref{Exp:1loop}) can give
$\langle \bar\zeta_I^2 \rangle$ or $\langle \delta \bar{\gamma}^2_I \rangle$. 
These variances can appear also from the second and third terms, when
$\gz_3$ includes terms with two $\zeta_I$s or two $\delta \gamma_{ij\,I}$s without 
differentiation. To make our discussion compact and
transparent, here, we pick up only the potentially divergent
contributions, which yield $\langle \bar\zeta_I^2 \rangle$ or $\langle \delta \bar{\gamma}^2_I \rangle$.  
We introduce the symbol $$\Approx$$ to denote the 
approximate equality neglecting the terms which yield neither
$\langle \bar\zeta_I^2 \rangle$ nor 
$\langle \delta \bar{\gamma}^2_I \rangle$ at the one-loop 
level~\cite{IRgauge_L, IRgauge}.

Now, we derive approximate equations of motion for $\zeta$ and 
$\delta \gamma_{ij}$. In the following, we will use
\begin{align}
  \cLRaI Q_I {\cal R}_x Q'_I
 \Approx Q_I \cLRaI {\cal R}_x Q'_I\,,  \label{Prop1}
\end{align}
with $\alpha=s, t$. Here, $Q_I$ and $Q'_I$ are either $\zeta_I$ or $ \delta \gamma_{ij\,I}$ and
${\cal R}_x$ is a derivative operator which suppresses the IR modes. Equation (\ref{Prop1}) can be 
proved as follows. The Fourier transformation of 
$\cLRsI Q_I {\cal R}_x Q'_I$ is proportional to
\begin{align*}
 & \int \dd^3 \bm{p}\int^t_{t_i} \dd t' \varepsilon_1(t') e^{3\rho(t')}
 \dot{\rho}^2(t') \{v_k(t) v_k^*(t') - v^*_k(t) v_k(t') \}
 Q_{I\,\sbm{p}}(t') \left( {\cal R} Q'_I \right)_{\sbm{k}-\sbm{p}}(t')\,,
\end{align*} 
where $Q_{I\,\sbm{k}}$ and $({\cal R} Q'_I)_{\sbm{k}}$ denote the Fourier
modes of $Q_I$ and ${\cal R} Q'_I$. 
Since
$\left( {\cal R} Q_I \right)_{\sbm{k}-\sbm{p}}(t')-
\left( {\cal R} Q'_I \right)_{\sbm{k}}(t')$ is suppressed and
$Q_{I, \sbm{p}}$ becomes time-independent
in the limit $\bm{p} \to 0$, the IR relevant piece of the 
integrand of the momentum integral can be recast into
\begin{align}
 & Q_{I\,\sbm{p}} \int^t_{t_i} \dd t' \varepsilon_1(t') e^{3\rho(t')}
 \dot{\rho}^2(t')  \{v_k(t) v_k^*(t') - v^*_k(t) v_k(t') \}
  \left({\cal R} Q'_I \right)_{\sbm{k}}(t')\,.  \label{Prop3}
\end{align}
Similarly, we can prove Eq.~(\ref{Prop1}) also for $\cLRtI$.
In the following discussion, we will also use the approximate identities
\begin{align}
 & \cLRaI f(x) \Approx 0\,,  \qquad \quad {\rm for}~~ f(x)
 \Approx 0\,.  \label{Prop2}
\end{align}

In the one-loop corrections to ${\cal R}_x \gz$, $\delta \gamma_{ij\,2}$ 
contributes only through $\gz_3$, and $\delta \gamma_{ij\,n}$ with 
$n \geq 3$ do not contribute. 
Since at least one of two interaction picture fields included in
$\delta \gamma_{ij\,2}$ is suppressed by ${\cal R}_x$, we find 
$\delta \gamma_{ij\,2} \Approx 0$. Then, we find 
\begin{align}
 & \delta \gamma_{ij}(x) \Approx \delta \gamma_{ij\,I}(x)\,,
\end{align}
and hence the one-loop corrections can be given without computing the
non-linear contributions in $\delta \gamma_{ij}$.

Next, we derive an approximate equation of motion for $\zeta$. 
Under the equality $\Approx$,
the non-linear action is reduced to 
\begin{align}
 &  S \Approx \Mp^2 \int  \dd t \dd^3 \bm{x} \,
e^{3(\rho+\zeta)} \varepsilon_1\! 
 \biggl[ (\partial_t \zeta)^2  -
 e^{-2(\rho+\zeta)} \left[ e^{- \delta \gamma}
 \right]^{ij}\! \partial_i \zeta 
 \partial_j \zeta  \biggr] \,,  \label{Exp:SIR_c}
\end{align}
where the terms with more than two fields with differentiation, 
which give neither $\langle \bar{\zeta}_I^2 \rangle$ nor
$\langle \delta \bar{\gamma}_{ij\,I} \delta \bar{\gamma}_{kl\,I} \rangle$, are
abbreviated. The variation of the above action gives 
the equation of motion as
\begin{align}
 & \cLs \zeta(x) = {\cal S}_{\rm NL}(x) \,, \label{EOMzeta}
\end{align}
with
\begin{align}
 & \cLs \equiv \partial^2_t + \left( 3 + \varepsilon_2 \right) \dot{\rho}\,
    \partial_t - e^{-2\rho}\partial^2\,,
\end{align}
and
\begin{align}
 &  {\cal S}_{\rm NL}(x) \Approx e^{-2\rho} 
\left(e^{-2\zeta} [e^{- \delta \gamma}]^{ij} - \delta^{ij} \right)  
 \partial_i \partial_j  \zeta(x)  -  \delta(t - t_i ) (e^{3\zeta} - 1) \partial_t \zeta(x) \,,  \label{SolRIR}
\end{align}
where the last term is added so that the solution 
satisfies the second condition in Eq.~(\ref{IC})~\cite{SRV1}.

\subsection{Computation of $\gz$}  \label{gz}
Here, we solve the equation of motion (\ref{EOMzeta}),
employing the initial conditions (\ref{IC}) and (\ref{IC2}). 
Expanding $\zeta$ as in Eq.~(\ref{Exp:zeta}), the
equation of motion (\ref{EOMzeta}) is recast into 
\begin{align}
 & \cLs \zeta_I = 0\,, \label{Eq:zeta1} \\
 & \cLs \zeta_2 \Approx -  (2\zeta_I \delta^{ij} + \delta \gamma^{ij}\!_I) \nabla_i \nabla_j \delta \zeta_I - 3
  \delta(t - t_i) \zeta_I \partial_t \zeta_I  \,, \label{Eq:zeta2}\\
 & \cLs \zeta_3 \Approx  - 2  
 \left( \zeta_2 \Delta \zeta_I + \zeta_I \Delta \zeta_2 - \zeta_I^2 \Delta \zeta_I
 \right)   \cr & \qquad \qquad \qquad \qquad
+ \frac{9}{2}  \delta(t - t_i)  \zeta_I^2 \partial_t \zeta_I
-  \Bigl\{ \delta \gamma^{ij}\!_I  \nabla_i \nabla_j \zeta_2  - \frac{1}{2}
 (\delta \gamma_I^2)^{ij} \nabla_i \nabla_j \zeta_I  \Bigr\} \,, \label{Eq:zeta30}
\end{align}
where we introduced
\begin{align}
 & \nabla_i \equiv e^{-\rho} \partial_i \,, \qquad \quad \Delta \equiv
 \delta^{ij} \nabla_i \nabla_j\,.
\end{align}
In deriving Eq.~(\ref{Eq:zeta30}), we used 
\begin{align}
 & \partial_t \zeta(t_i,\, \bm{x}) \Approx e^{- 3 \zeta_I(t_i,\,
 \sbm{x})} \partial_t \zeta_I(t_i,\, \bm{x})\,, 
\end{align}
which is derived from the initial conditions~(\ref{IC2}). 
Solving Eqs.~(\ref{Eq:zeta2}) and (\ref{Eq:zeta30}) formally, 
we obtain
\begin{align}
 \zeta_2 &\Approx - \bar{\zeta}_I \cLRsI \left[ 2 \Delta  + 3
 \delta(t - t_i) \partial_t \right] \zeta_I  -  \delta
 \bar{\gamma}^{ij}\!_I \cLRsI \nabla_i \nabla_j \zeta_I\,, \label{Sol:zeta2A} \\
 \zeta_3  & \Approx \frac{1}{2} \bar{\zeta}_I^2 \left[ 4 \cLRsI
 \Delta\, \cLRsI (2 \Delta  + 3
 \delta(t - t_i) \partial_t ) + 4 \cLRsI \Delta + 9 \cLRsI \delta(t - t_i) \partial_t
\right] \zeta_I  \cr
 & \qquad \qquad + \delta \bar{\gamma}^{ij}\!_I \delta
 \bar{\gamma}^{kl}\!_I 
\cLRsI \nabla_i \nabla_j \cLRsI \nabla_k \nabla_l \zeta_I 
 + \frac{1}{2} (\delta \bar{\gamma}^2_I)^{ij} \cLRsI \nabla_i \nabla_j \zeta_I
 \label{Sol:zeta3A} \,,
\end{align}
using the properties of the retarded integration given in
Eqs.~(\ref{Prop1}) and (\ref{Prop2}). Here, we replaced $\zeta_I$ and
$\delta \gamma_{ij\,I}$ with their superH contributions $\bar{\zeta}_I$
and $\delta \bar{\gamma}_{ij\,I}$, which contribute to the IRdiv and IRsec.

Next, using Eqs.~(\ref{Sol:zeta2A}) and (\ref{Sol:zeta3A}), we express
$\gz(t_f,\, \bm{x})$, defined in Eq.~(\ref{Exp:gz}), as
\begin{align}
 & \gz(t_f,\, \bm{x}) = \zeta\left( t_f,\, e^{-\Gbz(t_f)} [e^{-
 \Gbg(t_f)}]^i\!_j x^j \right)\,. \label{Exp:gzap}
\end{align}
Inserting Eq.~(\ref{Sol:zeta2A})
into Eq.~(\ref{Exp:gzap}), we can easily obtain 
\begin{align}
 & \gz_2(t_f,\, \bm{x}) \Approx - \bar{\zeta}_I\, \cDsx 
\zeta_I - \frac{1}{2}
 \delta \bar{\gamma}^{ij}\!_I \cDtx{ij} 
\zeta_I \,,  \label{Exp:gz2}
\end{align}
with
\begin{align}
 & \cDsx \equiv   2 \cLRsI \Delta + 3
 \cLRsI \delta(t - t_i) \partial_t + 
\bm{x} \cdot \partial_{\sbm{x}} \,,
\label{Def:calDs} \\
 & \cDtx{ij} \equiv 2 \cLRsI \nabla_i
 \nabla_j + x_i \partial_j\,. \label{Def:calDt}
\end{align}
The computation of $\gz_3$ is slightly
lengthy but straightforward. 
Using Eqs.~(\ref{Sol:zeta2A}) and (\ref{Exp:gzap}), we find 
\begin{align}
 \gz_3(t_f,\, \bm{x}) &\Approx \zeta_3 + \bar{\zeta}_I^2 \bm{x}  \cdot
 \partial_{\sbm{x}} \cLRsI  (2 \Delta + 3  \delta(t -
 t_i) \partial_t ) \zeta_I + \frac{1}{2}  \bar{\zeta}_I^2 (\bm{x}  \cdot
 \partial_{\sbm{x}})^2 \zeta_I \cr
 & \qquad + \frac{1}{2}\delta \bar{\gamma}^i\!_{j\,I} \delta
 \bar{\gamma}^{kl}\!_I x^j \partial_i \cLRsI \nabla_k \nabla_l
 \zeta_I  + \frac{1}{8} \delta \bar{\gamma}^i\!_{j\,I} \delta
 \bar{\gamma}^k\!_{l\,I} x^j \partial_i x^l \partial_k \zeta_I  \,.  \label{tempgz3}
\end{align}
To rewrite the terms with $x^i \partial_j \cLRsI$ in 
$\gz_3$ into a more tractable form, we use the identity
\begin{align}
 & x^i \partial_j \cLRsI = \frac{1}{2} \left( x^i \partial_j 
 \cLRsI + \cLRsI \cLs x^i \partial_j \cLRsI  \right) \,, \label{temp3}
\end{align}
which obviously holds if $\cLRsI \cLs$ can be replaced with unity. In general, for 
$$
\delta_R \equiv (1-\cLRsI \cLs) x^i \partial_j \cLRsI (\cdots)\,,
$$ 
we have $\cL_s \delta_R=0$, and hence $\delta_R$ is a homogeneous 
solution of the second order differential equation {\it i.e.}, 
$\cLs \delta_R =0$. Since $\delta_R$ and $\partial_t \delta_R$ are
both zero at the initial time, which is automatically satisfied by the
definition of the retarded integral $\cLRsI$, we can confirm
that $\delta_R$ vanishes for all $t \geq t_i$. Using
$$
 \left[ \cLs,\, x^i \partial_j \right]  = - 2 \nabla^i \nabla_j\,,
$$
the right hand side of Eq.~(\ref{temp3}) is further rewritten as 
\begin{align}
& x^i\partial_j \cLRsI
 =  \frac{1}{2} ( x^i\partial_j \cLRsI +
  \cLRsI x^i \partial_j )- \cLRsI
 \nabla^i \nabla_j \cLRsI \,. \label{temp0}
\end{align}
Using Eqs.~(\ref{Sol:zeta3A}), (\ref{tempgz3}), and (\ref{temp0}), we
obtain
\begin{align}
 & \gz_3(t_f,\, \bm{x}) \Approx \frac{1}{2} \bar{\zeta}_I^2 (\cDsx)^2 \zeta_I  +
 \frac{1}{8} \delta \bar{\gamma}^{ij}\!_I \delta \bar{\gamma}^{kl}\!_I 
\cDtx{ij}  \cDtx{kl} \zeta_I  \cr
 & \qquad \qquad \qquad - 3 \bar{\zeta}_I^2 \cLRsI \delta(t - t_i)
 \partial_t \cLRsI \Delta \zeta_I + \frac{9}{2}
 \bar{\zeta}_I^2  \{ \cLRsI
 \delta(t - t_i) \partial_t - ( \cLRsI
 \delta(t - t_i) \partial_t )^2 \} \zeta_I  \,.
\end{align}
Noticing that the definition of $\cLRI$ implies
\begin{align}
 & \cLRsI \delta(t - t_i)
 \partial_t \cLRsI \Delta \zeta_I=0\,, \qquad  \{
 \cLRsI 
 \delta(t - t_i) \partial_t - ( \cLRsI
 \delta(t - t_i) \partial_t )^2 \} \zeta_I = 0\,,
\end{align}
we obtain
\begin{align} 
 & \gz_3(t_f,\, \bm{x}) \Approx \frac{1}{2} \bar{\zeta}_I^2
  \cDsx{}^2 \zeta_I  +
 \frac{1}{8} \delta \bar{\gamma}^{ij}\!_I \delta \bar{\gamma}^{kl}\!_I 
\cDtx{ij} \cDtx{kl} \zeta_I \,. \label{Exp:gz3}
\end{align}
In the above expressions (\ref{Exp:gz2}) and (\ref{Exp:gz3}),
$\bar\zeta_I$ multiplied by the delta function $\delta(t-t_i)$ in
$\cDsx$ should be understood as $\bar{\zeta}_I(t_i)$.

\subsection{One loop corrections}
Using Eqs.~(\ref{Exp:gz2}) and (\ref{Exp:gz3}) into
Eq.~(\ref{Exp:1loop}), we obtain the one loop corrections to 
${\cal R}_x \gz(t_f,\, \bm{x})$ as 
\begin{align}
  &\langle {\cal R}_{x_1}\! \gz(t_f,\, \bm{x}_1) {\cal R}_{x_2}\! \gz(t_f,\, \bm{x}_2)
  \rangle_{1 {\rm loop}} \cr
  & \Approx \frac{1}{2} \langle \bar{\zeta}^2_I(t_i) \rangle {\cal F}_{\rm IRdiv}^s(x_1,\,x_2) +  \frac{1}{2} \langle
 \{\bar{\zeta}_I(t_f) - \bar{\zeta}_I(t_i)  \}^2 \rangle {\cal F}_{\rm IRsec}^s(x_1,\,x_2) + \frac{1}{8}  \langle \delta \bar{\gamma}^{ij}_I(t_f)
 \delta \bar{\gamma}^{kl}_I(t_f) \rangle {\cal F}^t_{ijkl} (x_1,\, x_2)\,,  \label{Exp:1loop2}
\end{align} 
with
\begin{align}
  {\cal F}_{\rm IRdiv}^s(x_1,\,x_2) &\equiv  {\cal R}_{x_1}  {\cal
 R}_{x_2} \Big\langle  2 
 \cDsxj{1} \zeta_I(x_1)\cDsxj{2} \zeta_I(x_2)  + 
 \cDsxj{1}\!\!^2 \zeta_I(x_1) \zeta_I(x_2)  + \zeta_I(x_1) \cDsxj{2}\!\!^2 \zeta_I(x_2)  
 \Big\rangle \,, \\
 {\cal F}_{\rm IRsec}^s(x_1,\,x_2) &\equiv  {\cal R}_{x_1}  
{\cal R}_{x_2} \Big\langle  2 \cDsxj{1}\!\!' \zeta_I(x_1)
 \cDsxj{2}\!\!' \zeta_I(x_2)  + {\cDsxj{1}\!\!'}^2 \zeta_I(x_1)
 \zeta_I(x_2)  + \zeta_I(x_1) {\cDsxj{2}\!\!'}^2  \zeta_I(x_2)  
 \Big\rangle \,, \\
 {\cal F}^t_{ijkl}(x_1,\,x_2) &\equiv  {\cal R}_{x_1} {\cal R}_{x_2} \Big\langle
  2 \cDtxj{1}{ij} \zeta_I(x_1) \cDtxj{2}{kl} \zeta_I(x_2)
 \cr & \qquad \qquad \qquad \qquad \quad   + 
 \cDtxj{1}{ij} \cDtxj{1}{kl} \zeta_I(x_1) \zeta_I(x_2)  +
 \zeta_I(x_1)  \cDtxj{2}{ij} \cDtxj{2}{kl} \zeta_I(x_2)
 \Big\rangle\,,
\end{align}
where we introduced $x_a\equiv (t_f,\, \bm{x}_a)$ for $a=1,\,2$ and
\begin{align}
 & \cDsx{}' \equiv   2 \cLRsI \Delta  + 
\bm{x} \cdot \partial_{\sbm{x}} \,,
\end{align}
which agrees with the trace of $\cDtx{ij}$. The first term in
Eq.~(\ref{Exp:1loop2}) can yield the IRdiv of the curvature perturbation, which can be
removed only if ${\cal F}_{\rm IRdiv}^s(x_1,\,x_2)$ vanishes. The second term,
which accompanies 
\begin{eqnarray*}
  \langle \{\bar{\zeta}_I(t_f) - \bar{\zeta}_I(t_i)  \}^2 \rangle 
     \simeq \int_{1/L_{t_i} \leq k \leq 1/L_{t_f}} \frac{\dd^3
     \bm{k}}{(2\pi)^3} P^s(k) \propto \ln \biggl\{ \frac{e^{\rho(t_f)}
     \dot{\rho}(t_f)}{e^{\rho(t_i)} \dot{\rho}(t_i)} \biggr\} \,,
\end{eqnarray*}
appears to yield the IRsec due to the 
curvature perturbation. This term can be removed only if
${\cal F}_{\rm IRsec}^s(x_1,\,x_2)$ vanishes. The third term appears to yield
the IRdiv and IRsec due to the graviton perturbation, 
which can be removed only if ${\cal F}^t_{ijkl}(x_1,\,x_2)$ vanishes. 

\subsection{IR regularity condition on the mode function}
Next, we discuss a condition which eliminates the IRdiv and IRsec due to the
curvature perturbation and the graviton perturbation. 
One may think that if the conditions, 
\begin{align}
 & \cDsx \zeta_I(x) =0\,, \label{Cond:regheu0} \\
 & \cDsx\,\!' \zeta_I(x) =0\,, \label{Cond:regheu02} \\
 &\cDtx{ij} \zeta_I(x) = 0 \label{Cond:regheu2} 
\end{align}
were fulfilled, ${\cal F}_{\rm IRdiv}^s(x_1,\,x_2)$,
${\cal F}_{\rm IRsec}^s(x_1,\,x_2)$ and ${\cal F}^t_{ijkl}(x_1,\,x_2)$
would vanish, 
and hence the IR regularity can be guaranteed without
imposing any further conditions. 
However, these conditions immediately contradict when we
insert the mode expansion of $\zeta_I$, given in (\ref{Exp:zetaI}), into
Eq.~(\ref{Cond:regheu0}). 
Operating $\bm{x} \cdot \partial_{\sbm{x}}$ on a Fourier mode 
$e^{i\sbm{k}\cdot\sbm{x}}$ yields  
the factor $(\bm{x} \cdot \bm{k}) e^{i\sbm{k}\cdot\sbm{x}}$, which
cannot be canceled by the remaining two terms in
Eq.~(\ref{Cond:regheu0}), since the retarded integral
$\cLRsI$~\cite{SRV1} acting on $e^{i\sbm{k}\cdot\sbm{x}}$ leaves it 
proportional to $e^{i\sbm{k}\cdot\sbm{x}}$. 
Similarly,
Eqs.~(\ref{Cond:regheu02}) and (\ref{Cond:regheu2})
cannot be compatible with the Fourier mode decomposition, as long as we
use the solution with the retarded Green function $\cLRsI$,
fixed by the initial condition (\ref{IC}) and (\ref{IC2}).

Here, following Ref.~\cite{SRV1}, we look for a simple alternative way
to remove the IRdiv and IRsec of the curvature and graviton perturbations. In Ref.~\cite{SRV1}, we pointed out that when
\begin{eqnarray} 
 \cDsx \zeta_I(x)=
\int{\dd^3 \bm{k} \over (2\pi)^{3/2}} \left(
  a_{\sbm{k}} D e^{i\sbm{k}\cdot\sbm{x}} v_k^s+ ({\rm h.c.}) \right)
\label{bettercondition}
\end{eqnarray}
is satisfied, where $D$ is defined as 
\begin{eqnarray}
 & D \equiv k^{-3/2} e^{-i\phi(k)}
   \bm{k} \cdot \partial_{\sbm{k}} k^{3/2}e^{i\phi(k)}\,, 
\end{eqnarray} 
and $\phi(k)$ is an arbitrary phase function $\phi(k)$, 
${\cal F}^s_{\rm IRdiv}(x_1,\,x_2)$ can be summarized in the total
derivative form as 
\begin{align}
 &  {\cal F}^s_{\rm IRdiv}(x_1,\,x_2) = {\cal R}_{x_1} {\cal R}_{x_2}  \int \frac{\dd (\ln k) 
\dd \Omega_{\sbm{k}}
}{(2\pi)^3} \partial^2_{\ln k}
 \left\{ k^3 |v_k^s|^2 e^{i \sbm{k}\cdot (\sbm{x}_1-\sbm{x}_2)}\right\}\,,
\end{align}
where $\int \dd \Omega_{\sbm{k}}$ denotes the
integration over the angular directions of $\bm{k}$. Then, since the integral of a total
derivative vanishes, the IRdiv can be eliminated. Using the mode
expansion (\ref{Exp:zetaI}), the condition (\ref{bettercondition}) can
be recast into a condition on mode functions as
\begin{equation}
  \cLRkI \left( -2 (k e^{-\rho})^2 + 3
 \delta(t - t_i) \partial_t \right)  v_k^s=D v_k^s\,,
\label{Cond:GI}
\end{equation}
where $\cLRkI$ is the Fourier mode of 
$\cLRsI$. Similarly, we can also eliminate the IRsec of the curvature
perturbation, by requesting 
\begin{eqnarray}
 \cDsx\,\!' \zeta_I(x)=
\int{\dd^3 \bm{k} \over (2\pi)^{3/2}} \left(
  a_{\sbm{k}} D e^{i\sbm{k}\cdot\sbm{x}} v_k^s+ ({\rm h.c.}) \right)\,, 
\label{bettercondition2}
\end{eqnarray}
which leads to a slightly different condition from Eq.~(\ref{Cond:GI}) as
\begin{equation}
  - 2\cLRkI (k e^{-\rho})^2 v_k^s=D v_k^s\,.
\label{Cond:GI2}
\end{equation}

Next, we will derive the IR regularity condition for the graviton loop. To compute ${\cal F}^t_{ijkl}(x_1,\,x_2)$, we first rewrite  
$\delta \bar{\gamma}^{ij}_I \cDtx{ij} \zeta_I(x)$ as
\begin{align}
 & \delta \bar{\gamma}^{ij}_I \cDtx{ij} \zeta_I(x) \cr
 & = \delta \bar{\gamma}^{ij}_I \int \frac{\dd^3 \bm{k}}{(2\pi)^{3/2}}
 a_{\sbm{k}} e^{- i\phi(k)} \left[  \frac{\partial}{\partial k^i} k_j e^{i \phi(k)} v_k^s e^{i \sbm{k}
 \cdot \sbm{x}} - e^{i \sbm{k} \cdot \sbm{x}} \frac{\bm{k}_i
 \bm{k}_j}{k^2} ( \cLRkI 2 (k e^{-\rho})^2
 + \bm{k} \cdot \partial_{\sbm{k}} ) e^{i \phi(k)} v_k^s  \right] + (\rm{h.c.})\,,
\end{align}
where the terms multiplied
by $\delta_{ij}$ in the square bracket vanish, being contracted with
$\delta \bar{\gamma}^{ij}$. 
Noticing that 
$\partial/\partial k^i v_k^s= (k^i/k) \partial/\partial k v_k^s$ since $v_k^s$
does not depend on the direction of $\bm{k}$, we find that the terms which potentially yield IRdiv and IRsec due to 
the graviton vanish as 
\begin{align}
 &  \langle \delta \bar{\gamma}^{ij}_I(t_f)
 \delta \bar{\gamma}^{kl}_I(t_f) \rangle {\cal F}^t_{ijkl} (x_1,\, x_2) \cr
 &\,\, =  \langle \delta \bar{\gamma}^{ij}_I(t_f)
 \delta \bar{\gamma}^{kl}_I(t_f) \rangle\, {\cal R}_{x_1} {\cal R}_{x_2}
 \int \frac{\dd^3 \bm{k}}{(2\pi)^3} \frac{\partial}{\partial k^i}  k_j
 \frac{\partial}{\partial k^k} k_l \{ |v_k^s(t_f)|^2 e^{i \sbm{k} \cdot
 (\sbm{x}_1 - \sbm{x}_2)} \}=0 \,, 
\end{align}
if the mode function satisfies
\begin{equation}
  - 2\cLRkI (k e^{-\rho})^2 v_k^s= e^{- i \phi(k)}\, \bm{k} \cdot
   \partial_{\sbm{k}}\, e^{i \phi(k)} v_k^s\,. 
\label{Cond:GI3}
\end{equation}
Thus, if we require 
Eq.~(\ref{Cond:GI3}), we can eliminate the IRdiv and IRsec due to the
graviton loops.

In the case with the isotropic graviton spectrum, 
the IR regularity can be guaranteed if 
the mode function satisfies Eq.~(\ref{Cond:GI2}). 
In fact, when we request the condition~(\ref{Cond:GI2}), 
we find 
\begin{align}
 & \delta \bar{\gamma}^{ij}_I \cDtx{ij} \zeta_I(x) = \delta
 \bar{\gamma}^{ij}_I \int \frac{\dd^3 \bm{k}}{(2\pi)^{3/2}} 
 a_{\sbm{k}} e^{- i \phi(k)} \left[ k^{-3/2}  \frac{\partial}{\partial
 k^i} k^{3/2} k_j e^{i \phi(k)} v_k^s e^{i \sbm{k} \cdot \sbm{x}} \right] + (\rm{h.c.})\,,  
\end{align}
and then the one-loop contribution from the graviton is given by
\begin{align} 
 &  \langle \delta \bar{\gamma}^{ij}_I(t_f)
 \delta \bar{\gamma}^{kl}_I(t_f) \rangle {\cal F}^t_{ijkl} (x_1,\, x_2) \cr
 &\, =  \langle \delta \bar{\gamma}^{ij}_I(t_f)
 \delta \bar{\gamma}^{kl}_I(t_f) \rangle\, {\cal R}_{x_1} {\cal R}_{x_2}
 \int \frac{\dd^3 \bm{k}}{(2\pi)^3} k^{-3} \frac{\partial}{\partial k^i}
 k_j  \frac{\partial}{\partial k^k} k_l \{k^3 |v_k^s(t_f)|^2 e^{i \sbm{k} \cdot
 (\sbm{x}_1 - \sbm{x}_2)} \}\,.  \label{gravitonloop}
\end{align}
Using the following relations
\begin{align}
 & \left[ k^{-3},\, \frac{\partial}{\partial k^i} \right] = 3
 \frac{k_i}{k^5}\,, \qquad 
  \left[ \frac{k_i k_j}{k^5},\, \frac{\partial}{\partial k^l} \right] = 5 \frac{k_i k_j
 k_l}{k^7} - \frac{\delta_{il} k_j + \delta_{jl} k_i }{k^5}\,,
\end{align}
we can rewrite Eq.~(\ref{gravitonloop}) as
\begin{align}
 &  \langle \delta \bar{\gamma}^{ij}_I(t_f)
\delta \bar{\gamma}^{kl}_I(t_f) \rangle {\cal F}^t_{ijkl} (x_1,\, x_2) \cr
 &\, =  \langle \delta \bar{\gamma}^{ij}_I(t_f)
 \delta \bar{\gamma}^{kl}_I(t_f) \rangle\, {\cal R}_{x_1} {\cal R}_{x_2}
 \int \frac{\dd^3 \bm{k}}{(2\pi)^3}  \left\{  \frac{\partial}{\partial
 k^j} \frac{k_i}{k^3} \frac{\partial}{\partial k^l} k_k + 3
 \frac{\partial}{\partial k^l} \frac{k_i k_j k_k}{k^5} \right\}
\{k^3 |v_k^s(t_f)|^2 e^{i \sbm{k} \cdot (\sbm{x}_1 - \sbm{x}_2)} \} \cr
 & \qquad + 3 \langle \delta \bar{\gamma}^{ij}_I(t_f)
 \delta \bar{\gamma}^{kl}_I(t_f) \rangle\, {\cal R}_{x_1} {\cal R}_{x_2}
 \int \frac{\dd^3 \bm{k}}{(2\pi)^3} \frac{k_k}{k^4} \{ 5 k_i k_j k_l -
 k^2 ( \delta_{il} k_j + \delta_{jl} k_i )\} |v_k^s(t_f)|^2 e^{i \sbm{k} \cdot (\sbm{x}_1 - \sbm{x}_2)} \,.
  \label{gravitonloop2}
\end{align}
Using Eq.~(\ref{Exp:spectrumdg}), we can show that the terms in the last
line cancel among them. Then, since the terms in the second
line, which are total derivatives, vanish, we find
that the condition (\ref{Cond:GI2}) can ensure the IR regularity of 
graviton loops as well.

As is pointed out in Ref.~\cite{SRV1}, no mode function can consistently satisfy the IR
regularity conditions (\ref{Cond:GI}) and (\ref{Cond:GI2}), suggesting the necessity to modify the
initial condition (\ref{IC}) and (\ref{IC2}). Apart from that, it is
shown that the same conditions as Eqs.~(\ref{Cond:GI}) and
(\ref{Cond:GI2}) are derived from the requirement that the quantum
states selected operationally in the same way in terms of two different 
canonical variables related by the dilatation transformation 
should agree with 
each other. This is in harmony with our claim that choosing the
Euclidean vacuum which guarantees Eq.~(\ref{Exp:EV}) is crucial for the
IR regularity.

In our previous work~\cite{IRgauge}, we computed the one-loop
contribution of the graviton in 
the two-point function of $\gR(x)$, which can be expressed in the
form ${\cal R}_x \gz(x)$ by neglecting the terms which do not contribute
to the IRdiv nor IRsec. Then, we claimed that
the one-loop contribution in the two-point function of
$\gR(x)$ becomes IR regular without restricting the mode function
$v_k^s$. 
However, in
Ref.~\cite{IRgauge}, to compute the graviton loop, we adopted 
\begin{align}
 & \zeta_2 \Approx\,  \cdots \,  -  \cLsI \delta \bar{\gamma}^{ij}_I
 \nabla_i \nabla_j \zeta_I \Approx\,  \cdots \, + \frac{1}{2} \delta \bar{\gamma}^{ij}_I x_i
 \partial_j \zeta_I   \label{Sol:zeta2ph}
\end{align}
as the solution for Eq.~(\ref{EOMzeta}), 
where ellipses represent 
the terms which do not include $\delta \gamma_{ij\,I}$. 
Notice that in Eq.~(\ref{Sol:zeta2ph}), the solution which satisfies
$$
 2 \cLsI \delta \bar{\gamma}^{ij}_I
 \nabla_i \nabla_j \zeta_I = - \delta \bar{\gamma}^{ij}_I x_i
 \partial_j \zeta_I\,,
$$
is selected. Based on the discussion after Eq.~(\ref{Cond:regheu2}), 
we find that this
solution cannot be obtained by using the retarded Green
function, $\cLRsI$, with the 
initial conditions~(\ref{IC}) and (\ref{IC2}). 
Therefore, in order to 
eliminate the IRdiv and IRsec
from the graviton loops for an arbitrary mode function $v_k^s$, we need
to abandon the initial conditions~(\ref{IC}) and (\ref{IC2}). 
Then, however, there is no guarantee any more that 
the standard commutation relations hold also for the
interaction picture fields.

\end{document}